\documentclass[twocolumn]{aastex631}
\usepackage{CJK}
\usepackage{enumitem}
\usepackage{lipsum}
\usepackage{amsmath}
\usepackage{physics}
\usepackage{tablefootnote}
\usepackage{supertabular}
\usepackage{pifont}
\usepackage{multirow}

\newcommand{\Ni}{$^{56}$Ni}
\newcommand{\Co}{$^{56}$Co}

\newcommand{\Ms}{M$_{\odot}$}

\newcommand      \grays       {$\gamma$-rays}

\newcommand{\til}{$\sim$}

\newcommand{\mdot}{$\dot{\mathrm{m}}$}

\def\gsim{\mathrel{\raise.5ex\hbox{$>$}\mkern-14mu
             \lower0.6ex\hbox{$\sim$}}}

\def\lsim{\mathrel{\raise.3ex\hbox{$<$}\mkern-14mu
             \lower0.6ex\hbox{$\sim$}}}

\newlength\mylen
\newlist{mycases}{enumerate}{1}
\setlist[mycases,1]{label=\textbf{Case~\arabic*.}, 
labelwidth=\dimexpr-\mylen-\labelsep\relax,leftmargin=0pt}         

\newlist{newcases}{enumerate}{1}
\setlist[newcases,1]{
  label={\textbf{Case~\arabic*}.},
  leftmargin=*,
  align=left,
  labelsep=1mm,
  itemindent=\dimexpr\labelsep+\labelwidth+10pt\relax
}

\begin{document}
\title{Dust Production in a Thin Dense Shell in Supernovae with Early Circumstellar Interactions}
\shorttitle{Dust in interacting supernovae}
\author{Arkaprabha Sarangi}
\affiliation{DARK, Niels Bohr Institute, University of Copenhagen, Jagtvej 128, 2200 Copenhagen, Denmark}
\email{sarangi@nbi.ku.dk}

\author{Jonathan Slavin}
\affiliation{Center for Astrophysics | Harvard \& Smithsonian, 60 Garden Street, Cambridge, MA 02138, USA}



\begin{abstract}

In supernovae (SNe), where the light curves show evidence of strong and early interaction between the ejecta and the circumstellar matter (CSM), the formation of new dust is estimated to take place in a dense shell of gas between the forward (FS) and the reverse shock (RS). For the first time, in this study, the mechanism of dust formation in this dense shell is modeled. A set of 9 cases, considering variations of the  ejecta mass, and the pre-explosion mass-loss rates is considered, accounting for the diverse nature of interactions reported in such SNe. For a single main sequence mass, the variation of ejecta mass was manifested as a variation of the H-shell mass of the star, lost due to pre-explosion mass-loss. 
We find that the dust masses in the dense shell range between 10$^{-3}$~\Ms\ and 0.8~\Ms, composed of O-rich and C-rich grains, whose relative proportions are determined by the nature of interaction. Dust formation in the post-shock gas is characterized by a gradual production rate, mostly ranging from 10$^{-6}$ to 10$^{-3}$ \Ms\ day$^{-1}$, which may continue for a decade, post-explosion.
A higher mass-loss rate leads to a larger mass of dust, while a smaller ejecta mass (smaller left-over H-shell) increases the efficiency of dust production in such SNe.  
Dust formed behind the RS, as in our calculations, is not subject to destruction by either the FS or RS and is thus likely to survive in larger proportion than dust formed in the ejecta. 


\end{abstract}

\section{Introduction}
\label{intro}

Among the subclasses of core-collapse supernovae (SNe), the Type IIn SNe, identified by bright and narrow Balmer lines of H in their spectra, and the Type Ibn, identified by narrow lines of He and relatively weak H lines, are broadly categorize as interacting SNe \citep{fil97,smi08}. Specifically, this SN category is characterized by the interaction of the fast-moving ejecta with a slow circumstellar medium (CSM) of sufficient density so that the resultant energy generated by the interaction dominates the SN light curve \citep{smith_2017}. They are associated with very high pre-explosion mass-loss rates, which can range between $\sim$ 10$^{-1}$ to 10$^{-4}$ \Ms\ ~yr$^{-1}$ \citep{fox11, tad13, mor14}.

As in all SNe, the interaction of the expanding ejecta with the CSM generates a forward shock (FS) passing through the CSM and a reverse shock (RS) propagating inward through the ejecta \citep{nymark_2006, reynolds_2017, chevalier_2017, tsuna_2019, milis_2008}. In the standard case of supernova remnant (SNR) evolution, both the FS and RS are fast adiabatic shocks. In contrast, for the so-called interacting SNe, the density is so high that radiative cooling is important at early times and a thin dense shell forms in the rapidly cooling gas between the FS and the RS at the interaction front \citep{chu04, moriya_2013, sarangi2018}.

Dust has commonly featured in IR observations of interacting SNe over the past two decades \citep{poz04, smith_2009,fox11, szalai_2019}. This dust could have formed in the winds during the pre-explosion phase of the star and survived the shock-breakout at the time of explosion. Alternatively, it could be newly formed dust, that has condensed either in the SN ejecta or in the CSM, after the explosion. 
Confirmed evidence of new dust formation after the explosion has been reported in objects like SN 2006jc, SN~2005ip, SN~1998S, SN~2010jl, SN~2006tf, SN~1980K, SN~1999em and others \citep{smith_2008, smith_2009, poz04, gal14, smith_2017, bevan_2019, fox11, mauerhan_2012, bevan_2020}. 

The masses of dust in these SNe vary widely between 10$^{-5}$ and 0.1 \Ms\ \citep{fox11}. In the case of SN~2005ip, SN~2006jc, or SN~2010jl, the formation of dust is estimated to commence within the first 100 days \citep{smith_2008, mattila_2008, smith_2009, gal14}, generally attributed to have formed in a dense shell. In contrast, the rebrightening of SN~2013ej, between 700 and 1000 days is also assumed to be correlated to dust formation in a dense shell of gas behind the RS \citep{mauerhan_2017}. Late time strengthening of He and O lines in SN~1998S around day 5000, along with more
pronounced mid-IR features are attributed to late time dust formation \citep{mauerhan_2012}. Accompanying this late variation are contrasting rates of decline of the mid-IR fluxes over time, for such SNe \citep{szalai_2019}.

From a theoretical perspective a core-collapse SN remnant, evolving in a typical ISM density, evolves through several phases starting with free-expansion, followed by an adiabatic Sedov-Taylor phase and a radiative snow-plow phase
\citep{weiler_1988, micelotta2018}. The analytical model of \cite{tru99} connected the evolution of the non-radiative SN remnant from the free-expansion and through the Sedov-Taylor phase, based on three parameters: the energy of the explosion, the mass of the ejecta and the density of the ambient medium.
For the standard case of SNR evolution, the Sedov-Taylor  phase ends when the shock speed slows enough that radiative cooling becomes important. At that point, the remnant enters the radiative or snowplow phase. For evolution in a typical interstellar medium (ISM) density, this transition does not occur until thousands of years after the explosion. 
The kinetics of the radiative phase were addressed by \cite{mckee_1977} and subsequently by \cite{cioffi_1988} modeling the evolution of the SNR through the pressure-driven and momentum conserving phase.  

SNR shocks have long been recognized as the primary destroyers of dust grains in the ISM \citep{dra79,Jones_etal_1994,Jones_etal_1996,Slavin_etal_2015}. At the same time, SNe are known to produce dust grains in their dense ejecta. A still unresolved question is how much of the newly formed grains can survive destruction by the reverse shock and escape into the ISM \citep{Micelotta_2016,bis16,florian_2019,slavin_2020}. Grains that are produced between the reverse shock and forward shock could thus play an important role, since they are not subject to shock destruction shortly after formation.


The formation of dust in between the forward and reverse shock does not occur in remnants expanding into a typical ISM density or a low density CSM because the gas in that region is hot and relatively low density. However, Type IIn or Ibn SNe evolve in a very high density CSM, and the ejecta-CSM interaction is immediately followed by a high rate of radiative energy loss. As a result, the region between the RS and FS quickly becomes cold and dense. In this case, the adiabatic conditions required for the self-similar solutions as described in \citet{tru99} are not achieved. Using the conservation of momentum, \cite{moriya_2013} addresses the evolution of the FS and RS in such SN that leads to the formation of a thin, dense shell. 

Modeling the evolution of the FS through the dense CSM in Type IIn SN~2010jl, \cite{sarangi2018} showed that evolving physical conditions in the post-shock gas provide new and efficient pathways to produce dust. Analyzing the IR emission from the newly formed dust, it was estimated that the rapidly cooling shocked gas is more likely to be the site of new dust formation, compared to the unshocked ejecta. 

In this paper, we present the first complete model for dust formation in the thin dense shell formed between the FS and the RS in interacting SNe, consisting of the shocked metal-rich ejecta (by RS interaction) and the shocked CSM (by FS interaction). As the output of the study, we calculate the masses of different dust components as a function of post-explosion time, and their corresponding rates of formation. 

To account for the large contrast in interaction characteristics \citep{smith_2017}, leading to the variation in dust masses, dust forming epoch, and dust composition \citep{fox11, szalai_2019}, we present 9 different cases, with variation in the mass of the ejecta and the pre-explosion mass-loss rate. 


The organization of this paper is as follows. 
In Section 
\ref{sec_geometry}, the geometry of the CSM-ejecta system is introduced. 
Section \ref{sec_shock_evol} explains the evolution of the SN ejecta and how the FS and RS evolves in these SNe, 
while in Section \ref{sec_parameters} the parameters relevant for this study are listed. 
Then in Section \ref{sec_cooling}, we present the cooling and heating mechanisms of the shocked gas, and the evolution of the physical conditions on that basis. 
Section \ref{sec_chemistry} addresses the formalism used to model the chemistry of dust condensation.  
After that, in Section \ref{sec_dustmasses}, the evolution of dust masses is presented as a function of post-explosion time for all 9 cases, with focus on the dust formation rates (Section \ref{sec_dustformationrate}) and the dependence on the parameter space (Section \ref{sec_massloss}, \ref{sec_ejectamass}).  
Finally, in Section \ref{sec_summary} we summarize the results and make some concluding remarks. 

\section{The SN ejecta and the inter-shock region}
\label{sec_geometry}

\begin{figure}
\vspace*{0.3cm}
\centering
\includegraphics[width=3.5in]{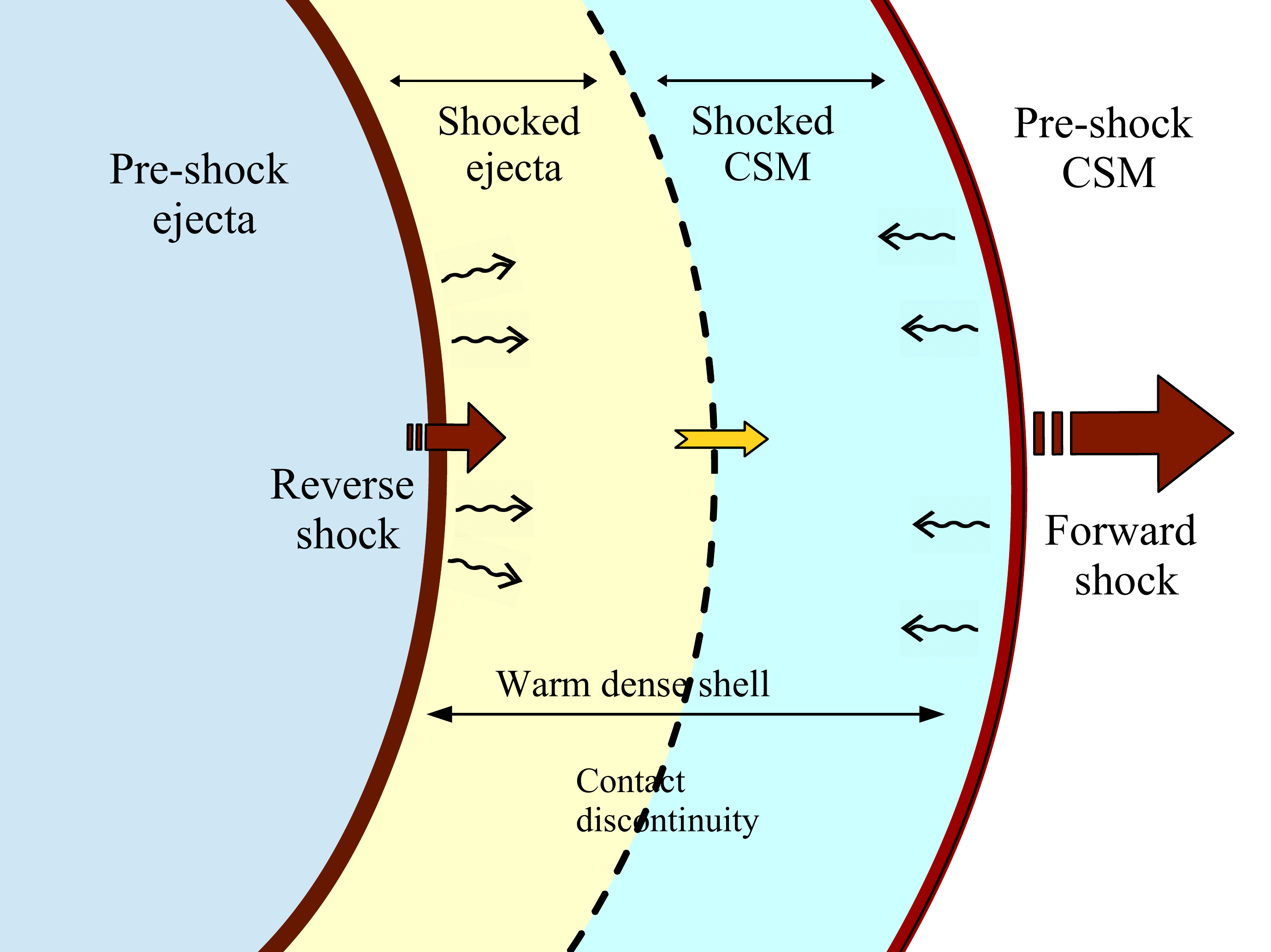}
\caption{\label{fig_geometry} A schematic diagram of the geometry of an interacting SNe is shown in the frame of the ISM. The remnant is stratified (outside to inside) into the pre-shock CSM, the shocked CSM gas, the shocked SN ejecta, and the freely expanding pre-shock ejecta. The FS traverses through the CSM while the RS travels through the ejecta. The reverse shock is moving outward in the rest-frame (rest-frame is the unperturbed ISM) as the figure shows, but it is moving inward relative to the expanding ejecta, so the region marked in yellow, is actually already behind the RS, even though the direction of the RS in the rest-frame shown by the arrow might create some confusion. The wavy arrows indicate the presence of downstream radiation from the shock front towards the cooling shocked gas (see Section \ref{sec_cooling}), that is made up of the shocked ejecta and the shocked CSM, together forming a warm dense shell that is shown in the figure (see Section \ref{sec_evolTn}).   Note that the intershock region, that is marked as the dense shell, is exaggerated in width. In reality, cooling causes the region to collapse to a very thin dense shell.}
\end{figure}

A basic schematic picture of the interacting SN, for a given epoch, in presented in Figure \ref{fig_geometry}.  
The distinct layers in the shell-ejecta morphology, along the radially outward direction, are described as follows: 
\begin{enumerate}[label=(\alph*),noitemsep, leftmargin=*]
\item the freely expanding ejecta powered by the explosion energy of the SN, 
\item the RS traversing through the SN ejecta, 
\item the shocked ejecta and the post-shock cooling layers, 
\item the shocked CSM cooling radiatively behind the FS,
\item the FS propagating outwards through the CSM, 
\item the CSM gas that is yet to be crossed by the FS
\end{enumerate} 

In rest-frame (rest-frame is the unperturbed ISM) the RS is moving radially outwards, but more slowly than the unshocked ejecta (hence the arrow-directions in the figure). Therefore, in the frame of the ejecta, the RS is moving inwards. The ejecta and the CSM are separated by a contact discontinuity. However, observationally a clear demarcation does not exist, rather the rapidly cooling intershock gas (between the FS and RS) behaves like a single shell, often called the thin dense shell. The focus of the study is the chemistry of this post-shock gas. In this study, we have not included clumps or the Rayleigh-Taylor instabilities associated with the contact discontinuity \citep{buchler_1980, duffell_2016}. The ejecta is considered to be stratified into layers, based on the abundance of elements, as will be explained in the subsequent Section \ref{sec_parameters}. 



In such scenario, dust can be present in all four regions, i.e., (i) pre-existing dust in the unshocked CSM beyond the evaporation radius, (ii) newly formed dust in the shocked CSM, (iii) newly formed dust in the shocked ejecta, and (iv) newly formed dust in the unshocked ejecta. It is often difficult to assess the accurate origin for the dust, though asymmetries in emission lines or decline in optical fluxes can be used to differentiate between newly formed and pre-existing dust (see the review by \citealt{fox11}). 

\subsection{The evolution of the ejecta and CSM}
\label{sec_shock_evol}
In this section, the dynamics of the ejecta and CSM are explained with respect to the propagation of the FS and RS, based on the formalism by \cite{tru99} and \cite{moriya_2013}. For details of the mathematical formulation, please refer to the Section \ref{appendix} (Appendix). 

After the explosion, the SN ejecta undergoes homologous expansion ($v \cong r/t$). We assume the ejecta to be made up of a core, which is at a uniform density, and an envelope, where the density falls rapidly as a power law \citep{tru99, dwa07, moriya_2013, micelotta2018}. Following the treatment by \cite{tru99}, we define the density of the ejecta core ($\rho_c$) and ejecta envelope ($\rho_e$) as (see Equation~\eqref{eqn_f0_rho} for the derivation),  
\begin{equation}
\label{eqn_ejecta_density}
\begin{gathered}
\rho_{c}(v,t) = \frac{3}{4 \pi} \frac{n-3}{n} \frac{M_{ej}}{v_{c}^3} t^{-3} \\
\rho_{e}(v,t) = \frac{3}{4 \pi} \frac{n-3}{n} \frac{M_{ej}}{v_{c}^3} \Big(\frac{v_c}{v}\Big)^n t^{-3}
\end{gathered}
\end{equation}
where $M_{ej}$ is the mass of the ejecta, $n$ is the power-law exponent for the density of the envelope, and $v_c$ is the velocity of the outer boundary of the ejecta-core. This can also be expressed as density at any given radius $r$, where $r = vt$, $v$ being the velocity of the ejecta. 

For the density of CSM, $\rho_{csm}$, we assume a steady, constant velocity, mass-loss rate, therefore the density declines with radius as a power-law with exponent $s = -2$. Taking the velocity of the pre-explosion wind to be $v_w$, we can write the density at a given radius $r$ as, 

\begin{equation}
\label{csm_ejecta_density}
\dot{m} = 4 \pi r^2 \rho_{csm} v_w, \ \ \  \rho_{csm} = \frac{\dot{m}}{4 \pi v_w} r^{-2}
\end{equation}

The velocity of the ejecta-core, $v_c$ (shown by \citealt{tru99}), is given by (see Equation \eqref{eqn_corevelocity}), 
\begin{equation}
\label{eqn_ejectacore}
v_c = \Big[\frac{10(n-5)}{3(n-3)} \frac{E}{M_{ej}}\Big]^{1/2},
\end{equation}
where E is the explosion energy of the supernova. 

To study the evolution of FS and the RS through the CSM and the ejecta, we have adopted the formalism by \cite{moriya_2013}. In case of interacting supernovae, the shocked gas, made up of the adjacent layers of shocked CSM (by the FS) and the shocked ejecta (by the RS), together forms a very thin dense shell of rapidly cooling gas. Therefore, the radius of the FS and the RS can be denoted by a single quantity, $r_{FS} (t)$ \citep{moriya_2013}. This is also confirmed when we study the cooling of the post-shock gas in the following section, and calculate the actual radial distance between the two shocks. 

If the total mass of the shocked gas is $M_{sh}$ and the FS velocity is $v_{FS}$ (= d$r_{FS}$/d$t$), it can be shown (see Equation~\ref{eqn_momentum_csm}) from the conservation of momentum \citep{moriya_2013} that, 
\begin{equation}
\label{eqn_momentum}
M_{sh}\frac{d v_{FS}}{d\mathrm{t}} = 4 \pi r_{FS}^2 [\rho_{ej}(v - v_{FS})^2 - \rho_{csm} v_{FS}^2] 
\end{equation}
While the FS shock propagates through the CSM, the RS initially propagates through the ejecta-envelope and thereafter through the ejecta-core. The time at which the RS reaches the boundary of the core and envelope is denoted by $t_c$. As shown by \cite{moriya_2013}, for time $t\leq t_c$, that is when the RS is passing through the envelope, the solution for the momentum conservation (Equation~\ref{eqn_momentum}) is given by (see Equation~\ref{eqn_parts_of_differential}, \ref{eqn_omega_beta}, \ref{eqn_sh_radius}), 

\begin{equation}
\label{eqn_r_tc}
\begin{gathered}
r_{FS}(t) = \Big[\frac{6 M_{ej} v_c^{n-3}}{n(n-4)}\frac{v_w}{\dot{m}}\Big]^{\frac{1}{n-2}} t^{\frac{n-3}{n-2}} \\
t_c = \frac{6 M_{ej}}{n(n-4)v_c} \frac{v_w}{\dot{m}}
\end{gathered}
\end{equation}
In the same way following  \cite{moriya_2013}, for time $t > t_c$, Equation~\ref{eqn_momentum} can be expressed (see Equation~\ref{eqn_momentum_core}) as, 
 \begin{equation}
\label{eqn_core_conserv}
\begin{gathered} 
\left[\frac{\dot{m}}{v_w} r_{FS} + M_{ej} - \frac{n-3}{n} \frac{M}{v_c^3} \left(\frac{r_{FS}}{t}\right)^{3}\right]\dv[2]{r_{FS}}{t} = \\
 = \frac{3(n-3)}{n} \frac{M}{v_c^3} r_{FS}^{2} t^{-3} \left(\frac{r_{FS}}{t} -   \dv{r_{FS}}{t} \right)^2 - \frac{\dot{m}}{v_w} \left(\dv{r_{FS}}{t}\right)^2  
 \end{gathered}
 \end{equation}

Using the boundary condition of $r_{FS}(t = t_c)$ and $v_{FS}(t = t_c)$, we solve the above equation numerically, to find $r_{FS}(t > t_c)$. 

Owing to the radiative nature of shocks in interacting SNe, the kinetic energy generated by the interaction is efficiently transformed into the luminosity powering the light curve \citep{smith_2017}. Also, the luminosity generated by shock-CSM interaction in SN~2010jl matches well with an efficient conversion of the kinetic energy \citep{fra14, gal14, sarangi2018}. Based on that, we take the luminosity of the FS and RS as, 
\begin{equation}
\label{eqn_luminosity}
\begin{gathered}
L(t) = L_{\rm{FS}} (t) + L_{\rm{RS}} (t) \\
= \frac{1}{2} \dv{m_{csm}(t)}{t} v_{FS}^2(t) + \frac{1}{2} \dv{m_{ej}(t)}{t} v_{RS}^2(t) \\
= 2 \pi r_{{FS}}^2 \big[ \rho_{csm}(r_{{FS}}, t) v_{FS}^3(t) +  \rho_{ej}(r_{{FS}}, t) v_{RS}^3(t)\big] 
\end{gathered}
\end{equation}
where $\rho_{ej} (r, t)$ is the density of the ejecta at the front of
interaction, and the RS velocity $v_{RS} = v_{FS} - v$.

\subsection{Choice of parameters} 
\label{sec_parameters}

\begin{figure*}
\vspace*{0.3cm}
\centering
\includegraphics[width=3.5in]{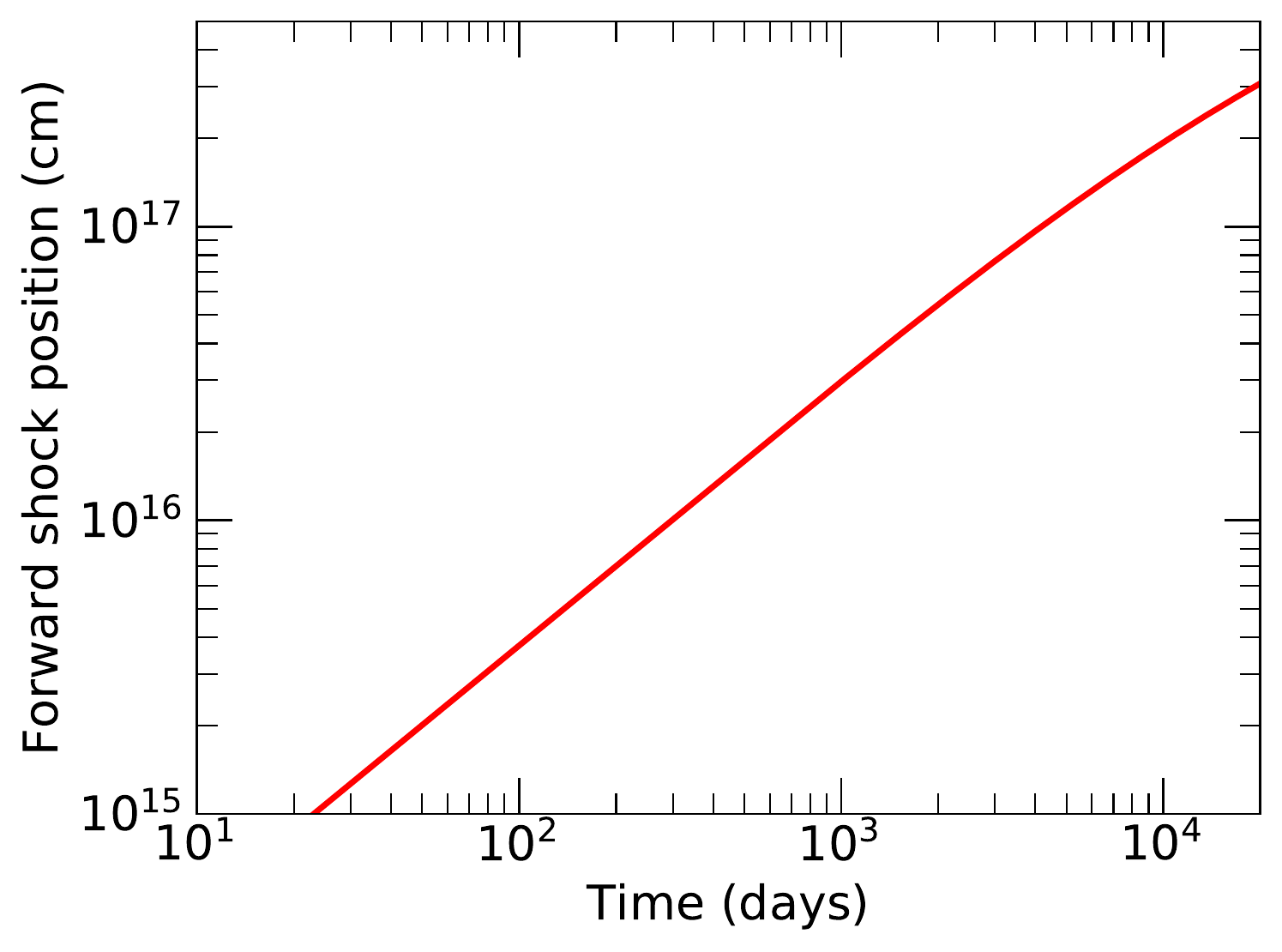}
\includegraphics[width=3.5in]{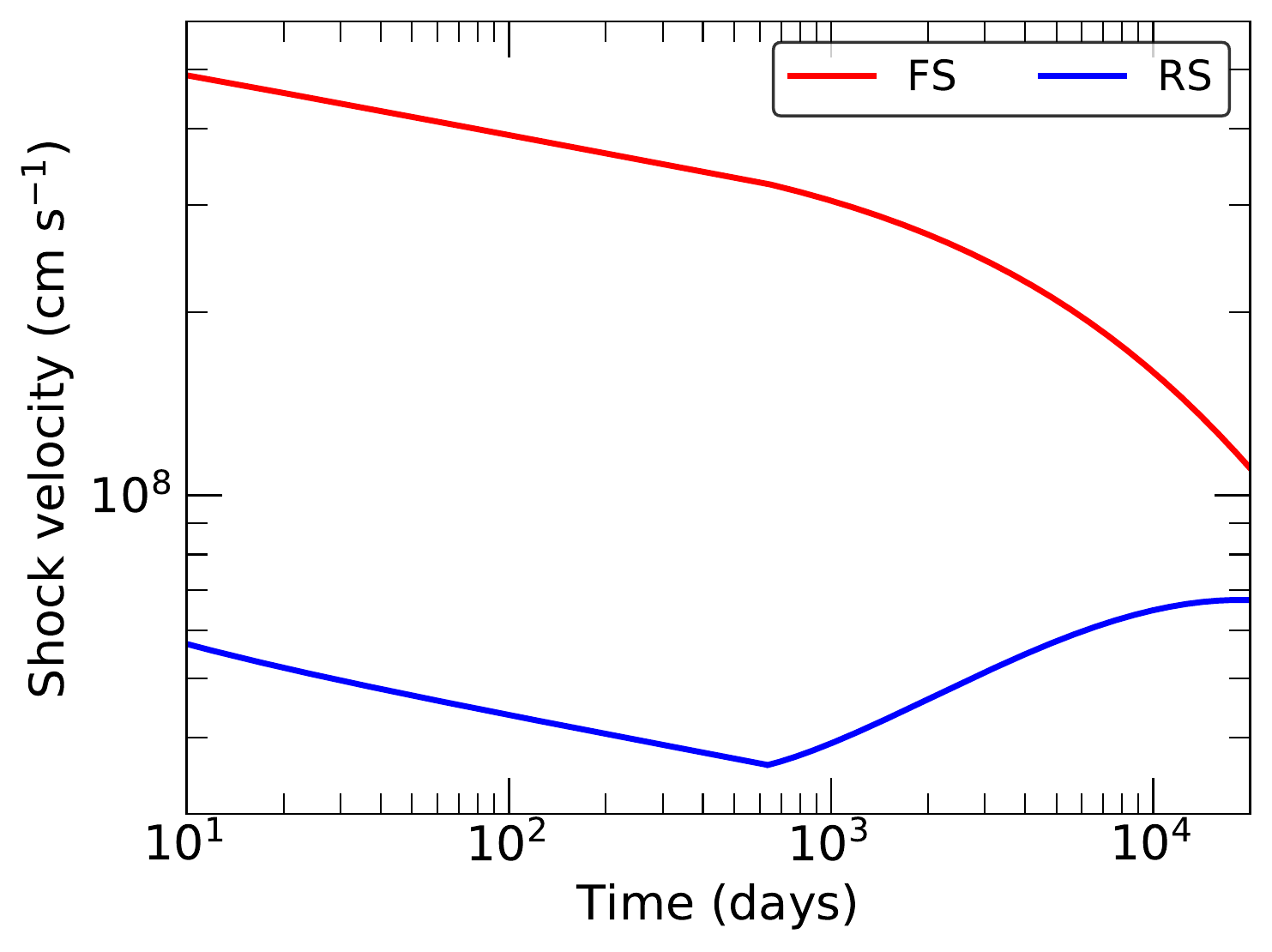}
\includegraphics[width=3.5in]{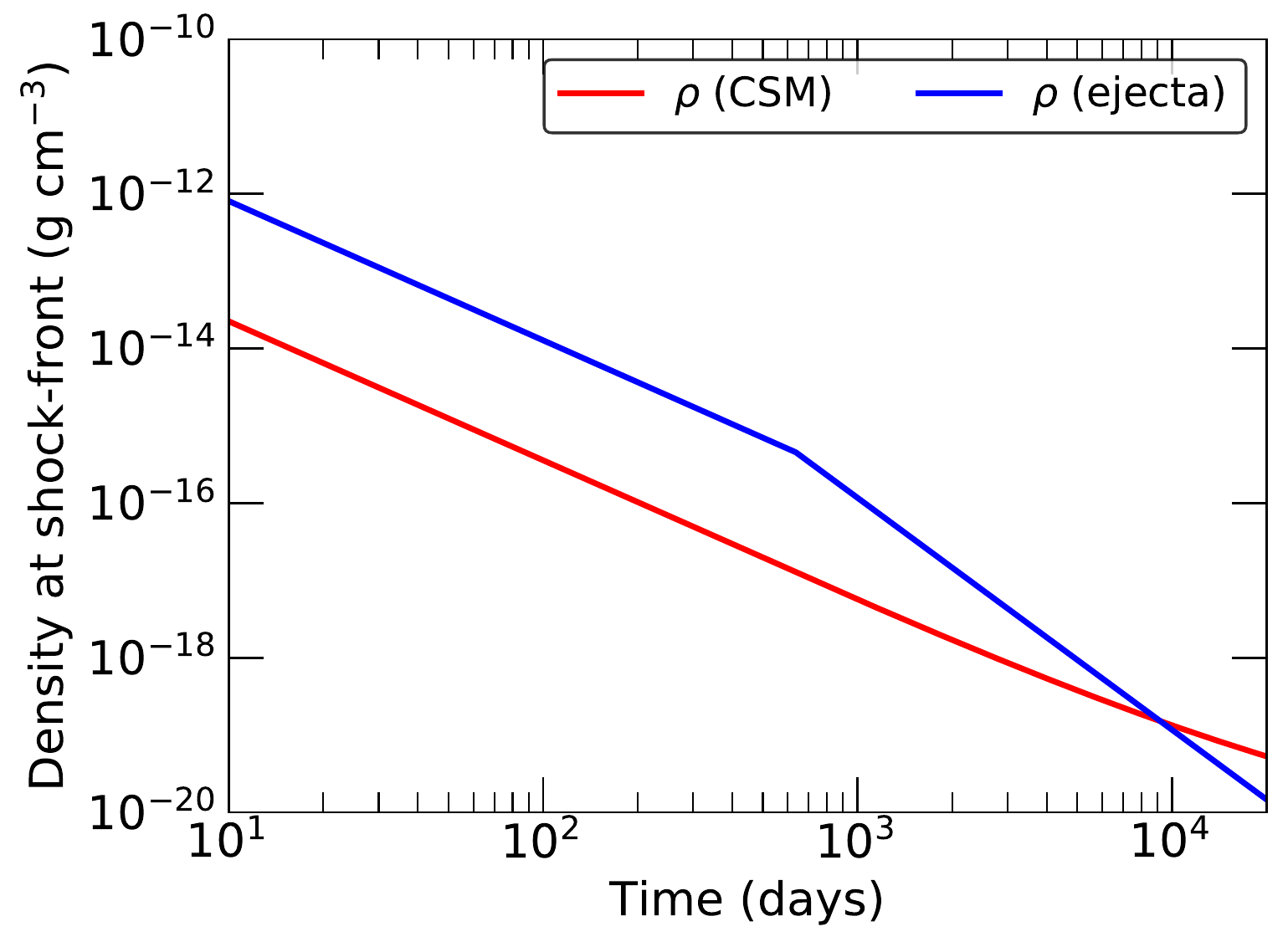}
\includegraphics[width=3.5in]{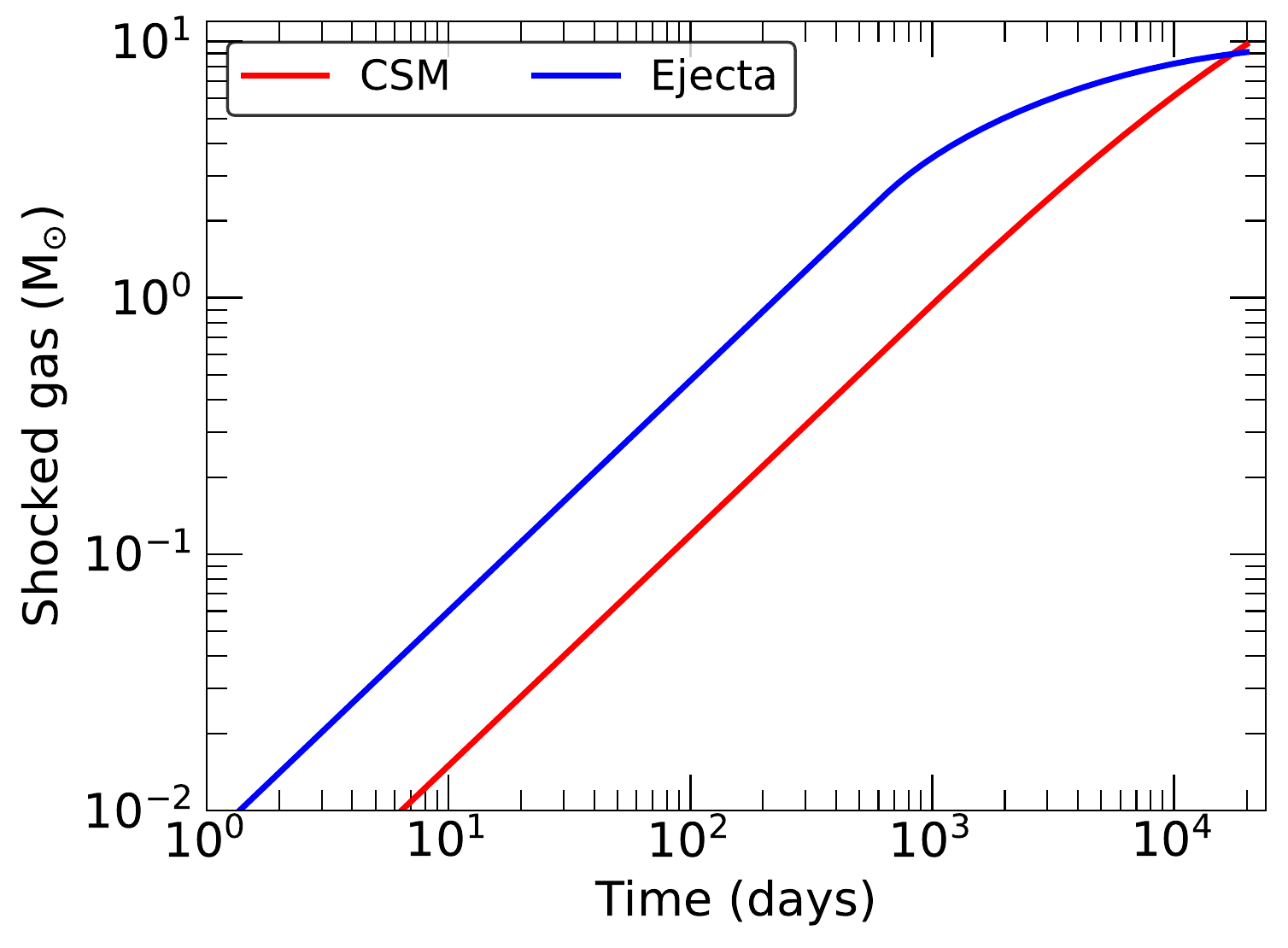}
\caption{\label{fig_R_cd} Based on the equations in Section~\ref{sec_shock_evol}, the following parameters are shown as a function of post-explosion time for the benchmark case (\mdot\ = 10$^{-2}$ \Ms\ yr$^{-1}$, M$_{ej}$ = 10 \Ms) In this scenario, the transition time $t_c$ (see Equation \ref{eqn_r_tc}) is day 632, which reflects to the time when the RS enters the core of the ejecta from the envelope.  \textit{Top-left:} The FS radius is shown, which is the same as the RS position, since the FS and RS are separated by a very thin shell only. \textit{Top-right:} The FS and RS velocity are presented, where the FS velocity, v$_{FS}$, is with respect to the rest-frame (ISM), while the RS velocity, v$_{RS}$, is the relative velocity with respect to the ejecta velocity at that given position and time. The change in the profile of the RS velocity at day 632 refers to the RS entering the core of the ejecta and the FS slowing down, so that the relative velocity with respect to the ejecta velocity increases.  \textit{Bottom-left:} The pre-shock gas densities of the CSM and the ejecta at the shock-front is shown \textit{Bottom-right:} The mass of the shocked ejecta and the shocked CSM gas is presented, which together forms the dense shell, that is the site of dust formation in our study.}
\end{figure*}

To study the evolution of the shocked gas, the properties of the shock itself and the pre-shock gas need to be quantified. Equation~\eqref{eqn_r_tc} and \eqref{eqn_core_conserv} indicate that these properties depend on the following free parameters only: $M_{ej}, \dot{m}, E, v_w$, and $n$ (assuming a steady pre-explosion mass-loss).

In this study, we have created an array of cases, showing how the interaction with the FS and RS can influence the formation of dust in different scenarios, addressing the diverse nature of observational findings stated in Section \ref{intro}. 

Specifically, we have chosen a single main sequence star, and altered two parameters, (a) how much mass did the star loose through its pre-explosion mass-loss activities, and, (b) at what rate was the mass lost. The mass of the ejecta at the time of explosion, $M_{ej}$ can be correlated to the mass that was lost prior to explosion, since mass at main sequence, $M_{ms}$, is the sum of the two (added to the residual neutron star). On the other hand, the rate of mass-loss is basically $\dot{m}$ in our formalism.

For a star of main sequence mass of 21~\Ms \citep{rau02}, 
we have modeled 9 different scenarios, with the variation of the mass of the ejecta as $M_{ej} = 6, 10, 15$ \Ms, and the pre-explosion mass-loss rates as $\dot{m} = 10^{-1}, 10^{-2}, 10^{-3}$ \Ms\ yr$^{-1}$, which are consistent with Type IIn-like SNe \citep{fox11, mor14}. The mass of the exploded He-core is about 5~\Ms, and a residual neutron star of 1.75 \Ms.

Importantly, the variation of the ejecta mass is manifested as the mass of hydrogen shell present on top of the star at the time of explosion. In other words, the differences in ejecta mass indicate the level of H stripping before the explosion. The total mass of the He-core remains unchanged at 5 \Ms\ for all the cases, while the mass of the H shell is chosen to be 1, 5 and 10 \Ms\ (for $M_{ej} = 6, 10, 15$ \Ms\ respectively). 

We use a coarse stratification of the ejecta into 4 layers or zones, based on the compositions from the nucleosynthesis models by \citet{rau02}, of a 21~\Ms\ progenitor, with mass of the exploded He core (Si/Fe zone, O/Si/C zone and He/C zone from Table \ref{table_initial}) of about 5.0 \Ms\ (and the remaining being the H shell), following a similar approach used in \citet{sar13, sar15}. The mass-fractions of the important elements are shown in Table \ref{table_initial}. There are, of course, other metals in the ejecta (of lesser abundances), like Al, Ni, Mg, S, etc., that play important roles in dust synthesis; for the list of all abundances please refer to the dataset in the footnote \citep{rau02}\footnote{\url{https://nucastro.org/nucleosynthesis/expl_comp.html}}. In this stratification, the entire CSM and the H/He zone in the ejecta have a C/O ratio smaller than 1, barring a small mass of $\sim 1$~\Ms\ of gas adjacent to the He-core which is mildly C rich. In other words, if  $M_{ej} = 15$ \Ms, it is made up of 5 \Ms\ of He core, and  10 \Ms\ of H/He shell, of which the inner 1~\Ms\ is C rich while the outer 9~\Ms\ is O rich. 

For all the cases, the explosion energy $E$ was taken as 10$^{51}$ ergs s$^{-1}$. The power-law exponent for the ejecta envelope $n$ was taken as 12 \citep{matzner_1999} and the pre-explosion wind velocity was taken as 100~km~s$^{-1}$ \citep{moriya_2013}. In addition, we have considered the CSM to extend up to a maximum mass of 10~\Ms, with reference to SN~2010jl \citep{sarangi2018, cha15}.  

\begin{table}
\centering
\caption{Ejecta compositions}
\label{table_initial} 
\begin{tabular}{lcc}
\hline \hline
Zone & Mass (\Ms) & Mass fractions (\%) \\
\hline
Si/Fe & 0.2 & 51 Si, 38 Fe, 11 O \\
O/Si/C & 3.8 & 85 O, 5 Si, 9 C, 1 Mg \\
He/C & 1.0 & 95.8 He, 4 C, 0.2 O \\
H/He & 10, 5, 1 & 71 H, 29 He \\
\hline \hline
\end{tabular}
\end{table}


\begin{figure}
\vspace*{0.3cm}
\centering
\includegraphics[width=3.5in]{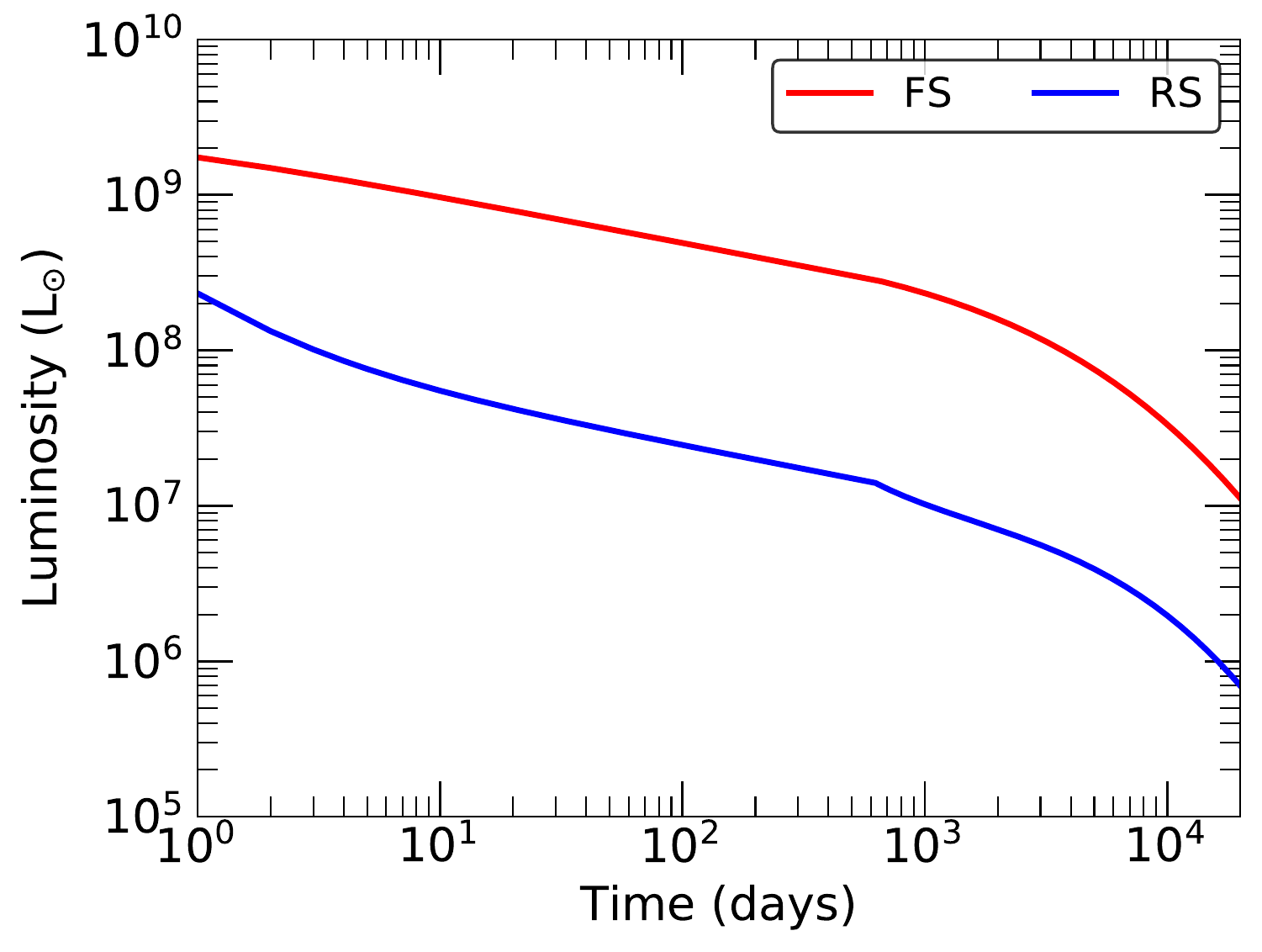}
\caption{\label{fig_luminosity}The figure shows the evolution of the FS and RS luminosity, generated by the interaction of the shock with the CSM or the ejecta, for the benchmark case, following Equation~\ref{eqn_luminosity}. } 
\end{figure}

Let us choose a case where, $M_{ej} = 10$ \Ms, and $\dot{m} = 10^{-2}$ \Ms\ yr$^{-1}$ as a benchmark case. Figure~\ref{fig_R_cd} presents the shock radius $r_{FS}$ (Equation \ref{eqn_r_tc}, \ref{eqn_core_conserv}), the FS and the RS velocity, and the density at $r_{FS}$ of the CSM and the ejecta, and the mass of the shocked ejecta and CSM, for the benchmark case. The total luminosity (Equation \ref{eqn_luminosity}) generated by the FS-CSM and RS-ejecta interaction is shown in Figure~\ref{fig_luminosity}. In this case, the transition time $t_c$ (when the reverse shock starts to penetrate into the ejecta core) is 632 days, which is evident from the figure. Based in these parameters, we now can parametrize the cooling of the post-shock gas, and derive the suitable conditions for the formation of the thin dense shell.

\begin{figure}
\vspace*{0.3cm}
\centering
\includegraphics[width=3.5in]{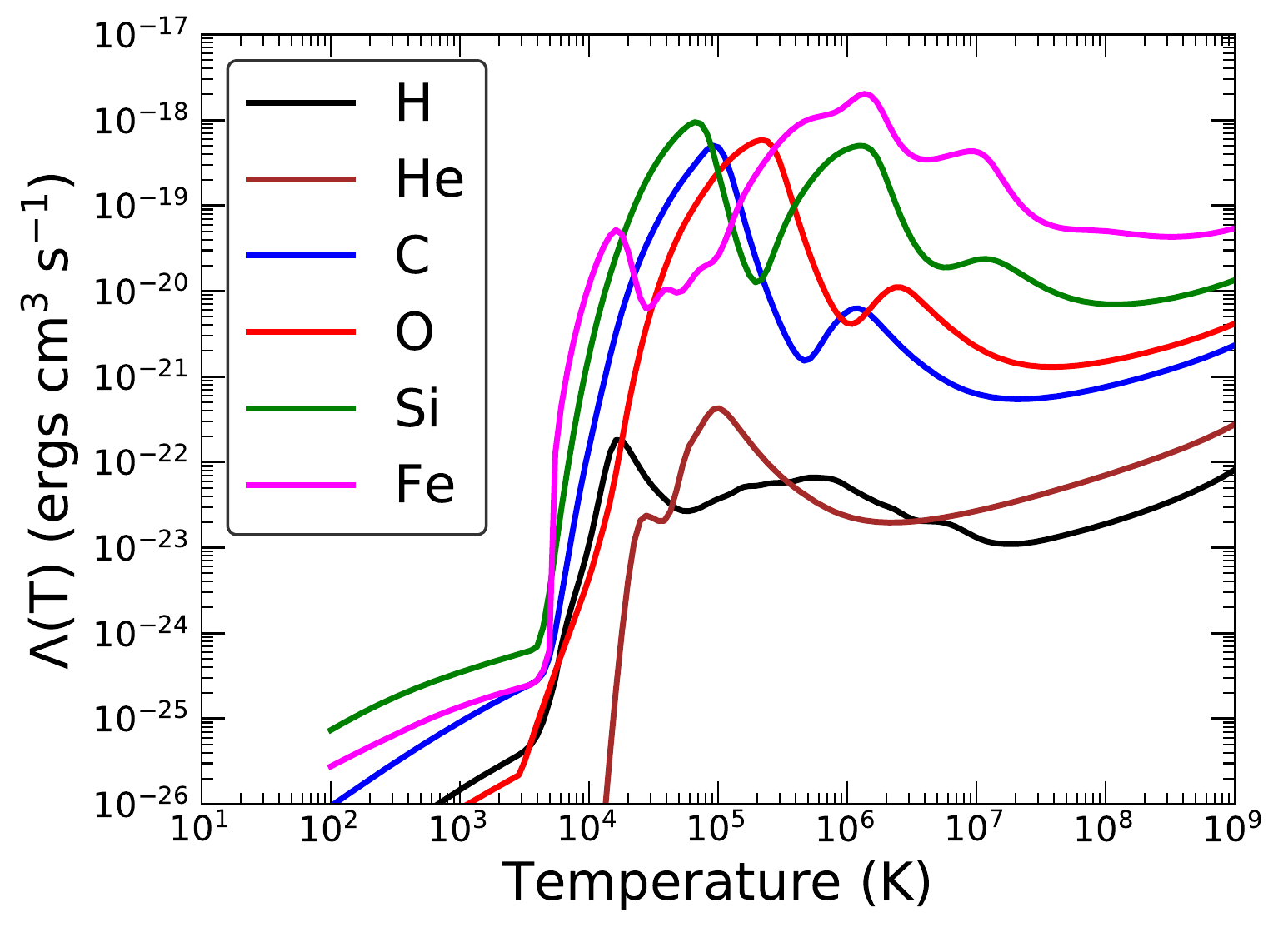}
\caption{\label{fig_coolingfunc} The individual cooling functions, $\Lambda_i$(T), are presented for H, He, O, C, Si and Fe, which are the main constituents of the SN ejecta. Please see Section \ref{sec_cooling} for the references and the details. }
\end{figure}

\begin{figure*}
\vspace*{0.3cm}
\centering
\includegraphics[width=3.5in]{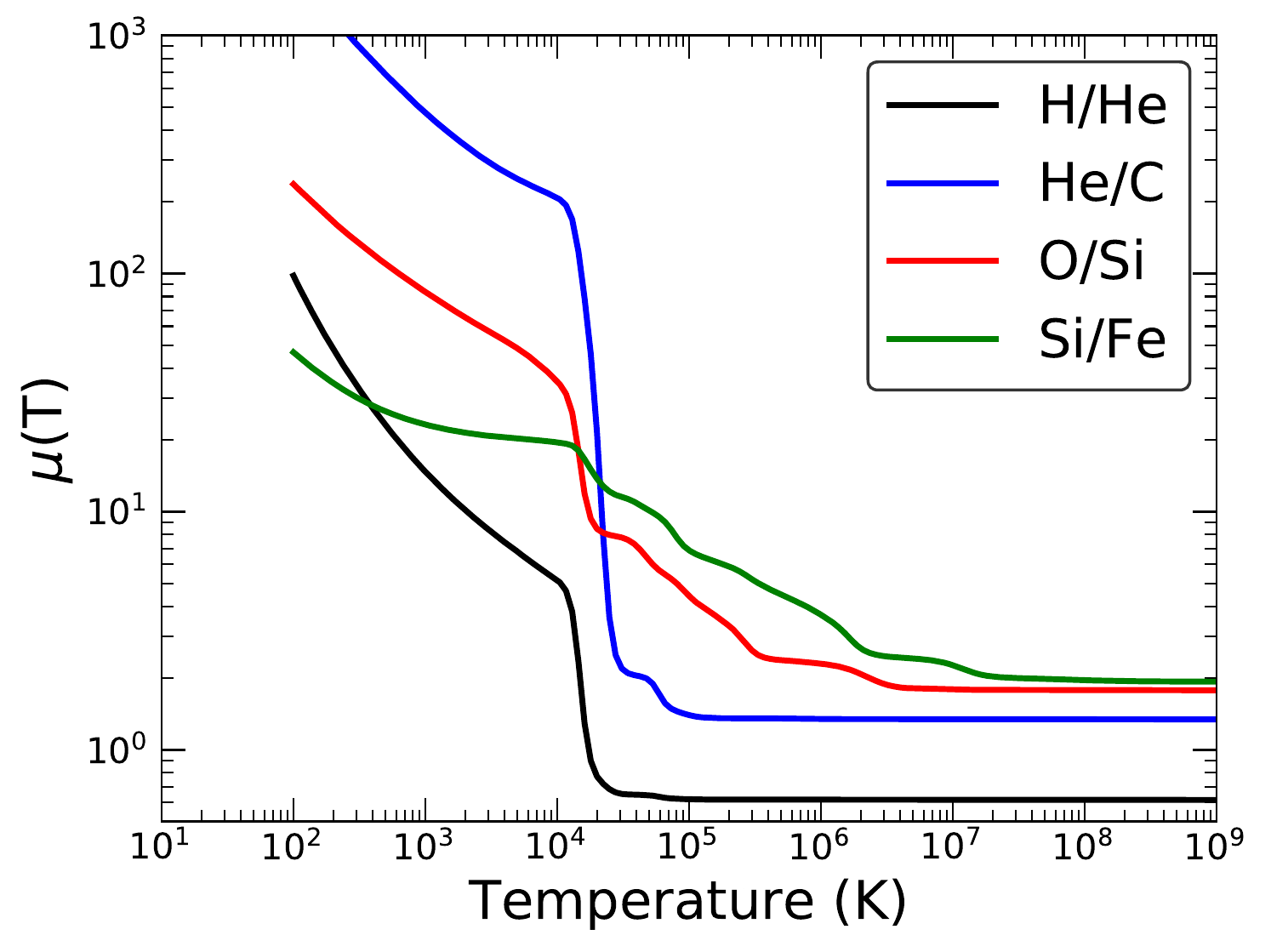}
\includegraphics[width=3.5in]{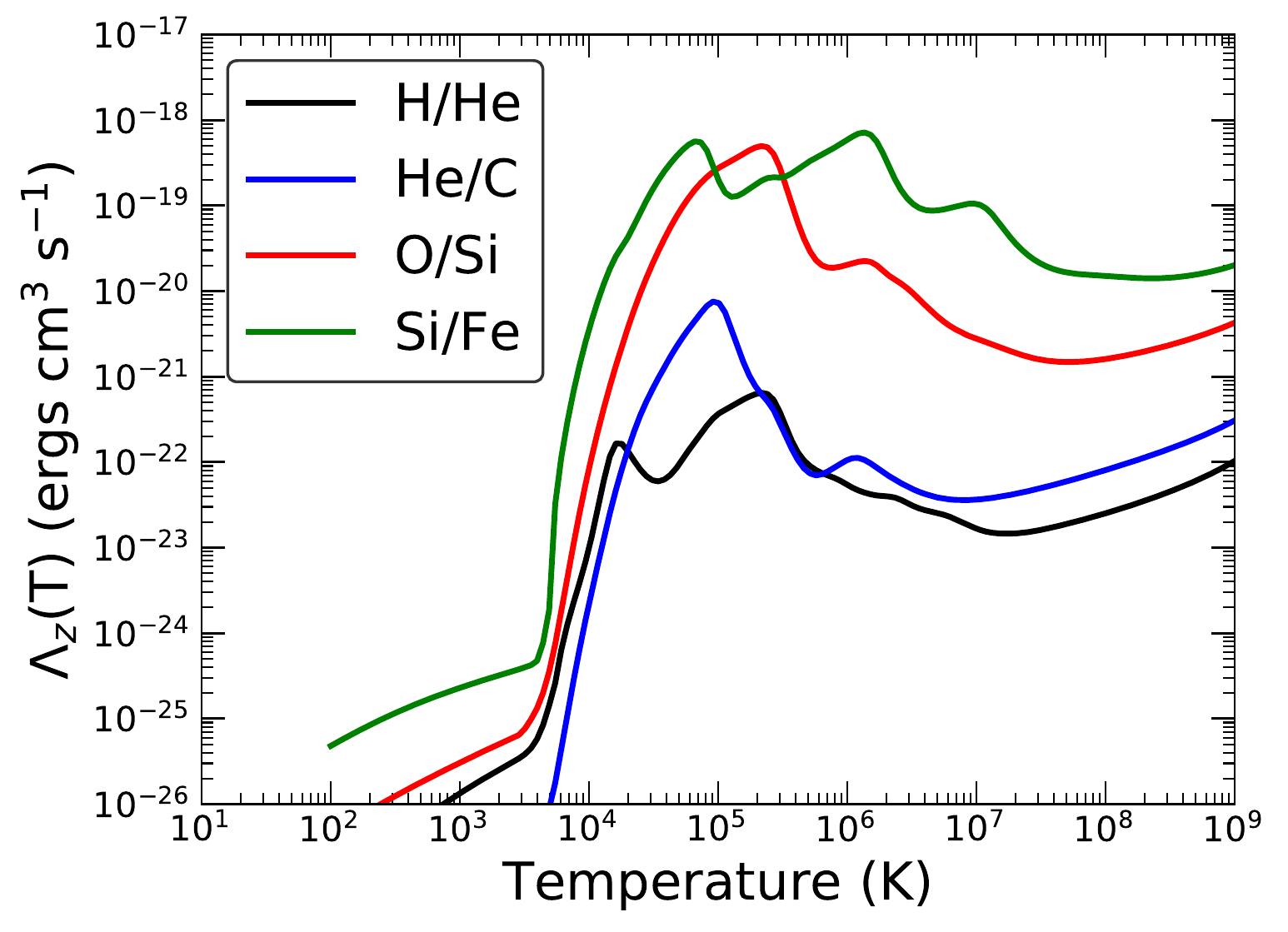}
\caption{\label{fig_mu} \textit{Left-panel:} The variation of the mean particle mass $\mu$ is shown as a function of temperature, for each zone of the ejecta (Table~\ref{table_initial}), with reference to Equation \ref{eq_Tj}. The mean particle mass increases with an increase in the metal concentration of the gas. The deviation is pronounced for lower temperatures, when the gas is partially ionized. \textit{Right-panel:} The cooling functions corresponding to each of the zones in the ejecta (Table~\ref{table_initial}) are shown, based on their chemical composition and ionization structure as a function of temperature. The H-rich CSM also follows a cooling function similar to the H/He zone.}
\end{figure*}

\section{Cooling of the shocked gas}
\label{sec_cooling}
In this section, we describe the dynamics of the post-shock gas, defining the physical conditions that control dust production.  

The nature of a SN shock, especially if it will be radiative or adiabatic, depends on three factors, (i) the density of the medium it travels through, (ii) the rate of cooling, i.e., the cooling function, and (iii) the velocity of the shock \citep{dopita_1976, dopita_1977, raymond_1979, plewa_1993, chevalier_2017, blinnikov_2017}. 
In interacting SNe, the density of the CSM is so high that the FS goes radiative well before the RS reaches the center of the remnant. This in turn triggers an early propagation of the RS into the ejecta which are still very dense at these times \citep{chu09, chandra_2018, smith_2017, nymark_2006}. The pre-shock ejecta cools through free expansion but after the encounter with the RS, radiative cooling becomes the dominant cooling mechanism. 

The dynamics of the RS were addressed by \citet{nymark_2006} with the focus on X-ray emissions. In this study, we follow the post-shock dynamics to much lower temperatures (typically of the order of a few hundred Kelvins), at which point the molecular chemistry becomes relevant. 

The rate, at which a parcel of gas cools down radiatively, can be expressed as a function of temperature, and is a property of the gas composition and gas conditions. 
In \citet{sarangi2018}, we studied the FS interaction with the dense CSM in the type~IIn SN, SN~2010jl, using a planar shock model, to derive the time-dependent temperature, density, and velocity profiles in the postshock shell. We adopt a similar formalism here to study the shocked ejecta gas behind the RS. 

By solving the 1-D shock-equation \citep{shull_1987,dopita_2003}, using the conservation of momentum and balancing the total energy to account for the radiative energy loss \citep{lacey_1988}, the evolution of density and temperature of the shocked gas can be expressed in terms of the radiative cooling rate per unit volume, $\eta(T)$, as,


\begin{gather}
\label{eq_dn}
\frac{\mathrm{d}\rho}{\mathrm{d}t} =  \frac{\rho \eta(T)}{\rho_0 V_R^2} \Bigg( \frac{5}{2}-4\frac{\rho_0}{\rho}\Bigg)^{-1}  \\
\frac{T}{T_0} = (1+ M^2) \frac{\rho_0}{\rho} - M^2 \frac{\rho_0^2}{\rho^2},\ \ \ M^2 = \frac{\rho_0 V_R^2}{p_0}
\end{gather}
where, the gas density and pressure at the shock-front are $\rho_0$ and $p_0$ respectively. Further, $\eta(T)$ for any gas composition of total density $n$ can be expressed as,
\begin{gather}
\label{eq_coolingrate}
\eta(T) = n_e \sum_i n_i \Lambda_i(T) = n_e n \ \Lambda_z(T) \\
\Lambda_z = \frac{\sum_i n_i \Lambda_i}{\sum_i n_i}; \ \ \ n_e = \sum_i \sum_l l f_{i,l} \ n_i
\end{gather}
where, $f_{i,l}(T)$ and $n_i$ are the fractional ionization and number density of the element $i$ ($i.e.$, H, He, O etc.) at ionization state $l$ and temperature T, while $\Lambda_i(T)$ is known as the cooling function \citep{dal72, sutherland_1995, schure_2009} discussed in the following text. For the SN ejecta, $\Lambda_z (T)$ represents the cooling function for any given zone (defined in Section \ref{sec_parameters}), characterized by the abundances of elements. The multiplication of $l$ represents the number of electrons per nuclei, per ionization state. 

Cooling functions applicable to metal-rich plasma, where the abundance of hydrogen is negligible, are not well documented in literature; moreover, the high-density limits relevant to this study adds to the complexity. 

For this paper, we have derived the cooling functions, $\Lambda(T)$, for individual elements (H, He, C, O, Si, Fe) using the spectral synthesis code CLOUDY \citep{ferland_2013}, using its inbuilt cooling routine. The cooling functions of these elements for temperatures from 10$^9$ to 100 K are shown in Figure \ref{fig_coolingfunc} in ergs cm$^3$ s$^{-1}$. 

When the gas densities are high, $\Lambda(T)$ can also vary with the variation of density. This variation is prominent mainly when the temperature is below 10$^5$ K \citep{wang_2014}. To define the cooling functions, we altered the densities between 10$^7$ to 10$^{13}$ cm$^{-3}$; however, we found the changes to be minor, and not significant for our study. The Figure presents the cases when the density is set at 10$^{10}$ cm$^{-3}$ for each element. The dependence of the cooling rates on various gas conditions will be the readdressed in the following Section \ref{sec_downstream}, when we discuss the heating of the gas by the radiation generated by the shock. 



Figure \ref{fig_mu} (right-panel) shows the cooling functions, $\Lambda_z (T)$, corresponding to each zone, in relation to Equation~\ref{eq_coolingrate}. They are comparable to the ones derived by \citet{nymark_2006}. 

When a parcel of gas encounters a shock, the immediate post-shock temperature, also known as the jump temperature, $T_J$ \citep{raymond_2018} is given by, 
\begin{equation}
\label{eq_Tj}
T_J = \frac{3}{16} \frac{\mu m_H}{k} V^2_R \ ; \ \mu(T) = \frac{\sum_i A_i n_i}{\sum_i \sum_l (l+1)f_{i,l}(T)n_i } 
\end{equation}

where, $\mu$ is called the mean particle mass, $m_H$ is the mass of a proton and $k$ being the Boltzmann constant. 
The mean particle mass $\mu$ is defined in terms of the mass number $A_{i}$ of the element $i$ and the number of nuclei and electrons present per unit volume of the gas. 

The ionization fractions of each element are simultaneously derived from CLOUDY \citep{ferland_2013}, when we determine the cooling functions. 
Figure \ref{fig_mu} shows the variation of $\mu$ as a function of temperature, for each zone of the ejecta. As evident from the figure, a larger abundance of heavy elements in the gas yields a larger value of $\mu$, and hence a larger jump temperature. In the regime between 10$^6$ and 10$^3$ K, due to partial ionization, $\mu$ is larger and the differences between the zones are more pronounced. It is worth mentioning, $\mu$(T) being a function of temperature, it is mathematically not feasible to use Equation \ref{eq_Tj} to correctly calculate $T_J$; however fortunately for this study, in almost all cases the jump temperatures being above 10$^{6.5}$ K, where $\mu$ remains almost constant (fully ionized) for a particular composition, this equation, therefore, is acceptable. 

\begin{figure*}
\vspace*{0.3cm}
\centering
\includegraphics[width=3.5in]{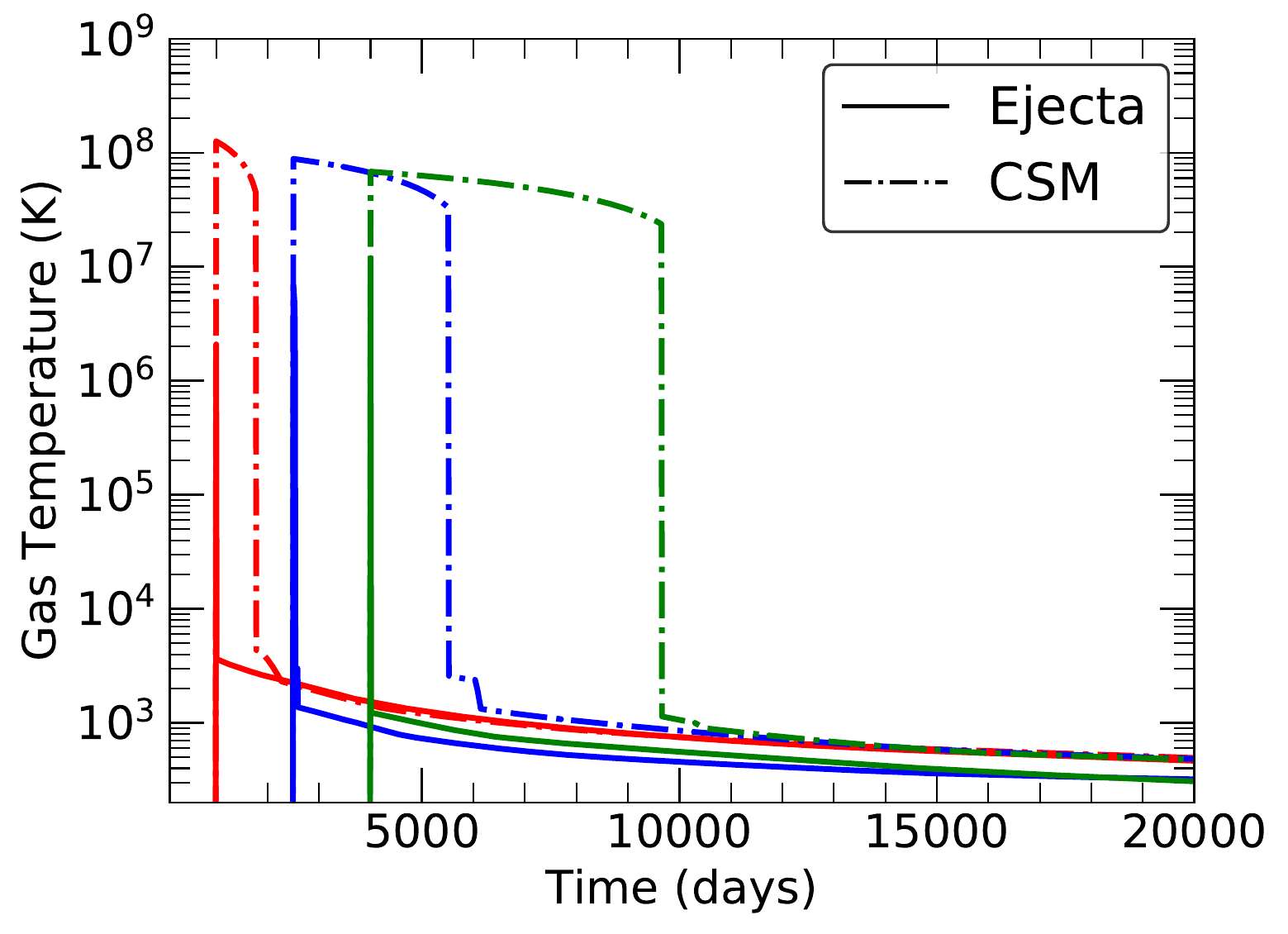}
\includegraphics[width=3.5in]{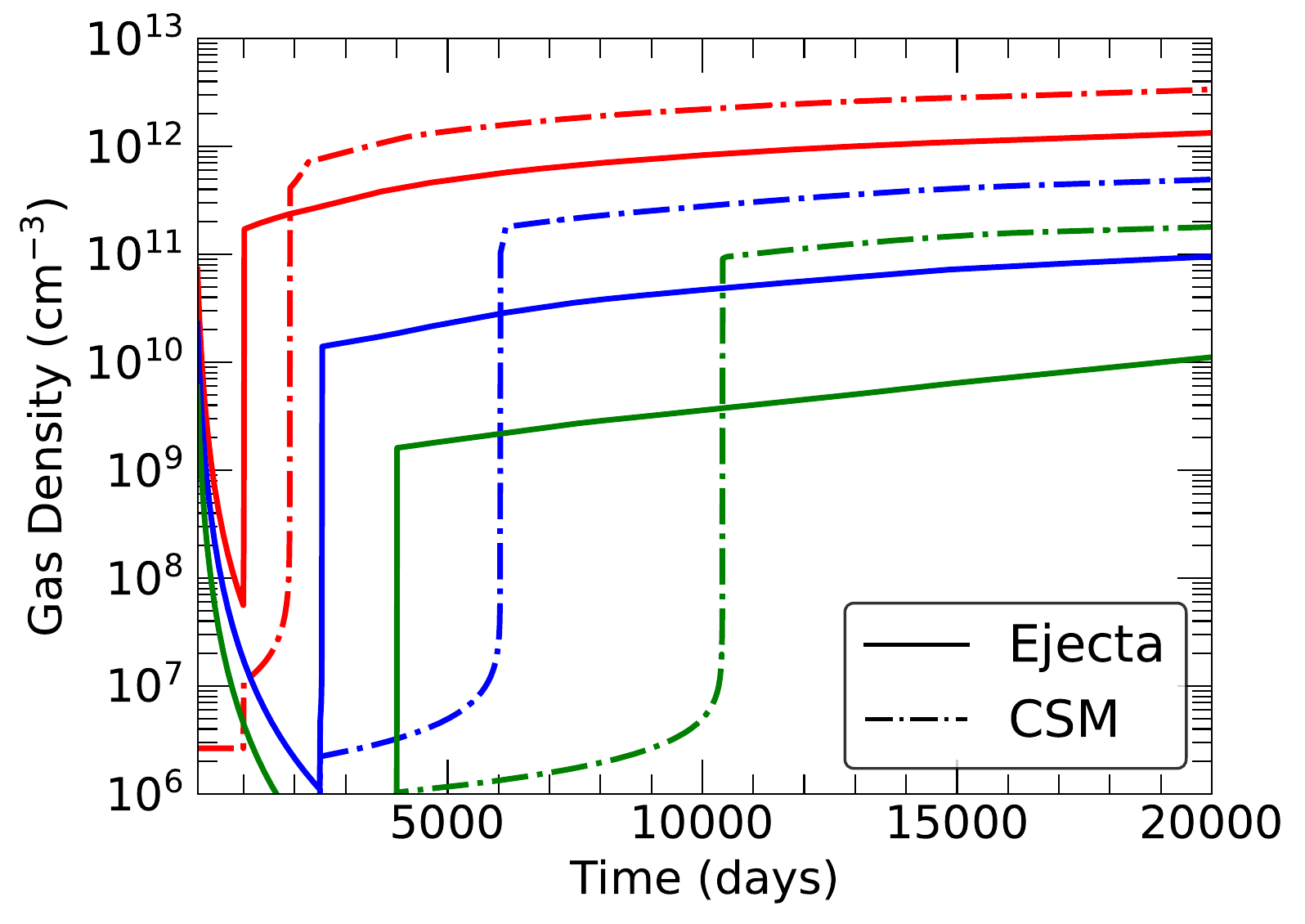}
\caption{\label{fig_Tdens} The evolution of temperature (\textit{left-panel}) and density (\textit{right-panel}) for the benchmark case. Six parcels of gas are shown, characterized by their epochs of FS and RS encounter, which are day 1000 (shown in red), 2500 (shown in blue), and 4000 (shown in green) respectively (one parcel in the CSM and one in the ejecta at each epoch). The solid-lines represent the ejecta parcels, while the dash-dotted lines are for the CSM. The ejecta-parcel shocked at day 1000 is in the H/He zone, the one shocked at day 2500 is in the He/C zone, and the one shocked at day 4000 is in the O/Si zone respectively. See section \ref{sec_evolTn} for the details.  } 
\end{figure*}

\begin{figure}
\vspace*{0.3cm}
\centering
\includegraphics[width=3.5in]{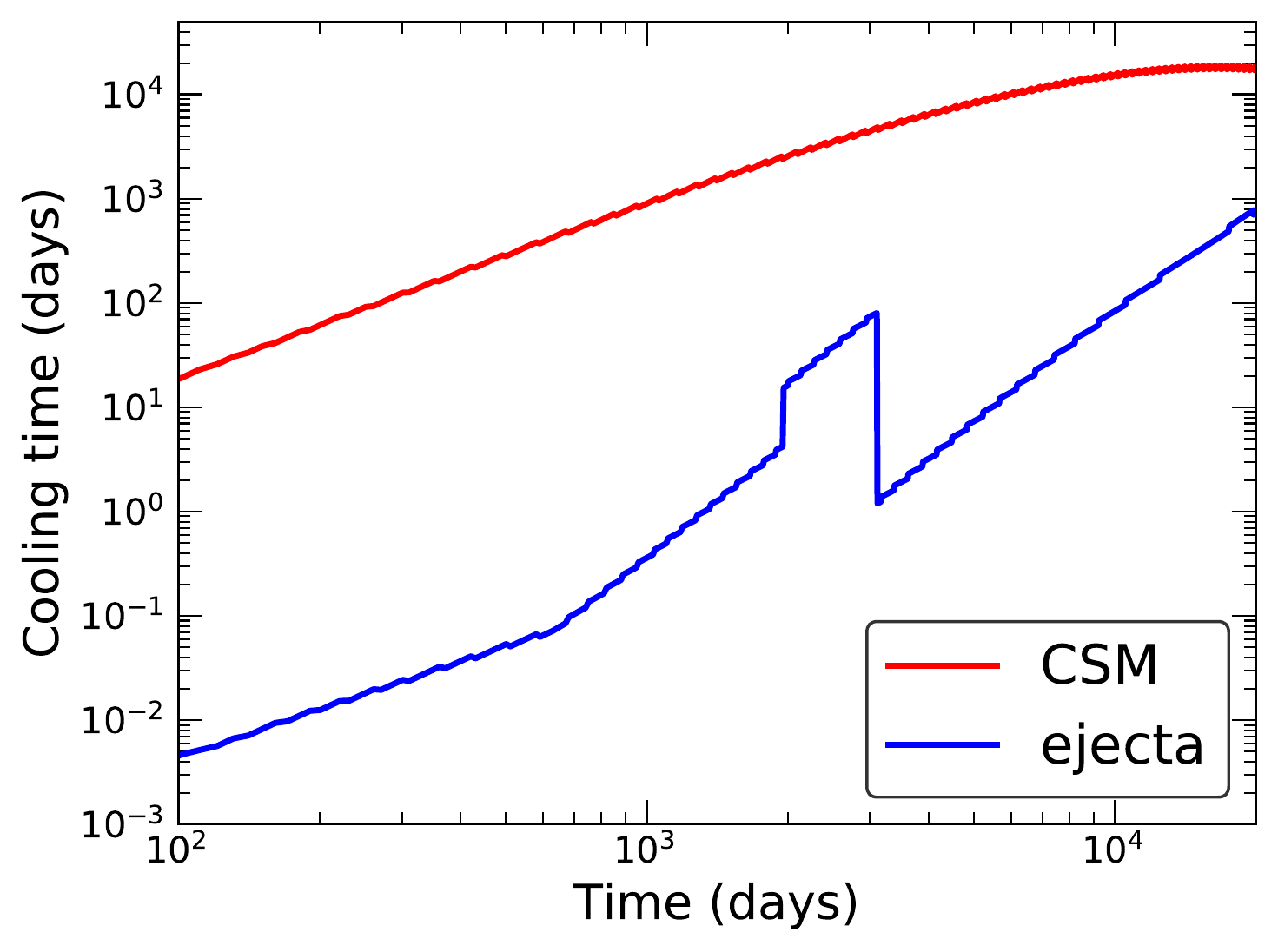}
\caption{\label{fig_coolingtime} The cooling time of individual parcels of gas for the benchmark case (\mdot\ = 10$^{-2}$ \Ms\ yr$^{-1}$, M$_{ej}$ = 10 \Ms). The x-axis represents the day at which the parcel has encountered the shock, while the y-axis indicates the cooling time for that parcel. }
\end{figure}


\subsection{Downstream radiation}
\label{sec_downstream}
For a high-density medium, especially when rich in metals, the cooling time is very short, typically of the order of hours to a few days (see Figure \ref{fig_coolingtime}). Behind a radiative shock, the pressure drops as the gas cools, leading to compression. Parcels then have post shock velocities that are low in the shock frame. The cooling gas is subject to ionization and heating by the radiation generated just behind the shock.
The heating rate, $H(T)$, for the gas modifies the cooling terms $\eta(T)$, as $S(T) \approx \eta(T) - H(T)$. 

This heating by the downstream radiation induces the formation of a warm dense shell (WDS), which is found to be vital for dust formation. To study the radiative energy balance, a formalism similar to \cite{sarangi2018} was set up, where we analyze a large grid of snapshot cases, along the evolution phase of the FS and RS. The energy balance of the post-shock cooling shell was studied for each case, using the code CLOUDY \citep{ferland_2013}. The FS is more energetic in most cases compared to the RS, so the heating by the downstream photons in the WDS is mostly controlled by the FS. 

The profile of the downstream radiation depends on the luminosity as well as the nature of the spectra. For SN~2010jl, \cite{fra14} showed that the photosphere could be modeled as a blackbody with the temperature ranging between 7000 and 9000 K. For this study, we choose a similar formalism, also used in \cite{sarangi2018}, where the heating source is represented by a blackbody-like photosphere at 8000 K, with the luminosity derived in Equation \ref{eqn_luminosity}. 

Unlike the cooling function, an analytical form for the heating rate cannot be generalized, given its dependence on the radiative transfer of energy between the evolving source and the evolving column of gas. 
When subjected to heating by a luminous source, the heating rate of a gas-parcel is defined by its composition and the flux of radiation incident on it. For the post-shock gas, the flux varies with time, as the energy generated by the FS and RS gradually declines, and the column density and number density of the post-shock gas also evolve. For the shocked ejecta, the composition of a given parcel depends on which pre-shock zone (see Table \ref{table_initial}) it belongs to. Importantly, at any given epoch, each parcel of the shocked gas is heated differently (even when they have identical abundances), determined by its pre-shock densities, and post-shock column density measured from the shock-front. 

 When the gas temperature is above \til\ 10$^4$~K, the highly ionized gas is characterized by large cooling rates (seen in Figure \ref{fig_coolingfunc}), and the effect of heating by the downstream photons is insignificant. For this regime, we can safely follow the evolution of the gas analytically, using the cooling function in Equation \ref{eq_dn}. 
 However, with a rapid drop in the cooling rates and ionization fractions below \til\ 10$^4$~K, the heating becomes extremely relevant, therefore we use CLOUDY to determine gas temperatures at these low temperature regimes. 
 
 Importantly though, CLOUDY \citep{ferland_2013} is not equipped to study the time evolution of the gas conditions, so we need to provide the initial densities and column densities to determine the temperature with the assumption that thermal equilibrium is attained. To do that, at first we use the analytical cooling function (Figure \ref{fig_mu}) in Equation \ref{eq_dn}, to determine the relation between the density and temperature. Then using that as an initial estimate, once the equilibrium gas temperature is calculated using CLOUDY, we correct the gas densities accordingly.  
 
To study dust formation, we are mostly interested in the chemistry of the gas at temperatures lower than 2000~K. At such low temperature regimes, line cooling by molecules are expected to significantly influence the cooling rate. The cooling functions for the ejecta zones (Figure \ref{fig_mu}) do not include cooling by molecules. Line emission by H$_2$ molecules are expected to be an important cooling mechanism for the H-rich gas in the CSM and the H-shell \citep{bourlot_1999}. CO molecules are one of the most abundant molecules formed in SN ejecta \citep{kam13}, which controls the rate of molecular cooling in the ejecta \citep{liljegren_2020}.  In our study, since we estimate the temperatures of the post-shock gas below $\sim$ 10$^4$ K by using the code CLOUDY \citep{ferland_2013}, the code itself is equipped to take into account the molecular cooling associated to the relevant abundances. We find that that cooling by H$_2$ molecules is reliably addressed by CLOUDY for all densities. However, for CO molecules, owing to the very high densities and concentration of metals in the shocked ejecta, along with the lack of H, the code often fail to account for CO cooling at low temperatures reliably, and encounters convergence failure (also the documentation of the code does indicate of such constraints). To correct for that, we refer to the comparison of the ejecta temperatures, with and without CO cooling, derived by \cite{liu92}, and use that as reference to estimate what should be the ejecta temperatures in a parcel of gas when CO cooling is taken into account, that reflects the CO abundances that we calculate at a given time. We acknowledge that in the future, calculating the effect of molecular cooling coupled to the complete molecular chemistry self-consistently, will be a major improvement to such models. 

In this way, combining the radiative cooling to the heating calculated using CLOUDY, the temporal profiles of the densities and temperatures are derived for each gas parcel in the post-shock gas that we discuss below. 

\subsection{The temperature and density and cooling time} 
\label{sec_evolTn}

For the benchmark case (\mdot\ = 10$^{-2}$ \Ms\ yr$^{-1}$, M$_{ej}$ = 10 \Ms), let us discuss the evolution of the gas conditions by looking at the parcels of gas that were shocked at three epochs, day 1000, 2500 and 4000 (let us call the parcels A, B and C respectively). Figure \ref{fig_Tdens} shows the evolution of the gas temperature and gas densities for these parcels of gas in the ejecta and in the CSM. 

For the shocked ejecta, A is in the H/He zone, B in the He/C zone, and C in the O/Si zone. 
The chronological phases of their dynamics are summarized here.  
\begin{enumerate}[label=(\alph*), noitemsep, leftmargin=*, align=left, wide = 0pt]
\item Before the encounter with the RS, the density and temperature follow free-expansion of the ejecta. 
\item After the interaction, the RS immediately heats the gas to temperatures of the range of 10$^{6}$ to 10$^{7}$ K. Owing to larger mean particle mass (Equation~\ref{eq_Tj}) in the inner cores, the jump temperatures increases from A to C. 
\item The initial phase of cooling is through free-free emission and hard X-rays, which is the bottleneck of the cooling process, being the slowest phase (Figure \ref{fig_coolingfunc}). 
\item Thereafter, a rapid cooling results in a rapid compression in density (Equation \ref{eq_dn}), leading to densities ranging between \til\ 5$\times$10$^{11}$ for A, to 5$\times$10$^{9}$ cm$^{-3}$ for C.
\item Finally the cooling is balanced by the heating through the downstream photons when $\eta$(T) $\approx$  $H$(T). The leads to the $d\rho/dt$ term in Equation \ref{eq_dn} to be close to zero. This marks the formation of the WDS. 
\item The heating rate H(T) of the dense shell is determined by the flux of radiation from the FS and RS. The luminosity decreases with time, and it moves further away from the gas-parcel, hence the steady temperature of the dense shell also decreases. We have chosen to call it WDS instead of the generally perceived cool dense shell, to distinguish the scenario from low density (large cooling time) cases where the downstream radiation is not significant; whereas here the temperature could well be as high as \til\ 10$^4$ at times (generally between 500 and 5000 K).
\end{enumerate}

The nature of cooling for the parcels of gas in the CSM follows an identical formalism (here A, B and C have the same abundances). Due to larger FS velocities (compared to the RS), the jump temperatures are close to 10$^8$~K, and the densities of post-shock gas in the WDS is also about one order of magnitude larger than the corresponding parcel in the ejecta. 

The cooling time of these parcels can be represented by the width of the T-spike in Figure \ref{fig_Tdens} (left-panel), which is the time taken by a parcel to assimilates into the WDS after it was shocked. 
The metal-rich ejecta, expectedly, cools much faster than the H-rich CSM, which is evident from the larger cooling rates shown in Figure \ref{fig_coolingfunc}. The RS velocity being much smaller than the FS velocity (Figure \ref{fig_R_cd}) also aids the rapid cooling. This can be confirmed by the extra-thin spikes of the solid-lines (which represents the ejecta) in Figure \ref{fig_Tdens}. 

The cooling times for the shocked CSM and the shocked ejecta gas are presented in Figure \ref{fig_coolingtime}.
It shows that the cooling time for the shocked ejeta ranges between 10$^{-1}$ day to a few 100 days, while, for the CSM, it is much longer. For both the CSM and the ejecta, the cooling time gradually increases, since the gas density at the shock-front decreases with time. The kinks at day 1956 and day 3109 for the cooling time of the shocked ejecta represent the time when the RS transits from the H/He zone to the He/C zone and from He/C zone to the O/Si zone respectively. When the RS enters the He/C zone from the outer H zone, a larger mean particle mass results in larger jump temperatures, therefore the gas takes longer to cool down (since the cooling functions for the two zones are comparable in the regimes between \til\ 10$^7$ and 10$^6$~K). On the contrary, the cooling rates being so much faster for the metal rich O/Si zone (see Figure~\ref{fig_mu}), the cooling time sharply drops when the RS enters this region. 

It is possible that, at late times, when the FS travels through a much lower density CSM gas, the cooling time of the shocked CSM is large enough, and the FS may turn adiabatic in nature. The RS will however still remain radiative. 

\begin{figure}
\vspace*{0.3cm}
\centering
\includegraphics[width=3.5in]{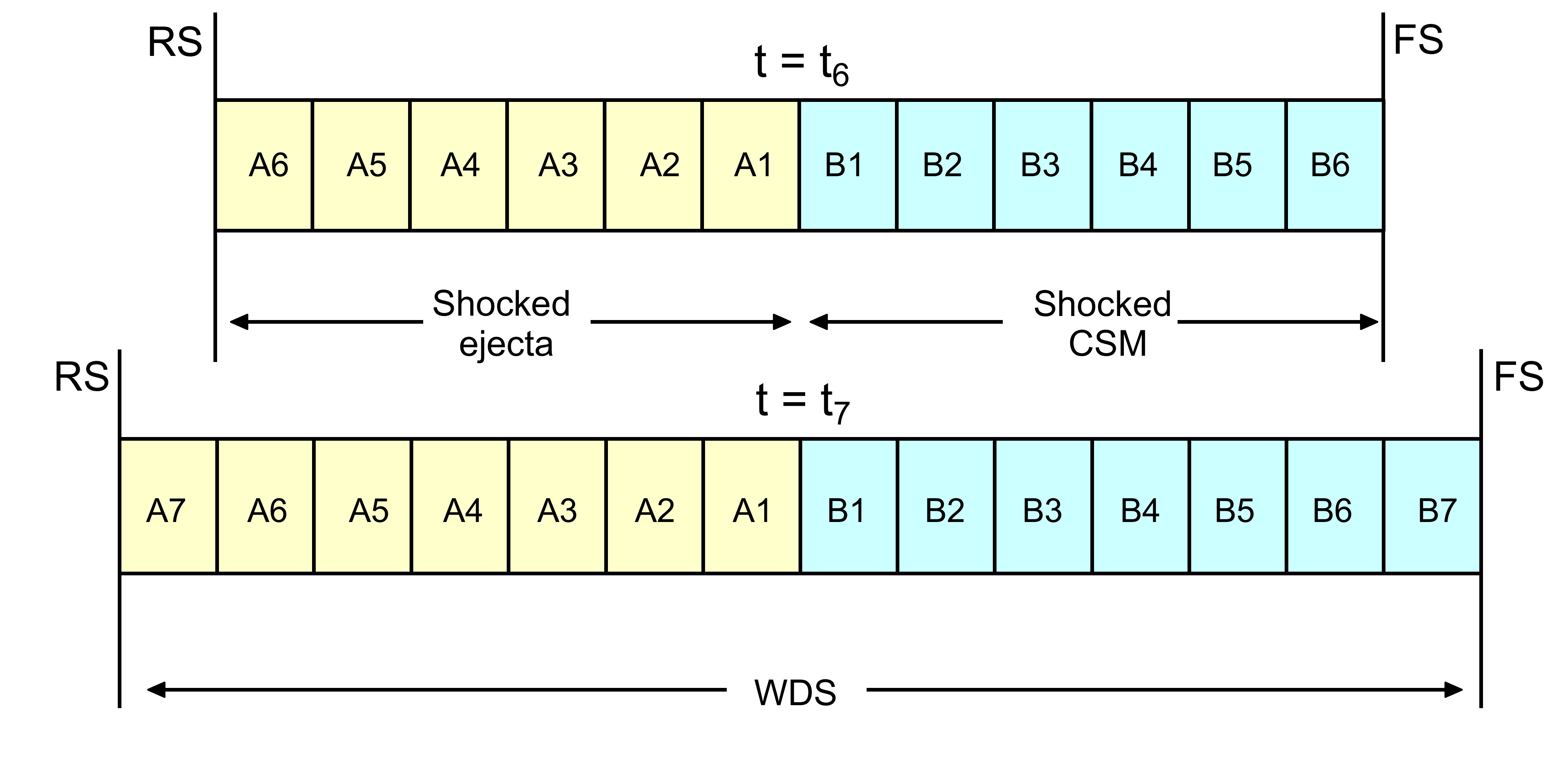}
\caption{\label{fig_shockparcel} The figure shows the schematics of how the chemistry of the WDS was addressed. The shocked ejecta and shocked CSM, together forming the WDS, were stratified into small parcels of gas. The figure shows the stratification for two arbitrary times, $t_6$ and $t_7$. The parcels A1 and B1 were shocked at $t = t_1$, A2 and B2 were shocked at $t = t_2$ and so on. The physical conditions in each parcel evolve uniquely with time, and we follow each of these parcels individually from the time it was shocked up to 20,000 days post-explosion. }

\end{figure}

\begin{figure*}
\vspace*{0.3cm}
\centering
\includegraphics[width=3.5in]{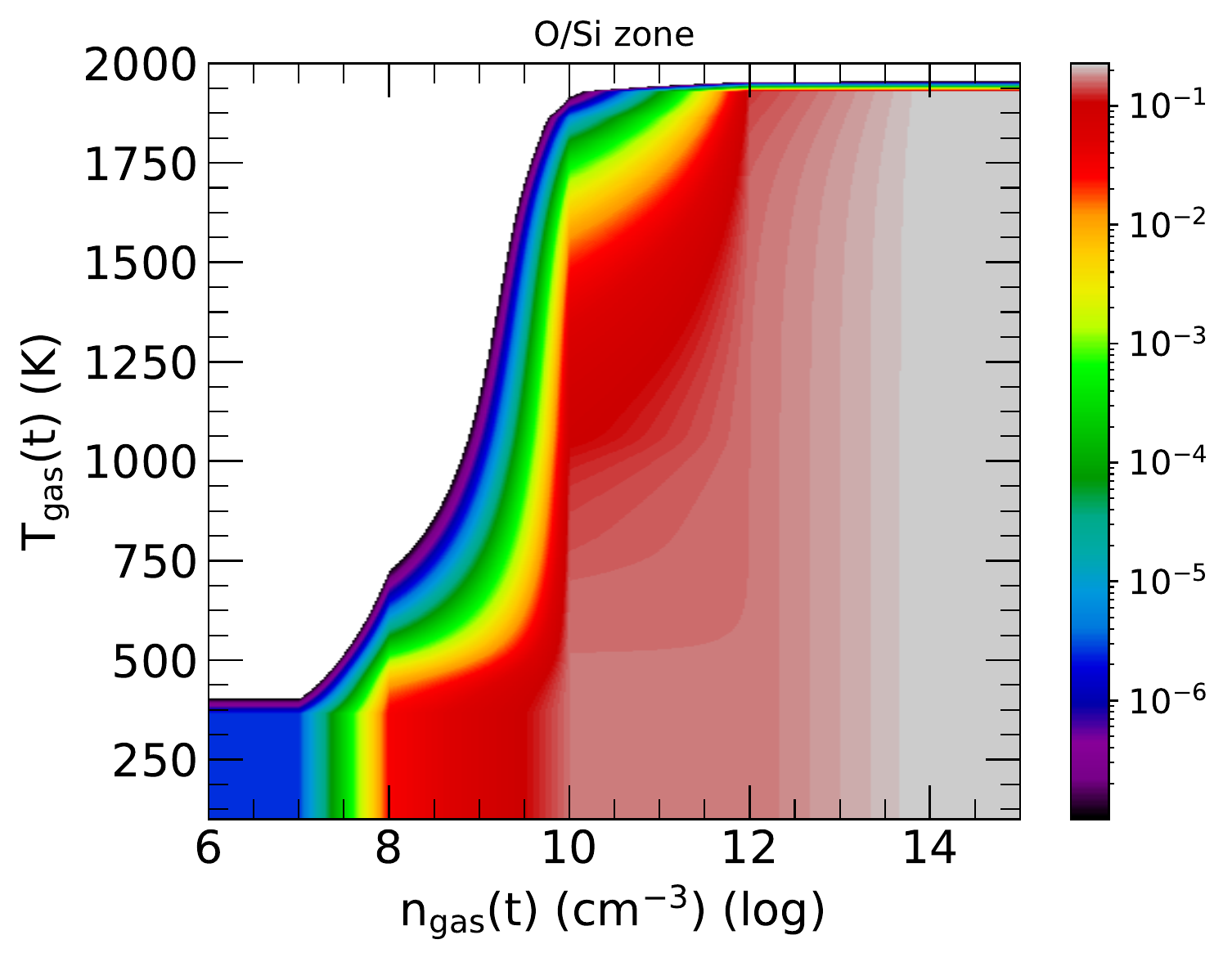}
\includegraphics[width=3.5in]{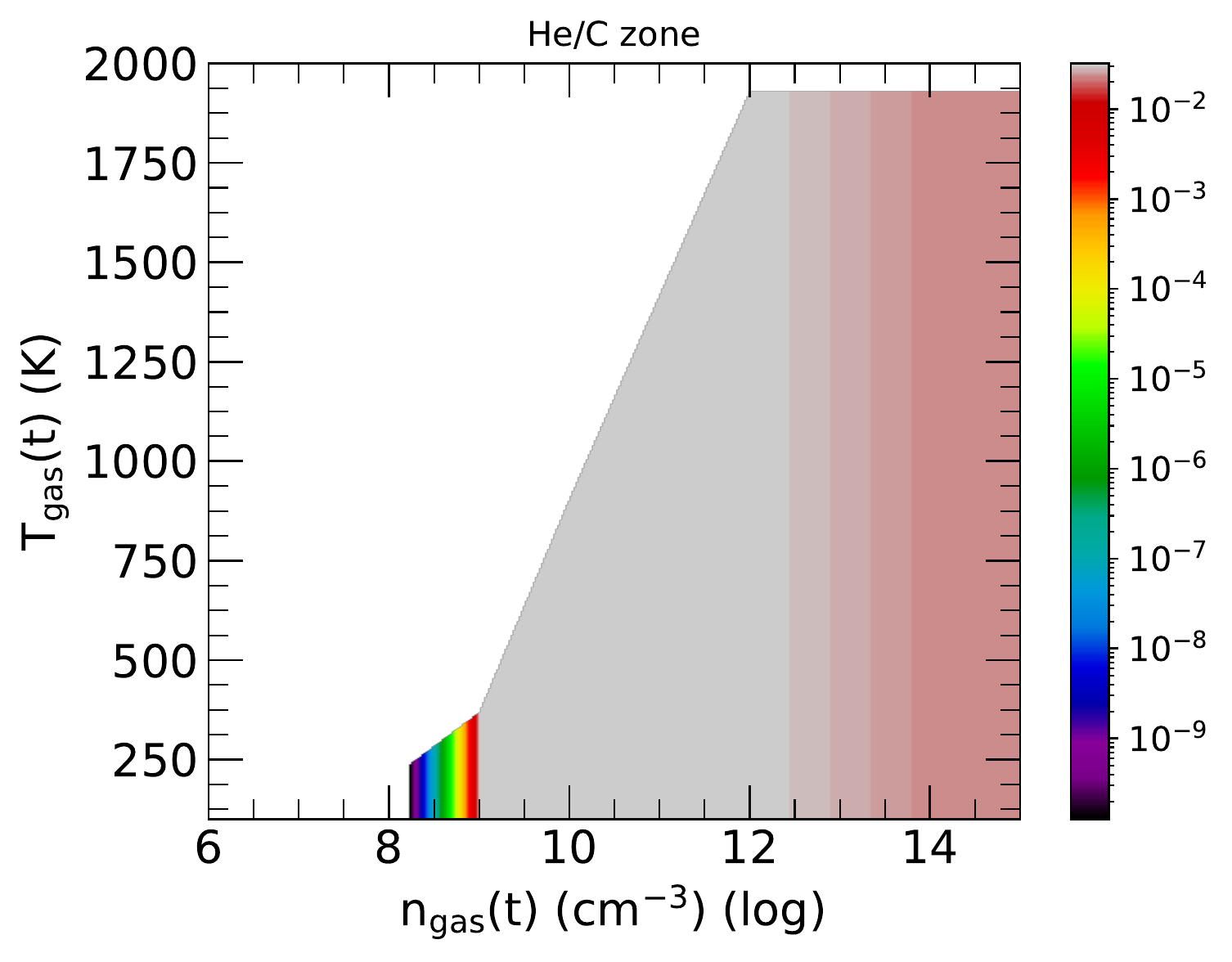}
\caption{\label{fig_chemistry}The efficiency (dust mass/gas mass of the zone) of dust formation is presented for the O/Si zone (\textit{Left-panel}), which forms mostly O-rich dust components (various silicates, alumina), and the  He/C zone (\textit{Right-panel}), where mainly the amorphous carbon and silicon carbide dust are formed. The efficiency is presented as a function of gas temperature and gas density, and the prevalent chemical pathways are that of the shocked SN ejecta described in Section~\ref{sec_chemistry}. Important to remember, this is not the gas to dust mass ratio of the entire ejecta, but only of the individual zones. The temperature and density are each functions of time. Each random parcel of gas in the WDS has its unique evolution of density and temperature (see Section \ref{sec_evolTn}). Moreover, it is worth mentioning that the efficiency is also a function of time, since how fast the physical conditions evolve can influence the favored chemical pathways. In the post-shock cooling ejecta, due to the very high densities of refractory elements, and the rather slow drop of temperature in the WDS, we can generalize the dust formation efficiency as function of gas density and temperature alone, ignoring the time dependence of the efficiency. As a word of caution, this figure however should not be taken as a general result for any environment, since the rate of evolution of physical conditions are known to impact the abundances of species in rapidly evolving circumstellar environments. } 
\end{figure*}

\section{Dust formation chemistry}
\label{sec_chemistry}
In this section, the chemistry of dust formation in the shocked ejecta is briefly discussed. 

The synthesis of dust grains is controlled by simultaneous phases of nucleation and condensation, where gas-phase nucleation chemistry acts as the bottleneck \citep{sarangi2018book}. In \citet{sar13, sar15}, the chemistry of SN ejecta was addressed, adopting a non-steady state, chemical kinetic approach, taking into account all possible chemical pathways for dust formation, involving the constituent elements in each zone.

The same formalism was used in this study. Most of the chemical network, involving atoms and molecules in their ground-state, remains applicable to the post-shock ejecta as well, a few ionization processes are altered based on the ionization fractions either obtained as explained in Section \ref{sec_cooling} or from CLOUDY simulations \citep{ferland_2013}. However, in the post-shock gas, the physical conditions that control dust formation evolve in a far more complex pattern, compared to that of the freely expanding ejecta. 

Due to the non-uniform (in other words non-analytical) nature of the heating and cooling rates, each parcel of gas follows its unique chemical signature, even when the initial abundances are the same. We address this by stratifying the column of the shocked gas into very small parcels, and following the temporal evolution of the physical conditions and abundances of molecules and dust grains in each such parcel; in Figure \ref{fig_shockparcel} we present a schematic picture of this. As the figure shows, the parcel A1 in the shocked ejecta and B1 in the shocked CSM is defined by the time $t = t_1$ at which they encountered the RS and FS respectively; similarly A2 and B2 encountered the shocks at time $t_2$, and An and Bn at time $t_n$. We have chosen the spatial bin ($t_{n+1} - t_n$) as 1 day. We study the evolution of each of these parcels individually from the time it encountered shock up to the post-explosion time of 20,000 days.



As the outcome of the above simulations, abundances of molecules, stable molecular clusters, dust precursors, and mass of dust species are obtained (for a complete list of species please refer to \citealt{sarangi2018book}). Here we have presented results in terms of O-rich dust species, constituting of silicates (Si$_n$O$_{n+1}$, (Mg$_2$SiO$_4$)$_n$), alumina (Al$_2$O$_{3}$)$_n$, iron oxides (FeO)$_n$ and magnesium oxides (MgO)$_n$, and C-rich dust species, consisting of amorphous carbon (C$_{28}$)$_n$ and silicon carbide (SiC)$_n$ (n here is a random positive integer generally greater than 4). Carbonaceous dust is only formed in the He/C zone, due to the overabundance of C compared to O. There are other dust components, such as metal clusters of Si, Mg etc., and FeS-clusters, which may form in smaller fractions. 

The formation of molecules, for example, SiO and CO, start in the partially ionized ejecta, when the gas temperatures are between 3000 and 4000 K. However, the depletion of SiO to form silicates transpires only when the temperatures are 1800~K or lower.

As evident from Figure~\ref{fig_Tdens}, the WDS is characterized by very high gas densities and a very gradual decline in temperature. This ensures that the rates of the important chemical pathways are faster than the rate of evolution of the gas conditions. In such special cases, it is possible to find the dust formation efficiency as function of the gas conditions only. In other words, for any random parcel of gas, if we can calculate the density and temperature of that parcel at a given time $t$, we can get an idea of how much dust will be present in that parcel. In the 2-D histogram shown in Figure~\ref{fig_chemistry}, we show the dust formation efficiencies of the O-rich dust species in the O/Si zone and the C-rich dust grains in the He/C zone, which is basically the dust to gas mass ratio of these zones. Important to realize, the resulting dust is the product of several intermediate chemical processes, which cannot be quantified by a single rate. Also, it should be noted that, such a representation of dust formation efficiency that is independent of time, is acceptable only for specialized environments like the WDS; for dust formation in the unshocked SN ejecta, this representation is not valid. 

Analyzing the figure, it can be understood that the efficiency is highly non-linear, where depending on the suitable density and temperature, the dust formation is a rather robust process. The temperature required for the onset of dust formation, for any given gas parcel, is directly proportional to gas densities, as expected due to larger recombination rates at higher densities (so basically if densities are high, dust can condense at higher temperatures and vice versa).  

In the C-rich zone, most of the carbon is locked up either in amorphous carbon dust or in silicon carbide. The relative mass of amorphous carbon to silicon carbide depends on the gas density, where the lower the density, the higher is the portion of silicon carbide. Moreover, there is a strict constraint on the density requirement for carbon dust formation in the He-zone, where below a gas density of \til\ 10$^9$~cm$^{-3}$, carbon dust formation is highly unfavorable.

The downstream radiation imparts additional heating and ionization of the gas, accordingly delaying the onset of dust formation in any parcel of gas. On the other hand, flux from the downstream photons invokes the formation of the WDS, allowing the gas enough time to complete the chemical routes leading to dust formation. Without its presence, the radiative cooling below 10$^5$ K would have been almost instantaneous and the reactions constrained by temperature requirements would have not transpired, making the production of dust highly inefficient. 

Our chemical network includes the pathways of destruction of molecules and molecular clusters by the collision with the fast Compton electrons generated by the degradation of \grays\ produced by the radioactive decay of \Co\ \citep{cherchneff2009, sar13}, however we find that the energy deposited by the radioactive species in the WDS in negligible compared to the downstream radiation from the shocks. 

We will refrain from repeating the description of all chemical paths and molecular routes, referring to the literature in \citealt{sarangi2018book}. 


\begin{figure}
\vspace*{0.3cm}
\centering
\includegraphics[width=3.5in]{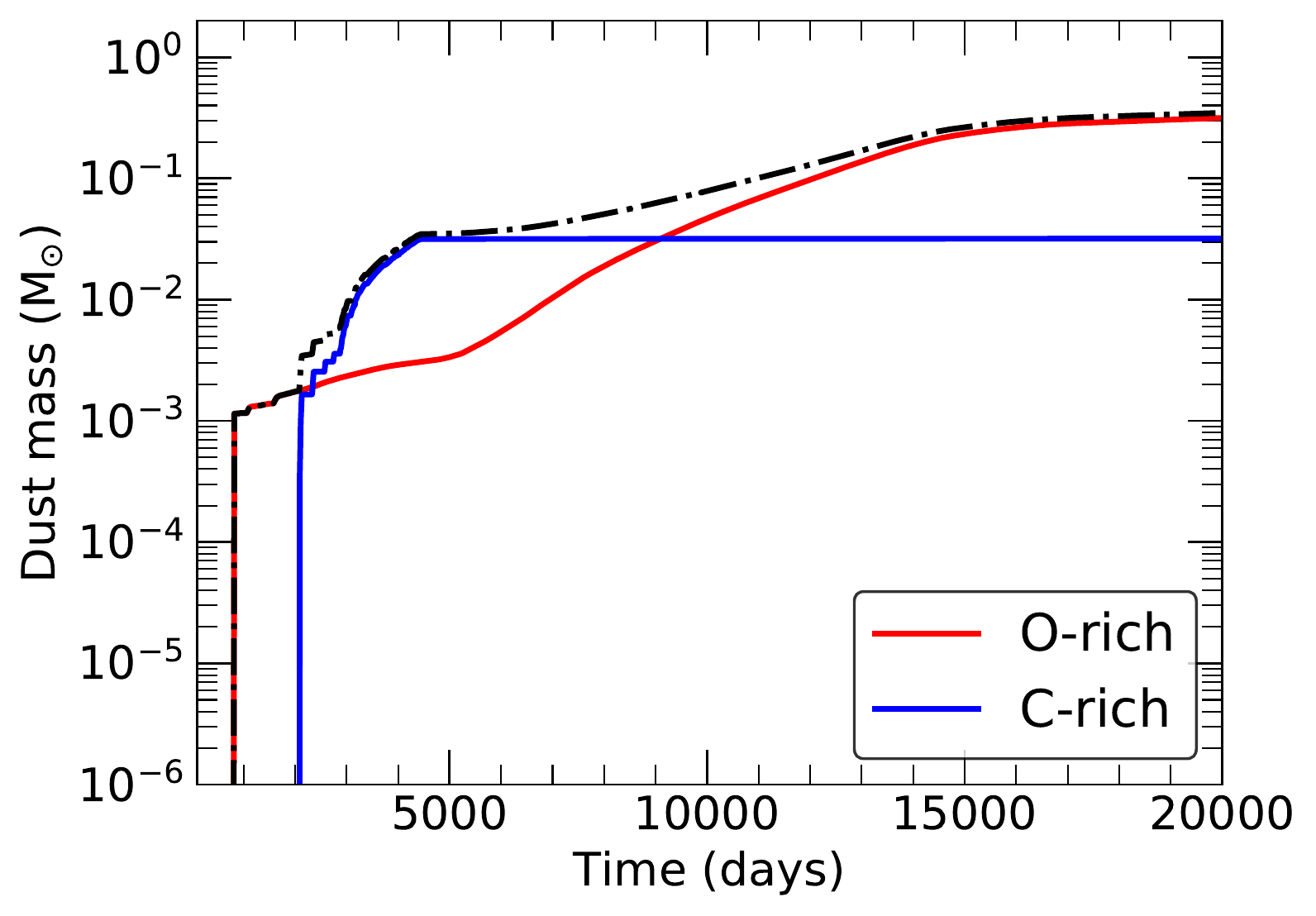}
\caption{\label{fig_dustmass}The mass of dust for the benchmark case (\mdot\ = 10$^{-2}$ \Ms\ yr$^{-1}$, M$_{ej}$ = 10 \Ms) is presented as a function of post-explosion time, following the chemistry described in Section \ref{sec_chemistry}. The masses are also listed as Case 5 in Table \ref{table_dustmass}. The O-rich and C-rich components include the dust present in both the CSM and the ejecta. }
\end{figure}

\subsection{Uncertainties}
The focus of this paper has been the variation of post-shock conditions in different cases of interacting SNe and how they are manifested in the chemistry of dust formation. In the following list, we acknowledge the uncertainties associated with the chemical network and its assumptions. 
\begin{enumerate}[label=(\alph*), noitemsep, leftmargin=*, align=left, wide = 0pt]
\item The favorable pathways relevant for the outer H-rich region of the ejecta require a more detailed analysis, since the previous studies have mostly focused on the He-core only. Some references are drawn from AGB stars \citep{gob16}, however, the hydrodynamic evolution being different and so being the elemental abundances, the need for a separate chemical network dedicated to the outer H-core of the SN ejecta is necessary. 
\item With respect to the same region, the abundances of elements in that outer zone (H/He) are also not well understood and may vary with progenitor mass, late evolution and mass-loss history of the progenitor. Many recent studies have reported of enhanced CNO processing \citep{fra14} in Type IIn SNe or their progenitors, which might alter the chemical pathways. 
\item The mass of silicate dust depends on the (uncertain) efficiency of condensation of molecules of the form Si$_n$O$_{n+1}$, which are formed in abundance in the gas phase and are structurally quite stable to act as dust precursors. 
\item The post-shock evolution of the gas parcel, and therefore its chemistry, depends on the history of its interaction with the ionizing source. The cooling function we have defined are not tested in such high density environments before. Moreover, even though we have included some impact due to line cooling by CO molecules, it should be more accurately determined and redefined in the cooling functions. 
\end{enumerate}


\begin{table*}
\caption{The dust masses (\Ms) of the O-rich (various silicates, alumina) and C-rich (amorphous carbon, silicon carbide) components, along with the total dust masses, at days 1000, 5000, 10000, and 20000 for all the cases. (For all the cases, M$_{ej}$ consists of 5~\Ms\ He-core and (M$_{ej}$ - 5)~\Ms\ of H-shell.) The $\dot{m}$ in the table is in \Ms\ yr$^{-1}$, and M$_{ej}$ in \Ms.}
\label{table_dustmass} 
\begin{tabular}{c | ccc | ccc | ccc}
\hline \hline
Time & O-rich & C-rich &  Total & O-rich & C-rich &  Total & O-rich & C-rich &  Total  \\
(Day) & (\Ms) & (\Ms) & (\Ms) & (\Ms) & (\Ms) & (\Ms) & (\Ms) & (\Ms) & (\Ms) \\
\hline
& \multicolumn{3}{l |}{Case 1, $\dot{\mathrm{m}}$ = 10$^{-1}$, M$_{ej}$ = 6} & \multicolumn{3}{l |}{Case 4, $\dot{\mathrm{m}}$ = 10$^{-1}$, M$_{ej}$ = 10} & \multicolumn{3}{l }{Case 7, $\dot{\mathrm{m}}$ = 10$^{-1}$, M$_{ej}$ = 15}  \\
\hline
1000 & 4.9(-1) & 2.3(-2) & 5.1(-1) & 2.2(-2) & 2.7(-2) & 4.9(-2) & - & - & - \\
5000 & 7.1(-1) & 2.3(-2) & 7.3(-1) & 5.5(-1) & 2.7(-2) & 5.8(-1) & 3.6(-1) & 3.3(-2) & 3.9(-1) \\
10000 & 7.2(-1) & 2.3(-2) & 7.4(-1) & 6.0(-1) & 2.7(-2) & 6.3(-1) & 4.4(-1) & 3.3(-2) & 4.7(-1) \\
20000 & 7.2(-1) & 2.3(-2) & 7.4(-1) & 6.0(-1) & 2.7(-2) & 6.3(-1) & 4.4(-1) & 3.3(-2) & 4.7(-1) \\
\hline
& \multicolumn{3}{l |}{Case 2, $\dot{\mathrm{m}}$ = 10$^{-2}$, M$_{ej}$ = 6} & \multicolumn{3}{l |}{Case 5, $\dot{\mathrm{m}}$ = 10$^{-2}$, M$_{ej}$ = 10} & \multicolumn{3}{l }{Case 8, $\dot{\mathrm{m}}$ = 10$^{-2}$, M$_{ej}$ = 15}  \\
\hline
1000 & 1.2(-6) & 1.8(-2) & 1.8(-2) & 1.2(-3) & - & 1.2(-3) & 1.7(-3) & - & 1.7(-3) \\
5000 & 3.3(-1) & 2.9(-2) & 3.6(-1) & 3.4(-3) & 3.2(-2) & 3.5(-2) & 3.8(-3) & - & 3.8(-3) \\
10000 & 4.2(-1) & 2.9(-2) & 4.5(-1) & 4.7(-2) & 3.2(-2) & 7.9(-2) & 6.1(-3) & 1.2(-3) & 7.3(-3) \\
20000 & 5.4(-1) & 2.9(-2) & 5.7(-1) & 3.2(-1) & 3.2(-2) & 3.5(-1)& 6.8(-2) & 3.1(-2) & 9.9(-2) \\
\hline
& \multicolumn{3}{l |}{Case 3, $\dot{\mathrm{m}}$ = 10$^{-3}$, M$_{ej}$ = 6} & \multicolumn{3}{l |}{Case 6, $\dot{\mathrm{m}}$ = 10$^{-3}$, M$_{ej}$ = 10} & \multicolumn{3}{l }{Case 9, $\dot{\mathrm{m}}$ = 10$^{-3}$, M$_{ej}$ = 15}  \\
\hline
1000 & 9.4(-6) & 3.5(-3) & 3.5(-3) & 1.2(-3) & - & 1.2(-3) & 1.7(-3) & - & 1.7(-3) \\
5000 & 3.3(-5) & 1.7(-2) & 1.7(-2) & 1.3(-3) & - & 1.3(-3) & 1.9(-3) & - & 1.9(-3) \\
10000 & 1.1(-2) & 3.1(-2) & 4.2(-2) & 1.4(-3) & - & 1.4(-3) & 2.0(-3) & - & 2.0(-3) \\
20000 & 7.2(-2) & 3.2(-2) & 1.0(-1) & 2.0(-3) & 9.2(-8) & 2.0(-3) & 2.5(-3) & - & 2.5(-3) \\
\hline

\end{tabular}
\end{table*}




\section{Dust mass yields}
\label{sec_dustmasses}

Now, we will describe the evolution of dust masses for all the 9 cases. 

\begin{figure}
\vspace*{0.3cm}
\centering
\includegraphics[width=3.5in]{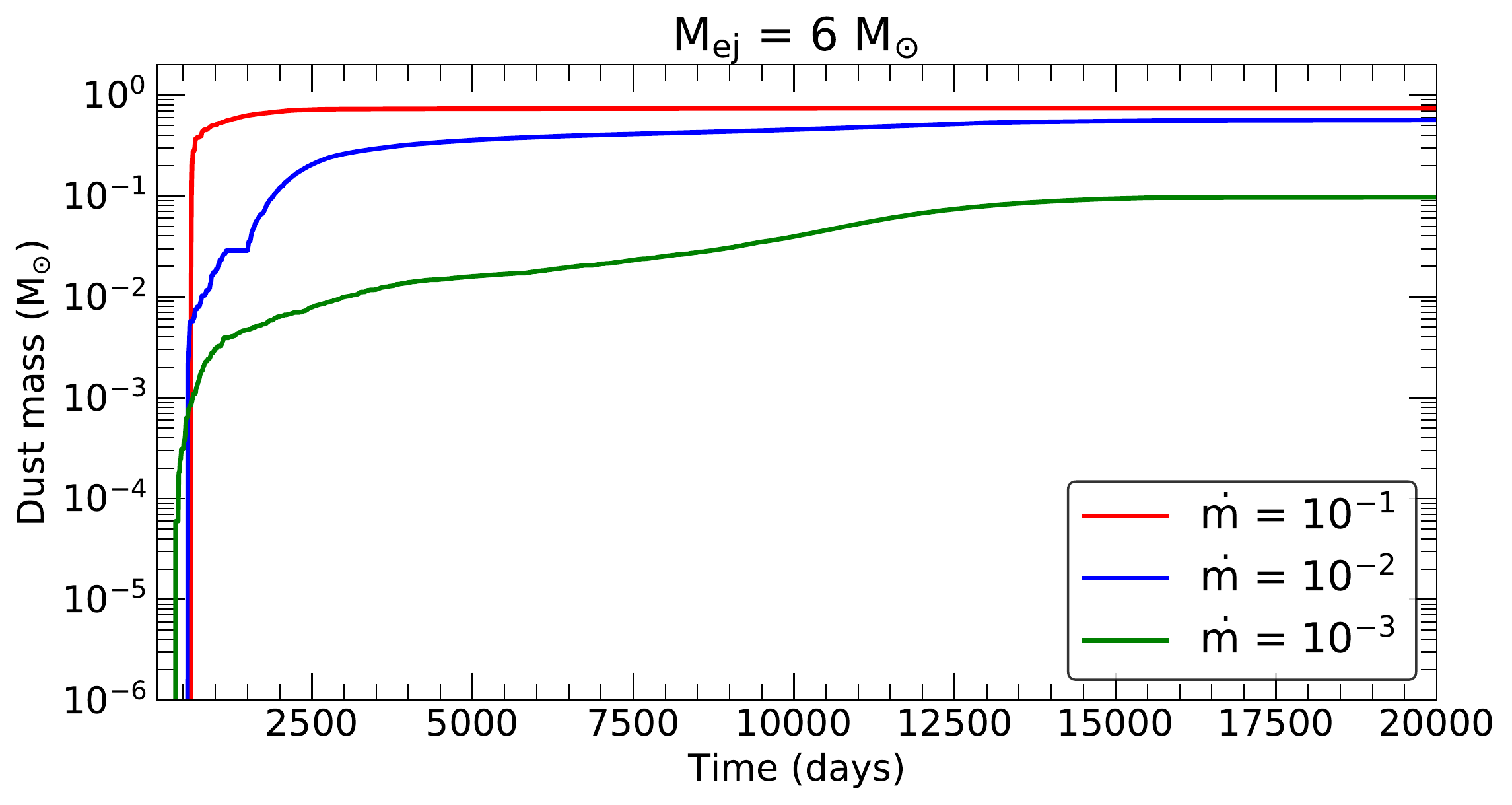}
\includegraphics[width=3.5in]{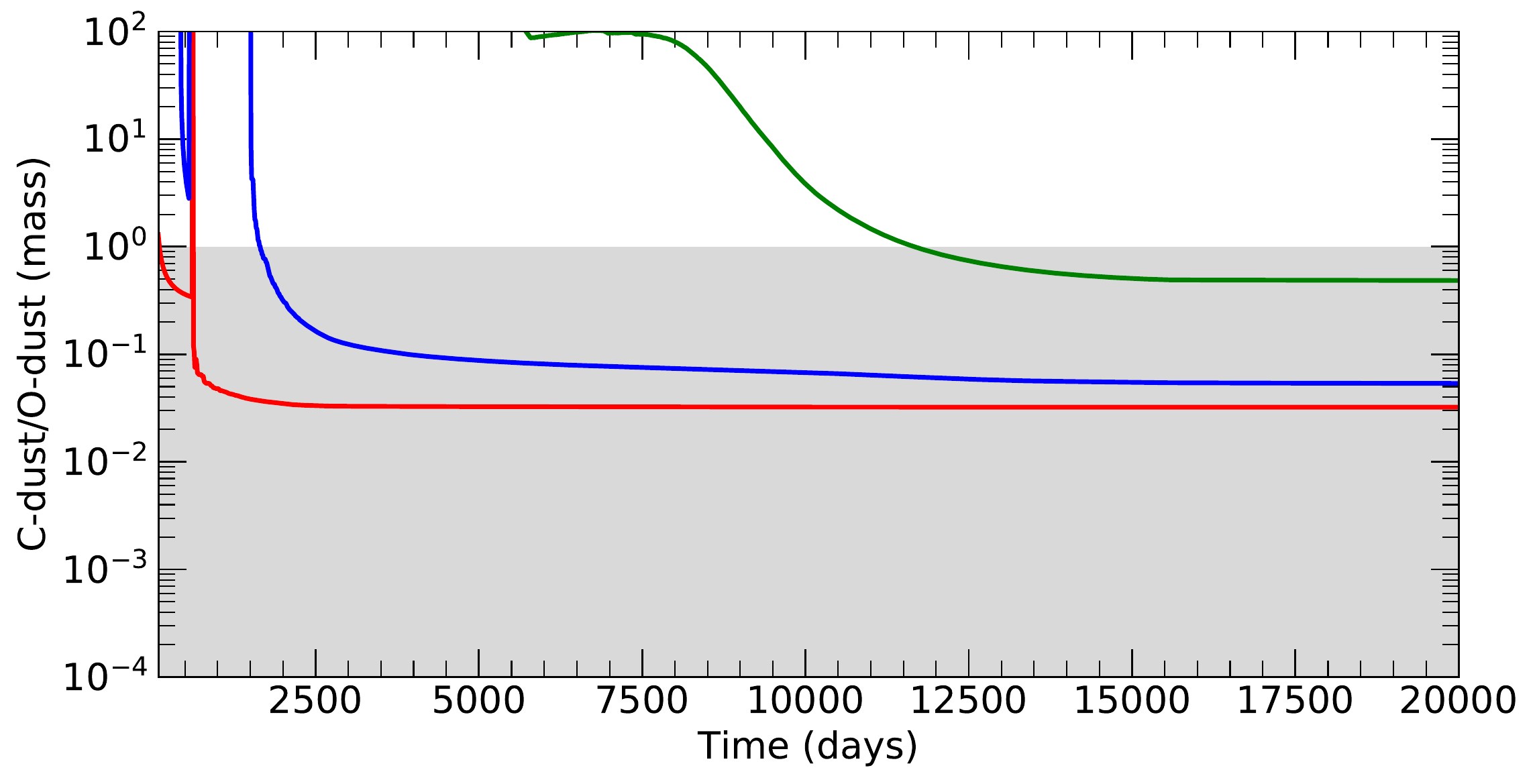}
\includegraphics[width=3.5in]{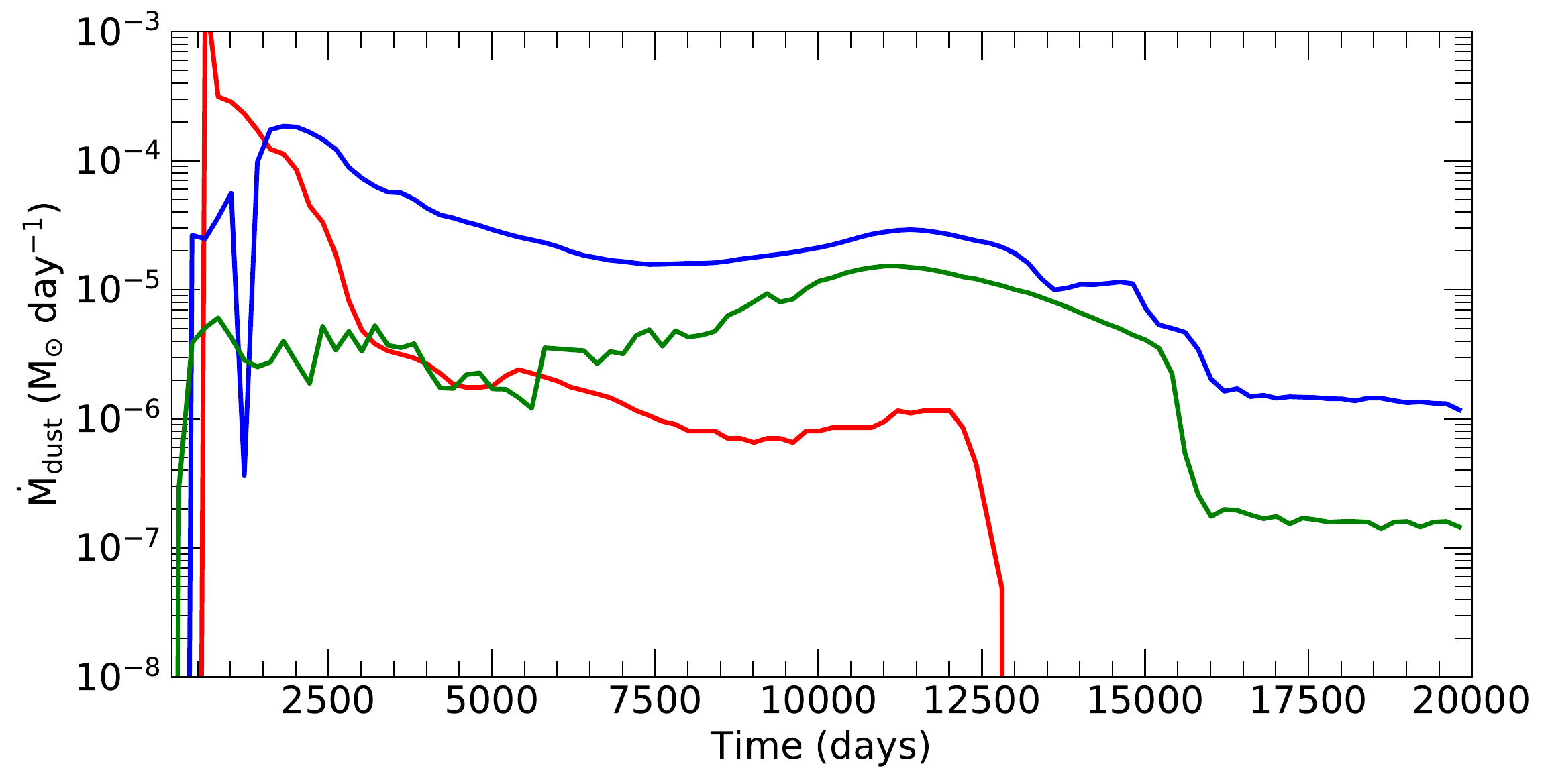}
\caption{\label{fig_M6}The following results for Cases 1, 2 and 3, all corresponding M$_{ej}$ = 6 \Ms\ (5~\Ms\ He-core, 1~\Ms\ of H-shell), and mass-loss rates varying as 10$^{-1}$, 10$^{-2}$, 10$^{-3}$ \Ms ~yr$^{-1}$, are presented as a function of post-explosion time: \textit{Top:} the evolution of total dust masses \textit{Middle:} the ratios of carbon-rich to oxygen-rich dust masses (the white region on the figure is C-rich and gray region is O-rich) \textit{Bottom:} the average rate of dust formation in per day. }
\end{figure}

\begin{figure}
\vspace*{0.3cm}
\centering
\includegraphics[width=3.5in]{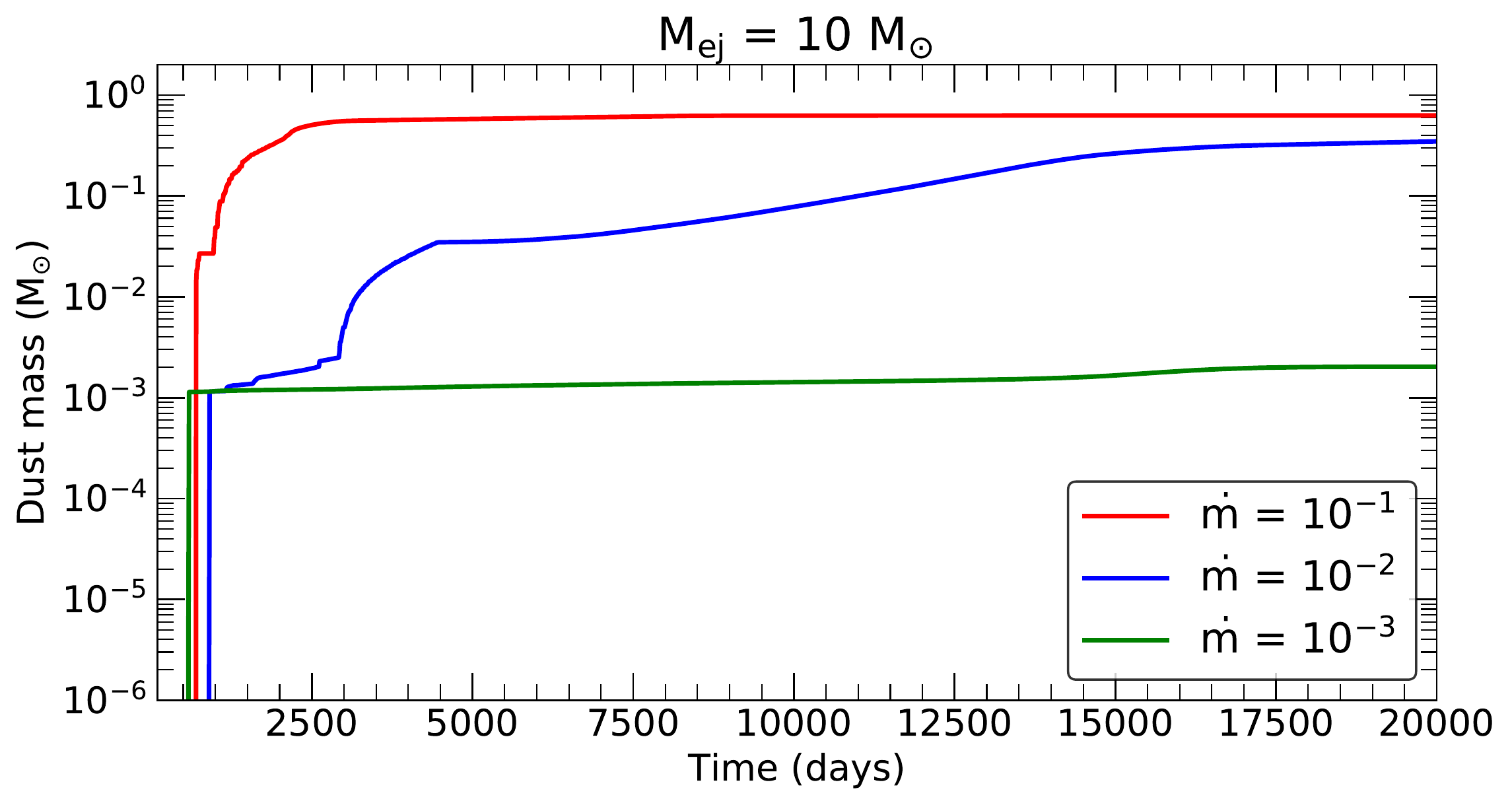}
\includegraphics[width=3.5in]{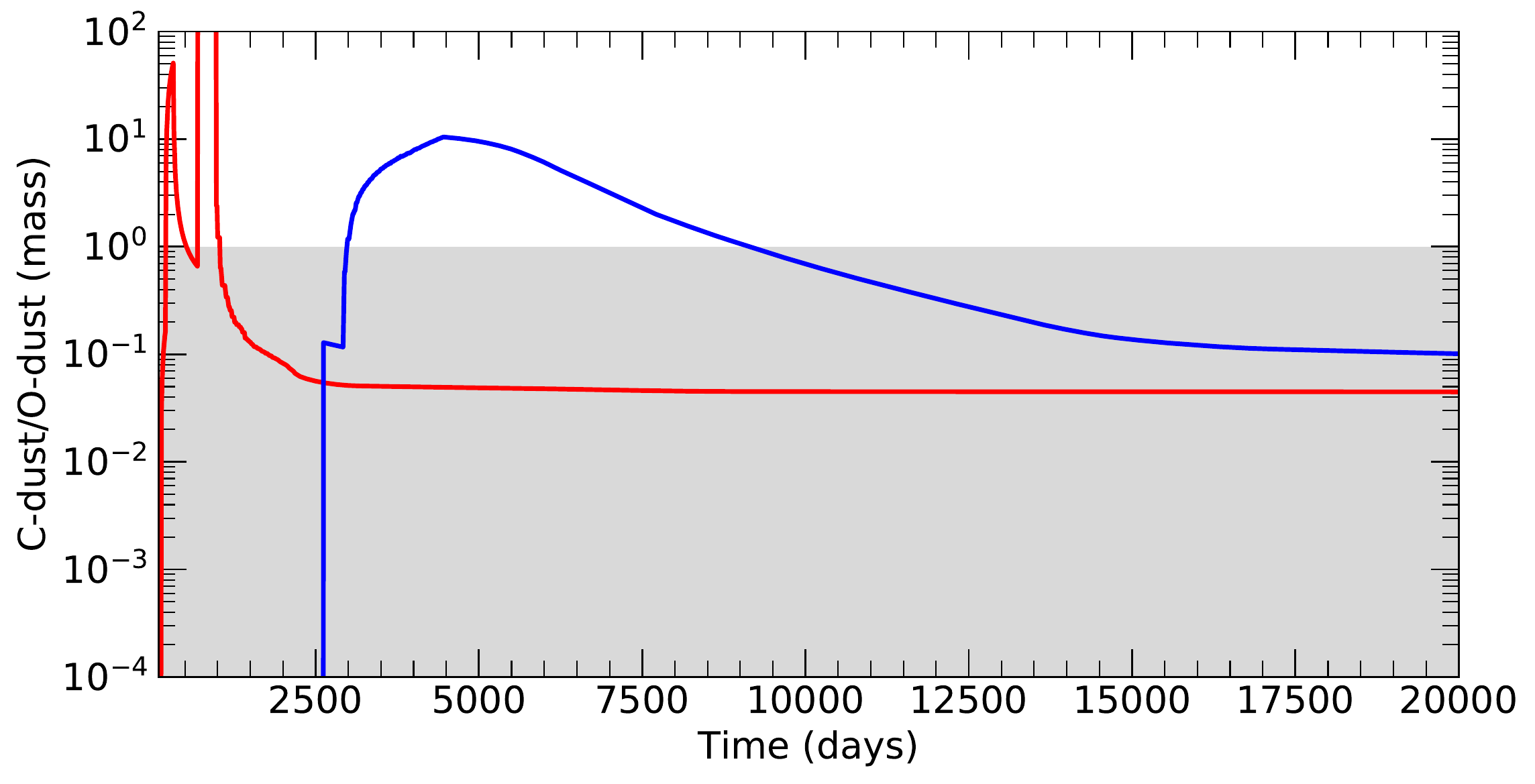}
\includegraphics[width=3.5in]{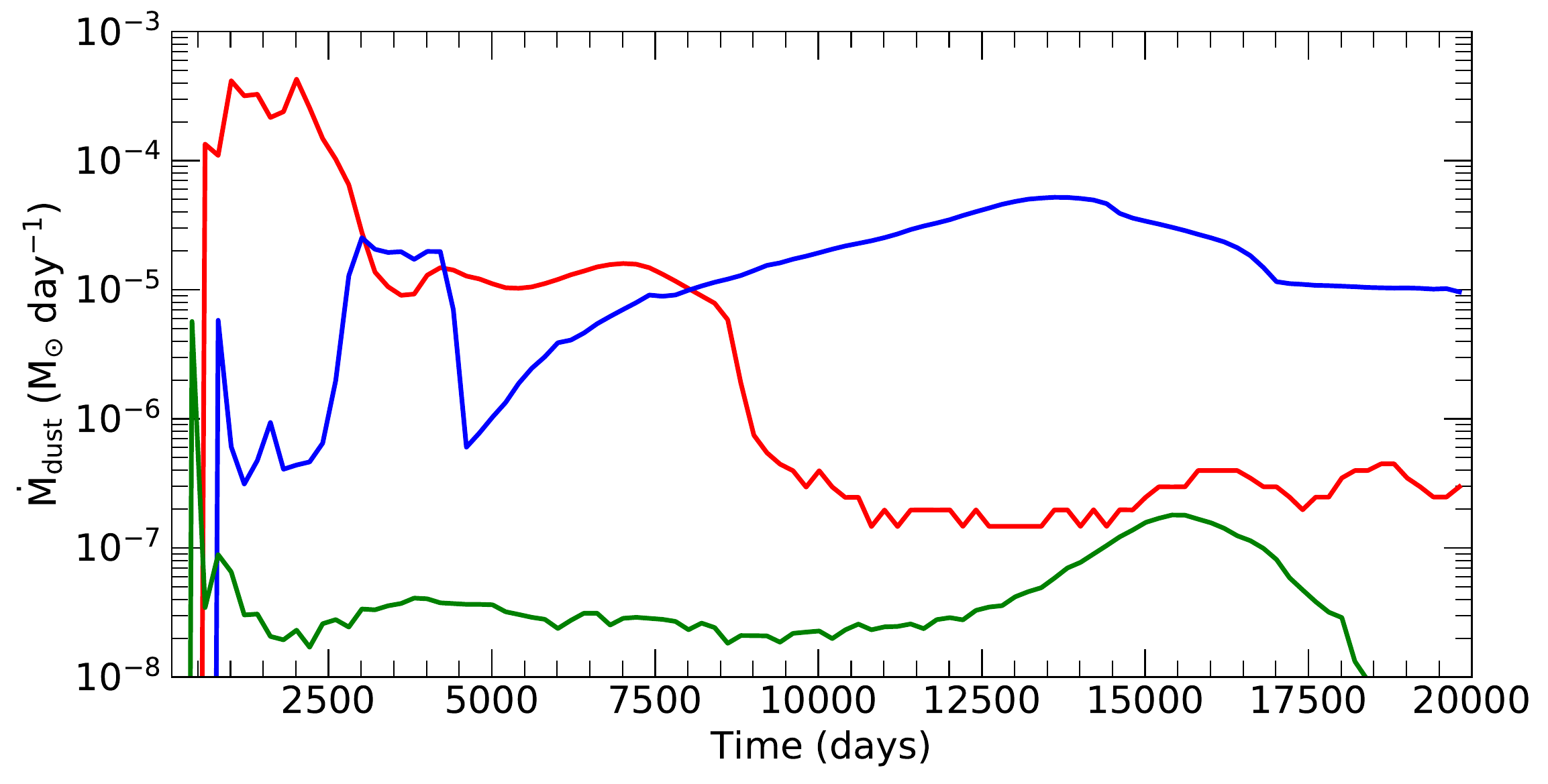}
\caption{\label{fig_M10}The following results for Cases 4, 5 and 6, all corresponding M$_{ej}$ = 10 \Ms\ (5~\Ms\ He-core, 5~\Ms\ of H-shell), and mass-loss rates varying as 10$^{-1}$, 10$^{-2}$, 10$^{-3}$ \Ms ~yr$^{-1}$, are presented as a function of post-explosion time: \textit{Top:} the evolution of total dust masses \textit{Middle:} the ratios of carbon-rich to oxygen-rich dust masses (the white region on the figure is C-rich and gray region is O-rich) \textit{Bottom:} the average rate of dust formation in per day. }
\end{figure}

\begin{figure}
\vspace*{0.3cm}
\centering
\includegraphics[width=3.5in]{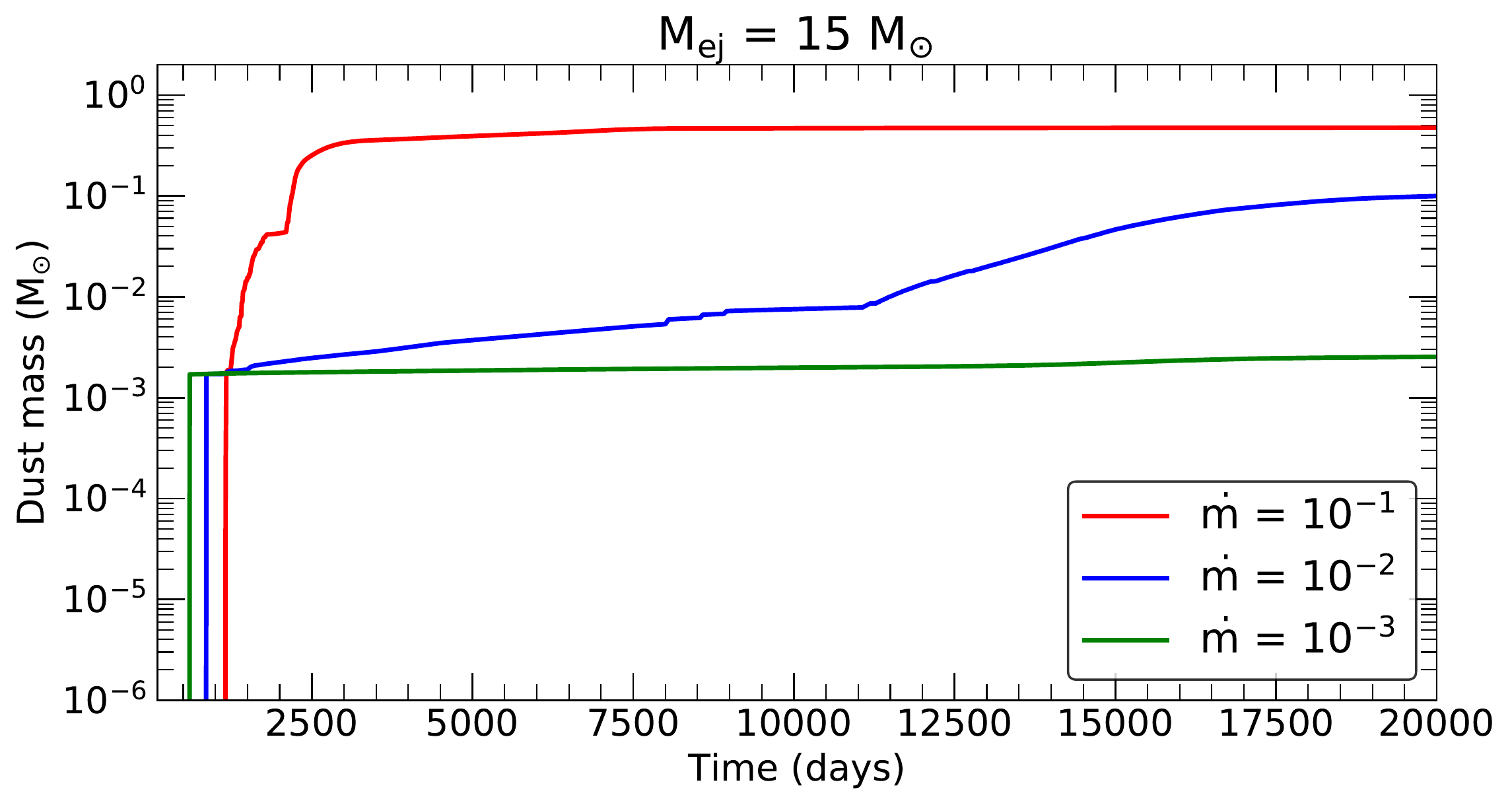}
\includegraphics[width=3.5in]{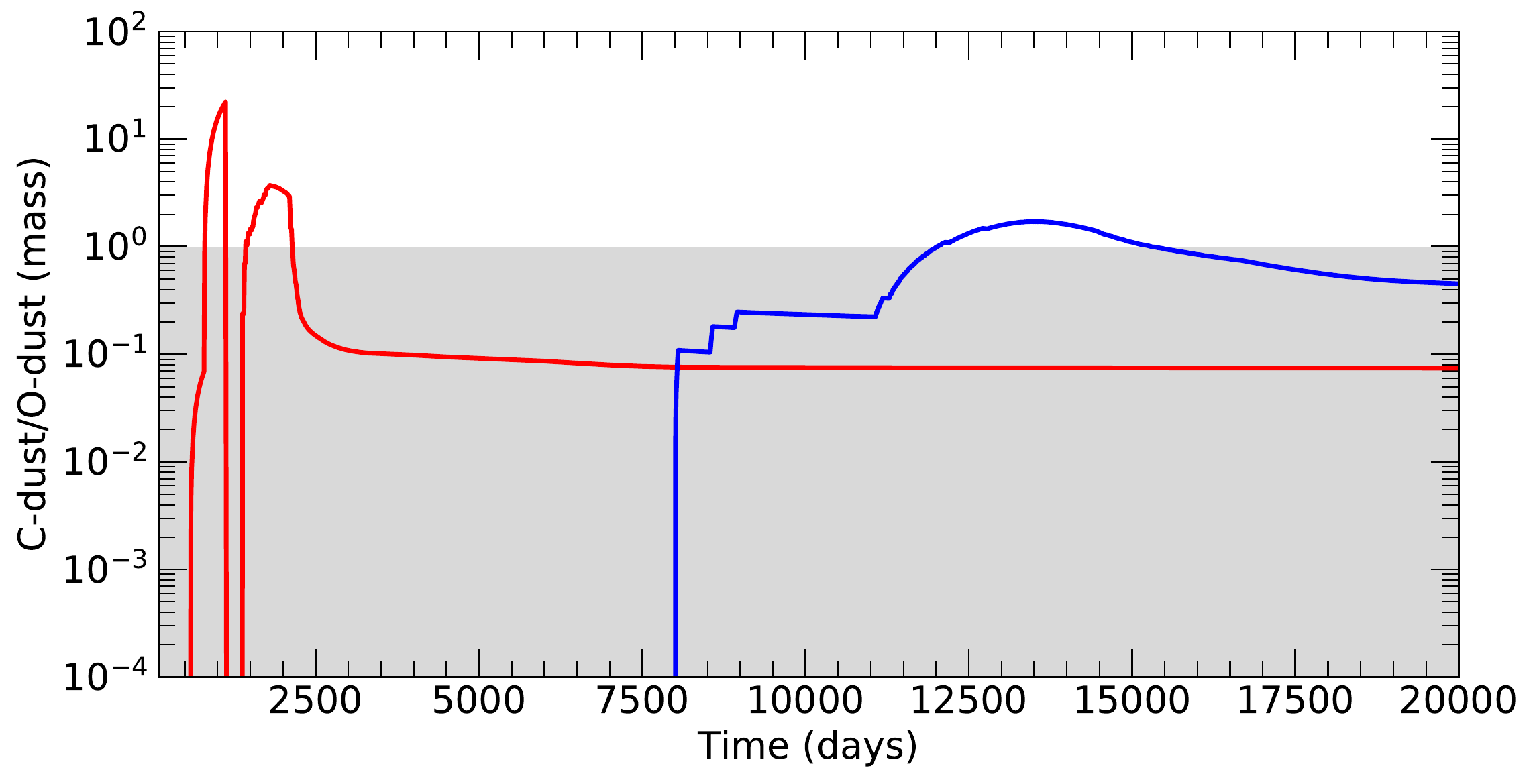}
\includegraphics[width=3.5in]{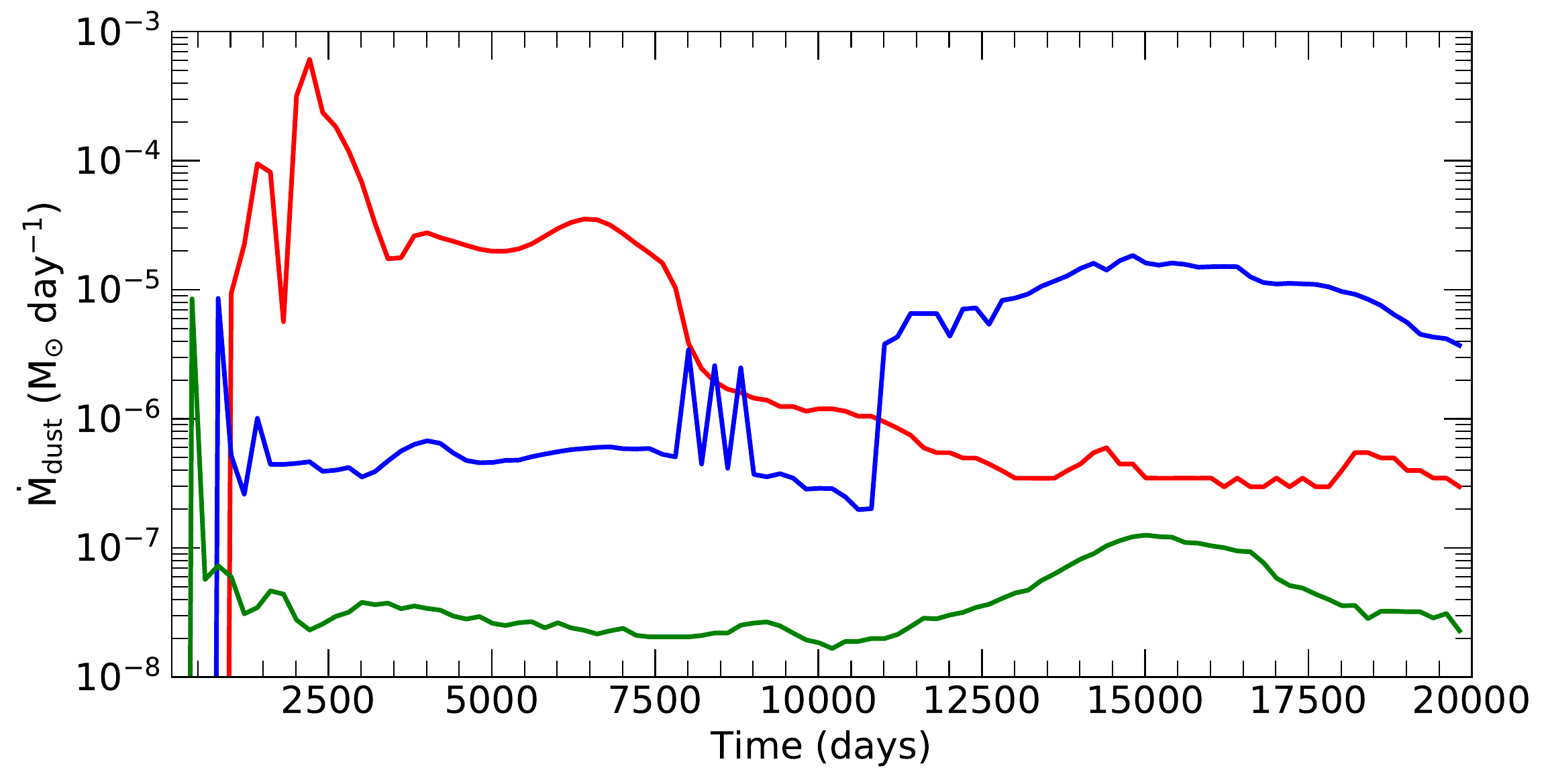}
\caption{\label{fig_M15}The following results for Cases 7, 8 and 9, all corresponding M$_{ej}$ = 15 \Ms\ (5~\Ms\ He-core, 10~\Ms\ of H-shell), and mass-loss rates varying as 10$^{-1}$, 10$^{-2}$, 10$^{-3}$ \Ms ~yr$^{-1}$, are presented as a function of post-explosion time: \textit{Top:} the evolution of total dust masses \textit{Middle:} the ratios of carbon-rich to oxygen-rich dust masses (the white region on the figure is C-rich and gray region is O-rich) \textit{Bottom:} the average rate of dust formation in per day. }
\end{figure}

The evolution of the total dust mass for each case, the mass ratios of C-rich to O-rich dust species, and the average dust formation rates for all the cases are shown in Figure \ref{fig_M6} for M$_{ej}$ = 6 \Ms,  Figure \ref{fig_M10} for M$_{ej}$ = 10 \Ms, and Figure \ref{fig_M15} for M$_{ej}$ = 15 \Ms, respectively. We trace the chemistry of WDS up to 20000 days from the time of explosion. Table \ref{table_dustmass} lists the masses of O-rich, C-rich and the total dust, present at day 1000, 5000, 10000 and 20000 for all the cases. The case numbers on the Table refers to the following individual cases ($\dot{m}$ in \Ms\ yr$^{-1}$, and M$_{ej}$ in \Ms). 

\begin{newcases}[noitemsep, nolistsep]
\item (\textit{$\dot{m}=10^{-1}, M_{ej} = 6$}) Dust formation in the shocked ejecta commences as early as day 620 after the explosion. The C-rich dust and O-rich dust species almost form simultaneously, however, is dominated by O-rich dust components all the time. Dust formation in the CSM begin around day 1200, and also is mostly O-rich in composition. The dust formation is fastest as in the first 1000 days, resulting in a mass of 0.5 \Ms, finally ending up at 0.74 \Ms\ after a decade. 
\item (\textit{$\dot{m}=10^{-2}, M_{ej} = 6$}) The RS travels through the He/C zone for the first 430 days. The formation of C-rich dust initiates in this zone around 570 days, and the formation of O-rich dust in the O/Si zone starts around day 950. The dust formation in the CSM commences after 1600 days. The total dust mass is about 0.02 \Ms\ at day 1000, which steadily increases to 0.36 \Ms\ by day 5000, and finally saturates at \til\ 0.57~\Ms. The WDS in this case is dominated by C-rich dust components until day 1700, and thereafter becomes O-rich.  
\item (\textit{$\dot{m}=10^{-3}, M_{ej} = 6$}) The RS passes through the He/C zone as long as 4000 days from the time of explosion. In this case, dust forms in the C-rich zone as early as day 400, however at a very controlled pace, which does not exceed 10$^{-5}$ \Ms\ day$^{-1}$. At day 1000 the mass of dust is 3$\times$10$^{-3}$ \Ms, which reaches 0.04 \Ms\ by day 10000 and \til\ 0.1 \Ms\ by day 20000. C-dust remains more abundant in the post-shock gas for almost the first 30 years of its evolution. 
\item (\textit{$\dot{m}=10^{-1}, M_{ej} = 10$}) In the WDS, dust appears around day 700 which has a C-rich composition. The rate of dust formation is fastest in the first 2500 days. O-rich dust forms in the O/Si zone around day 1000 and in the shocked CSM at \til\ day 1400. The mass of dust grows from 0.05 \Ms\ at day 1000 to 0.63 \Ms\ after a couple of decades. O-rich dust becomes the dominant dust component after day 1000.  
\item (\textit{$\dot{m}=10^{-2}, M_{ej} = 10$}) We have considered this as the benchmark case (Figure \ref{fig_dustmass}). Here, dust formation starts \til\ day 950 with O-rich dust species, and gains pace with the formation of C-dust after day 2200. The rate of formation varies between 10$^{-5}$ to 10$^{-4}$ \Ms\ day$^{-1}$. Formation of dust continue well after 10000 days. The mass of dust grows from \til\ 10$^{-3}$ \Ms\ at day 1000 to 0.035 \Ms\ at day 5000 to 0.35 \Ms\ at the end of 20000 days. The newly formed dust is dominated by C-rich components between day 2200 and day 9000, and O-rich components at other times. 
\item (\textit{$\dot{m}=10^{-3}, M_{ej} = 10$}) In this case, dust formation stars around 600 days in the shocked ejecta and around 900 days in the shocked CSM. However, the RS travels through the outer H-shell for most of the first 20000 days, so the formation of dust is limited to the H/He region of the ejecta and the CSM only. The final mass of dust here is 2.0$\times$10$^{-3}$ \Ms, almost completely made up of O-rich dust grains and a very small fraction of C-dust. 
\item (\textit{$\dot{m}=10^{-1}, M_{ej} = 15$}) Formation of O-dust commences around day 1100 days in the shocked ejecta gas, closely followed by the production of C-rich grains. Between 1500 and 2500 days, steady formation of C-dust leads to a dominance of C-rich dust species in the WDS. After 2500 days, however, the abundance of O-rich grains take the lead and remains so for the rest of the time. In this scenario, the WDS is dust free in the first 1000 days, however gradual yet steady growth after that, leads to a total mass of 0.47 \Ms\ of dust. 
\item (\textit{$\dot{m}=10^{-2}, M_{ej} = 15$}) Here, the formation of dust initiates in the H/He core of the shocked ejecta, as well as in the shocked CSM. Although it starts by day 850, the rate of synthesis remains below 10$^{-6}$ \Ms\ day$^{-1}$ until 10000 days, piling up only \til\ 7$\times$10$^{-3}$ \Ms\ of dust by that time. The rate of formation however boosts after that when dust formation commences in the He-core, finally leading to a mass of 0.1 \Ms. The WDS is found to be rich in C-dust components between 12500 and 15000 days. 
\item (\textit{$\dot{m}=10^{-3}, M_{ej} = 15$}) Similar to Case 6, dust production in the H/He shell starts as early as day 600. However, the final dust mass remains as low as 2.5$\times$10$^{-3}$ \Ms\ composed of O-dust grains only.

\end{newcases}



\subsection{Rate of dust formation}
\label{sec_dustformationrate}
The formation of dust, for any of the cases, is sporadic in nature, collectively depending on the shock, CSM, and the ejecta conditions. When suitable conditions are met in a parcel of gas, an immediate phase of rapid condensation is initiated; while at other epochs either the parcel remains devoid of dust or has already reached maximum condensation efficiency. For the WDS, the average rate of dust production in \Ms\ day$^{-1}$ is shown in Figures \ref{fig_M6}, \ref{fig_M10} and \ref{fig_M15}, corresponding to the mass of dust present at a given epoch. 

Unlike dust formation in ejecta, dust formation in the WDS in controlled by the evolution of the post-shock gas. As each parcel of gas evolves differently, at a given epoch only a small mass of gas attains conditions conducive to dust production. On that account, despite following similar chemical pathways, the overall formation rate of dust in interacting SNe is much more gradual. 


The figures indicate that when the mass-loss rates are higher, the formation of dust is more localized in time; in other words, dust forms rather rapidly in short phases, followed by periods of low dust formation. On the other hand, with an increase in the ejecta mass at the time of explosion, the rate of dust production in the WDS is more and more restrained, characterized by moderate rates for prolonged periods. 


As explained in Section \ref{sec_chemistry}, dust grains are newly synthesized when the gas temperatures are about 2000~K or lower. The spectra at the mid-IR wavelengths reflect to this newly formed dust, until it cools down to temperatures below the mid-IR range. 
Probing the same object in the far-IR or submillimeter wavelengths would reveal the dust that was formed earlier and has already cooled down. However, finding targets that are bright enough to be probed at such low energies with currently available instruments, is a bit of a challenge. Therefore, our observations of dust in interacting SNe are closely correlated to the dust production rate,  instead of the total dust mass at a given time. The formation rates, as shown in the figures, vary between 10$^{-7}$ and 10$^{-3}$ \Ms\ day$^{-1}$, which multiplied by the dust cooling time will then be reflected in the mid-IR spectra.
An accurate estimation of the dust temperatures and the cooling time requires an elaborate analysis, taking into account the radiative and collisional energy balance of the grains \citep{dwe87, hol79, sarangi2018}, which will be addressed in a subsequent study. 


Importantly, in SN~2010jl, the dust formed in the WDS was found to be thick enough to completely obscure the inner RS that is heating the dust up \citep{sarangi2018}. If enough dust is formed in the dense shell resulting in large optical depths at mid-IR wavelengths, then of course, the mid-IR data will be insufficient to estimate of the total dust present or the dust production rate at any given epochs.



The impact of additional cooling of the gas due to dust formation was not accounted for in this study, which has the potential to affect the chronology of dust formation derived here.

\subsection{Dependence on the mass-loss rate}
\label{sec_massloss}

The pre-explosion mass loss rates ($\dot{m}$) are chosen to be 10$^{-1}$, 10$^{-2}$ and 10$^{-3}$ \Ms\ yr$^{-1}$. Figures \ref{fig_M6}, \ref{fig_M10} and \ref{fig_M15} show the individual cases, which are summarized in Figure \ref{fig_dust_massloss}.  

\begin{figure}
\vspace*{0.3cm}
\centering
\includegraphics[width=3.5in]{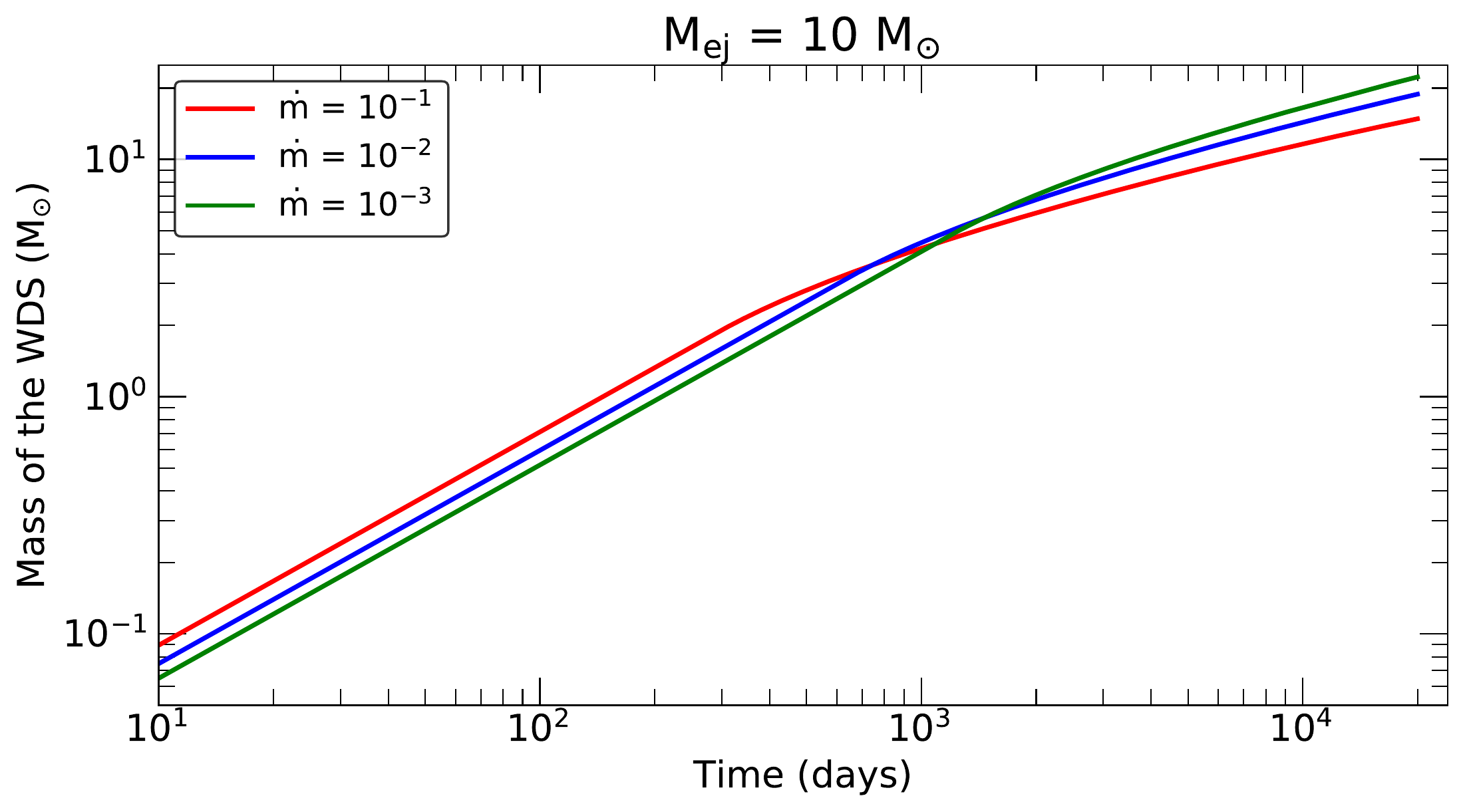}
\includegraphics[width=3.5in]{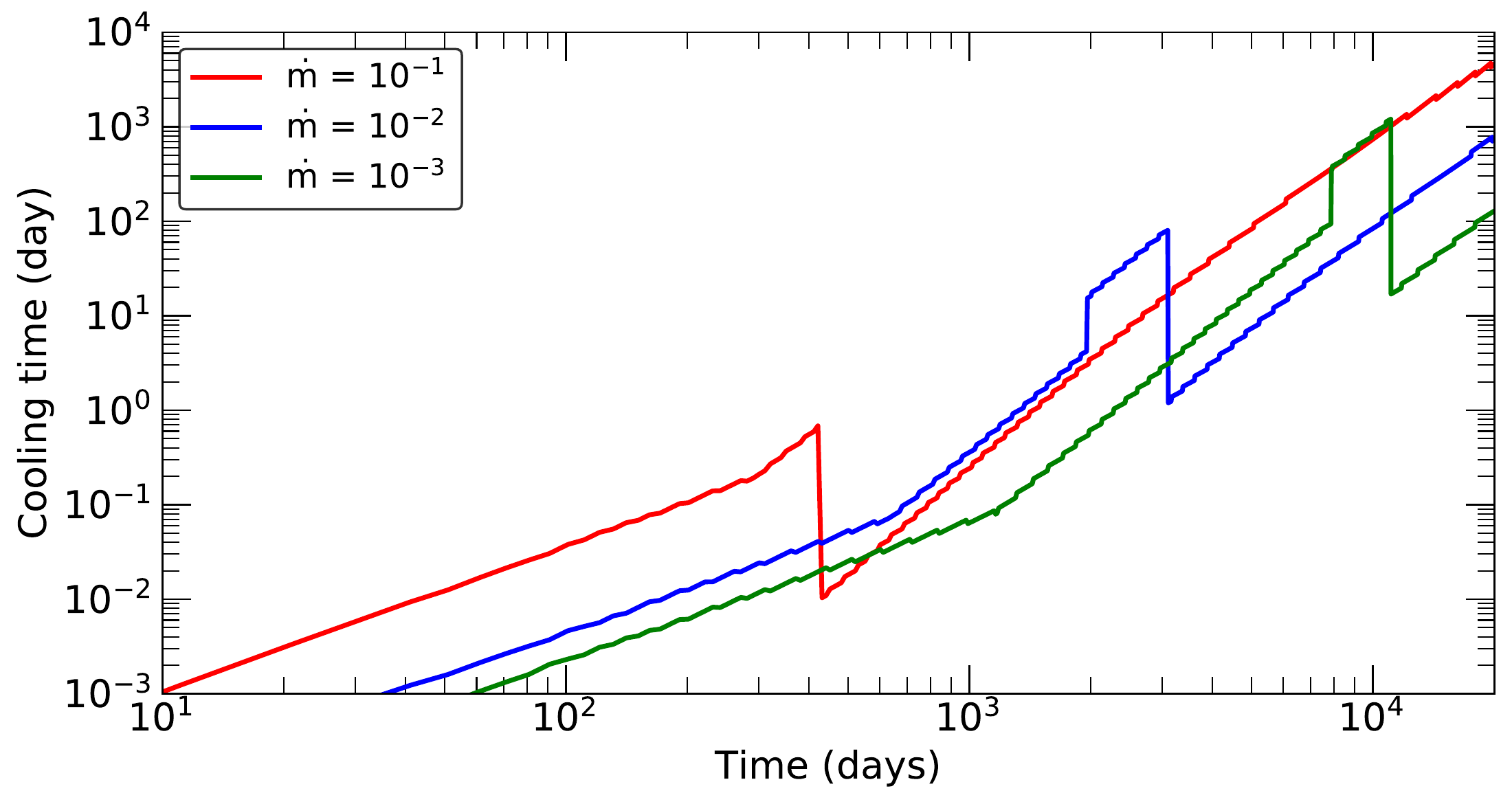}
\includegraphics[width=3.5in]{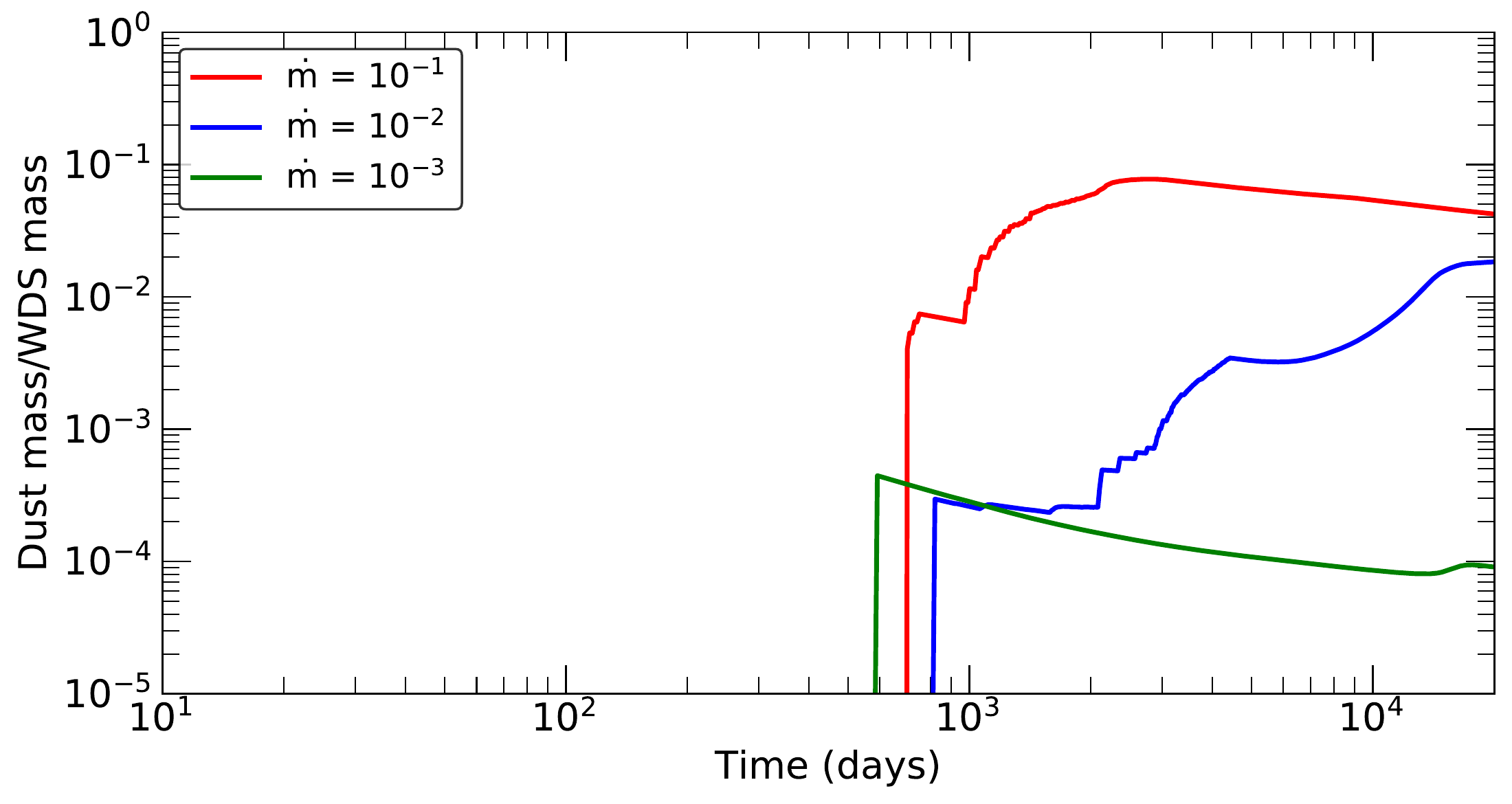}
\caption{\label{fig_depmassloss} The following results for M$_{ej}$ = 10 \Ms\ (5~\Ms\ He-core, 10~\Ms\ of H-shell), and mass-loss rates varying as 10$^{-1}$, 10$^{-2}$, 10$^{-3}$ \Ms ~yr$^{-1}$, are presented (Cases 4, 5, 6) as a function of post-explosion time: \textit{Top:} the mass of the post-shock gas in the WDS. \textit{Middle:} the cooling time of the shocked ejecta in the WDS (for the nature of the figure, please see Section \ref{sec_evolTn} and Figure \ref{fig_coolingtime}).  \textit{Bottom:} the ratios of dust to gas mass in the WDS}
\end{figure}

The dust mass yields can be characterized by the total mass that is produced, and the first epoch of dust formation. Figure \ref{fig_depmassloss} illustrates these factors with respect to a fixed ejecta mass of M$_{ej}$ = 10 \Ms, by showing the mass of the WDS and the cooling time, in the context of the dust to gas mass ratio. The epoch of dust formation is mainly a function of the cooling time, which is further a function of the pre-shock density, the shock velocity and the composition of the gas (see Section \ref{sec_cooling}). As we see in the figure (\ref{fig_depmassloss}), when the mass-loss rate is the lowest (10$^{-3}$~\Ms~yr$^{-1}$), the onset of dust formation is the earliest in that case, since the cooling time is small, owing to lower shock velocities. In the same line, as Figure \ref{fig_dust_massloss} shows, in the case of the highest $\dot{m}$ and M$_{ej}$ = 15 \Ms, dust formation does not start until later than day 1000, owing to the long cooling time; however once it commences, the rate of formation is rapid. 

In the opposite scenario, a higher mass-loss rate results in a denser CSM, that creates a slower FS and slower RS in the frame of the ISM. In the frame of the ejecta, therefore, the RS velocities are proportional to the mass-loss rates. That is reflected in the rate of increase in the mass of the WDS (which is proportional to the shock velocity), shown in the top-panel of Figure \ref{fig_depmassloss}. Now that the WDS piles up mass at a faster pace, the RS-interaction with the inner metal-rich part of the ejecta happens earlier, when the gas is denser, resulting in more efficient dust formation. As a consequence, the total amount of dust produced is always proportional to the pre-explosion mass-loss rate, as shown in Figure \ref{fig_dust_massloss}.

 Importantly, in the Figure \ref{fig_depmassloss}, the decrease in the dust to gas mass ratio at later times should not be confused as a decrease in the total mass of dust; it is actually indicative of the decrease in efficiency of dust production at later times, even though the mass of the WDS continue to increase.

\subsection{Dependence on mass of ejecta}
\label{sec_ejectamass}

The mass of the ejecta, M$_{ej}$, at the time of explosion, was taken to be 6, 10 and 15 \Ms, which was manifested in terms of how much of the H-shell was lost through mass-loss prior to the explosion. In other words, we have modeled SN ejecta with a 5 \Ms\ metal-rich He-core, and 1, 5 and 10 \Ms\ of H-shell respectively. 

With reference to Table \ref{table_dustmass} and the summary presented in Figure \ref{fig_dust_massloss}, it is evident that smaller the mass of the ejecta, or rather smaller the mass of H-shell present during the explosion, larger is the mass of dust produced, and vice-versa. 
So, even though the presence of a large H-shell in the outer ejecta, say of 10~\Ms, increases the total mass of ejecta by more than twice, it is found to substantially decrease the yield of dust. The two responsible factors behind this are the following: (i) The H-shell lies in the outermost layer of the ejecta, which first encounters the RS (ii) The abundance of the dust producing metals is less than 10$^{-2}$ (in mass) in that zone, unlike the other ejecta layers. 

Therefore, the RS spends a long span of its evolution phase passing through the large H-shell, yet can only produce a small amount of dust due to low metal abundances. By the time the inner He-core is reached, gas densities are already too low in the metal-rich core. This effect is compounded when the pre-explosion mass-loss rates are smaller. In those cases, the final dust mass remains of the order of 10$^{-3}$~\Ms\ only. 

In addition, the ejecta mass at the time of explosion also influences the ratio of C-rich to O-rich dust components, given that the extent of H-shell determines the time when the RS will pass through different zones of the inner ejecta. In other words, if the H-shell is small, then the RS reaches the inner O-rich zone rather quickly, leading to an efficient formation of O-rich dust. In the opposite scenario where the H-shell is large, by the time the gas in inner O-core is shocked by the RS, the efficiency of dust production declines, and hence the C-dust wins in relative proportion. 

Importantly, in this paper, we have assumed the He-core mass to be fixed at 5 \Ms\ corresponding to a 21 \Ms\ main sequence star. The mass of the He-core is known to be proportional to the main sequence mass of the star \citep{rau02, woosley_2019}. A change in the He-core mass will proportionately change the mass of the dust, however, the relative sizes of the He/C and the O/Si zones will determine the dust composition; we shall address this in a subsequent study.

\subsection{Other scenarios}



Apart from the 9 cases, there could be many other interesting scenarios. 

When expanding the parameter space, the alteration of explosion energy plays an important role in the evolution of the SN ejecta and its nature of interaction with the dense CSM. The explosion energy is often proportional to the mass of the progenitor star, and could also be related to the amount of \Ni\ produced during the explosion \citep{sar13}, which impacts the chemistry of the ejecta. However, given that the velocities depend on the ratio of E and M$_{ej}$, the evolution of the shock through the CSM will not be transformed significantly. 

In some extreme cases, the star could also lose a part of its He-core in the pre-explosion mass loss. In that scenario, the inner part of the CSM will likely to be C-rich, and will produce dust simultaneously with the shocked ejecta. 

We have assumed a steady mass-loss rate in this study. In reality, the mass-loss events are often sporadic and cannot be averaged by a single rate. Moreover, there is often an offset between the mass-loss events and the time of explosion, which creates a low density region between the CSM and the star \citep{gal14, sarangi2018}. That will influence the timeline of dust formation in the WDS profoundly, and depending on the distance of the CSM, the density of the pre-shock ejecta will be altered, which controls the efficiency of dust formation.

\section{Summary}
\label{sec_summary}

Let us now summarize the findings of this study. 
The evolution of dust in the WDS of interacting SNe are calculated for 9 different cases, characterized by the variation in the mass of the ejecta (M$_{ej}$) and the pre-explosion mass-loss rate ($\dot{m}$). The variation of ejecta mass was expressed in terms of the mass of the H-shell present in the ejecta at the time of explosion, and how much was lost due to rapid pre-explosion mass-loss.

\begin{figure}
\vspace*{0.3cm}
\centering
\includegraphics[width=3.5in]{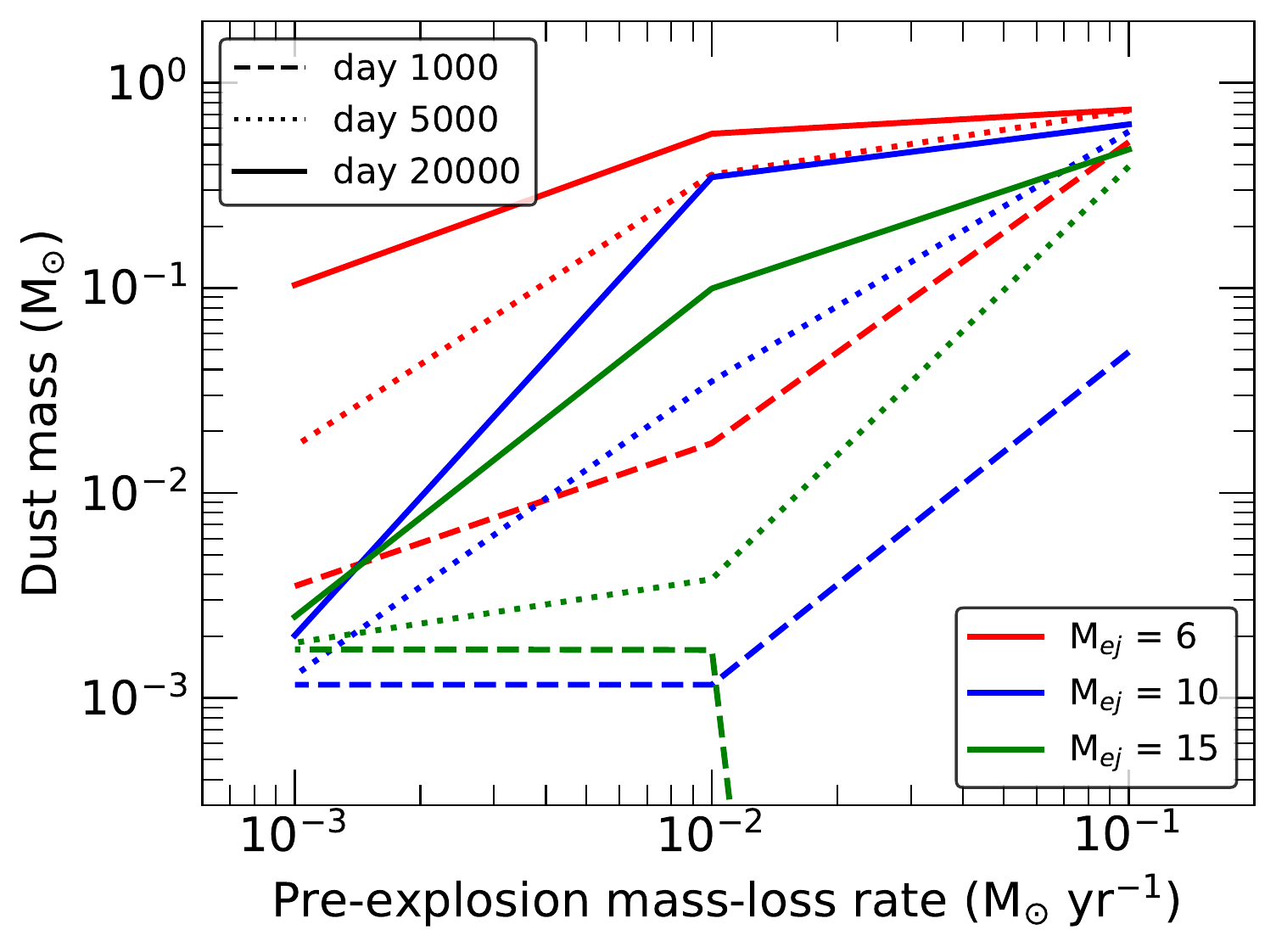}
\caption{\label{fig_dust_massloss} The figure shows the dust masses as function of pre-explosion mass-loss rate, for different ejecta masses at three different epochs. Please see Section \ref{sec_parameters} for the details of all the parameters used as initial conditions. The main sequence mass of the progenitor in all the cases is 21 \Ms, and the residual neutron star (NS) after the explosion is 1.75 \Ms. The mass of the ejecta, M$_{ej}$, is the exploded mass in each case. The rest the mass (21 - M$_{ej}$ - NS) is lost due to pre-explosion mass-loss, with the specified rates; and it forms the CSM. We have used an upper limit on the CSM mass to be 10 \Ms.} 
\end{figure}

In Figure \ref{fig_dust_massloss}, the dust masses at day 1000, 5000 and day 20000 are shown for all the cases, as the function of $\dot{m}$ and M$_{ej}$. The non-linear nature of the dust formation is evident from the figure. The trends suggest that with high mass-loss rates, the dust forms more efficiently, so all the curves converge closely. On the other hand, in the intermediate case, where $\dot{m}$ is 10$^{-2}$ \Ms\ yr$^{-1}$, there is a large spread between the lines, indicating that dust formation is heavily dependent on time and ejecta mass. In case of lower $\dot{m}$, dust only forms efficiently if most of the H-shell is lost in the pre-explosion stage. Moreover, the trend also indicates that when the mass-loss rates are even smaller, the WDS will not provide a favorable environment to produce dust. 

Let us summarize the important findings of this study in the following points, with reference to the Case numbers in Section \ref{sec_dustmasses}.

\begin{enumerate}[label=(\arabic*), noitemsep, leftmargin=*, align=left, wide = 0pt]
\item Final dust masses are found to vary widely; the largest mass of dust is 0.74 \Ms\ formed in Case 1,  where the mass-loss rate is the highest and ejecta mass is the lowest. In the opposite scenarios, when the ejecta has a large H-shell and $\dot{m}$ is small, the final dust masses formed in the WDS is only about 10$^{-3}$ \Ms.

\item We found that when suitable conditions prevail, dust formation can accelerate as late as 20 years from the time of the explosion. In many scenarios of interacting SNe, the rate of dust production is much more gradual compared to the ejecta of a Type II-P SN, since the dust formation here is associated with the slow evolution of the RS through the ejecta. In other words, only a small mass fraction of the total ejecta mass attains suitable conditions for dust synthesis at a given time. 

\item The gas in the shocked ejecta is much more efficient in producing dust owing to its large metal abundances, compared to the CSM. However, the CSM and the H-rich shell encounters the shock before the metal-rich He-core. So the site of the earliest dust formation basically depends on the cooling time of these zones.  

\item The larger the mean molecular weight of the gas, the faster is the cooling, which compensates for the lower densities as the metal-rich zones in the ejecta are shocked later (hence lower densities) compared to the outer H-rich zones. In interacting SNe the cooling time of the radiative RS is extremely short that ranges from less than a day to a maximum of up to few 100 days. 

\item A delayed encounter with the RS definitely reduces the productivity of the post-RS ejecta; in addition, it also alters the relative fractions of O-rich and C-rich dust components. This trend is reflected in a reduced net mass of dust, when the mass of the H-shell is larger. 

\item A high mass-loss rate leads to denser CSM, which translates into a slower FS. This scenario results in a piling-up of mass in the WDS at a faster pace, thereby also increasing the efficiency of dust synthesis.  

\item The formation of dust is controlled by the flux of downstream radiation generated at the shock-front, towards the cooling post-shock gas, which induces the formation of the WDS. This radiation, when too strong, increases the gas temperature and ionization, consequently delaying the synthesis of stable dust precursors. More importantly however, the radiation keeps the shell warm for a long duration, allowing enough time for the key nucleation processes to complete. 

\item This work also stresses that, whether the ejecta is dominated by C-rich or O-rich dust, is a function of the epoch it is observed. The ratio of the dust masses belonging to these two categories, shown in Figures \ref{fig_M6}, \ref{fig_M10} and \ref{fig_M15}, clearly indicates that the WDS can be either one at a given epoch, based on the initial conditions. The lower mass-loss rates generally leads to a larger concentration of C-rich grains, since the RS takes a long time to cross the He/C zone where the C-dust is formed. 

\item It is important to realize, the rate of dust production is a more observationally relevant quantity in such SNe, compared to the final dust masses, since observations are mostly limited to warm dust in mid-IR wavelengths, which is analogous to the newly formed dust.

\item Looking at Figure \ref{fig_dust_massloss}, it can be implied that, the later the interaction, or lower the mass-loss rate, the less efficient is the dust production in the WDS. A very late interaction between the RS and the ejecta, like in case of regular Type II-P SNe, will therefore not provide an environment, which is suitable enough to synthesize new dust of any significant amount in the post-shock gas.

\item In this paper, we have assumed that all the pre-explosion mass-loss is correlated to the mass of the dense CSM. However, in reality, observations often suggest that the dense CSM surrounding the exploding SN is thin, and only consists of a fraction of the mass that the star has lost prior to the explosion. This can lead to a change in the behavior of the FS, which will be addressed when we model individual SNe in subsequent studies.
\end{enumerate}

Based on the decreasing column densities of the unshocked ejecta and the presence of strong downstream radiation at the earlier epochs originating from the FS and the RS, we argue that the shocked gas in the WDS yields a more suitable environment for dust production, compared to the unshocked ejecta. Our previous study on SN~2010jl \citep{sarangi2018} also clearly advocated for dust to have formed in the shocked-gas. Moreover, the dust formed in the unshocked ejecta will be subjected to depletion as it encounters the RS. Empirically speaking, given the high densities of the ejecta and high velocities of the shock, added to the premature dust formation (early times, smaller grains), the survival probability of the ejecta dust will be seriously challenged. Having said that, there are evidences for the co-existence of dust in the shocked, along with the unshocked ejecta, like in SN~1998S \citep{mauerhan_2012}. This study has not accounted for dust destruction by the RS, but only considered the formation of new dust grains in a dust-free shocked ejecta in the WDS.  

Finally, as concluding remarks, we will now discuss, some important aspects this study can potentially address.

The RS is known to destroy a significant amount of dust formed in the SN ejecta \citep{Micelotta_2016, florian_2019, slavin_2020}; estimated survival rates are between 10-30\%. In interacting SNe, dust forms in the WDS which is already behind the RS, so this dust will not be subjected to destruction by the catastrophic RS. In this context, despite being a small fraction of all core-collapse SNe, their net contribution to the dust in the galaxies could be significant. A detail statistics of various types of interacting SNe and their respective properties are necessary to quantify this claim.

Massive Luminous blue variable (LBV) stars are generally perceived as parents of interacting SNe. In the low metallicity galaxies of the early universe, characterized by a top-heavy IMF \citep{marks_2012, martin_2015}, the LBVs are expected to be more common. Dust formation in a dense shell, following an interaction with a shock, could therefore be an important mechanism of dust formation in the SNe of the low-metallicity galaxies, as also pointed out by \cite{chevalier_2008}. 

One of the major puzzles of dust formation in core-collapse SNe in general, is to account for the very late onset (between 2 and 20 years after explosion) for the formation of presolar silicon carbide grains of SN origin, found in meteorites \citep{liu_2018, ott_2019}. Dust formation models of SN ejecta \citep{sar13, sluder2018} indicate that the formation of dust starts early and quickly saturates; at late times, the conditions are not viable to support new chemical pathways involving gas phase reactions. In this paper, we have presented results showing that in some scenarios, dust in these interacting core-collapse SNe can start to form in a dense shell at very late times. \cite{hoppe_2018} suggests that the isotopic signatures of presolar silicon carbide grains of SN origin, from the Murchison meteorite, matches well with a 25~\Ms\ progenitor model, which again increases the chances of the SN to be Type IIn-like.

\subsection{Improvements}
\label{sec_improvements}

This is the first model to address dust formation in the WDS of interacting SNe. We acknowledge that the assumption of a homogeneous density is not ideal. However, while the effects of clumping of the ejecta are very important for the case where the reverse shock is non-radiative (for example the shock can become radiative in the clumps), it is not expected to be so important when the shock is radiative. This is because, if the shock is radiative for both the clumps and the inter-clump smooth ejecta then, other than likely adding some waviness to the shock front, we expect less significant effects from the clumps. All of the ejecta will enter the RS, become radiative, and cool over a short distance. In addition, in SN~2010jl \citep{sarangi2018} we found that the dust formed in the shocked CSM, behind the FS, is optically thick (to UV-optical light), which clearly indicates to a relatively smoother gas. 

In this model, our 1D approach could not account for turbulences and mixing in the post-shock gas, which can potentially change the nature of cooling and dust formation. In the future, we aim to address these anomalies by developing a 2D hydrodynamics model for interacting SNe.

A crucial component of this calculation is the estimation of the radiative cooling rates. In the future, for better estimation, the history of heating and ionization of the ejecta by the shock should be accounted for. Importantly, in the future, the molecular cooling needs to be self-consistently addressed, coupled to the chemistry of the post-shock gas. The cooling of the ejecta by dust emissions, once the dust is formed in the WDS, was not considered in this model, and we acknowledge its importance. 


Finally, for a complete picture, dust formation in the ejecta and the WDS should be combined with the pre-existing dust in the CSM and the surviving dust behind the RS, which will then be able to reproduce the observed IR spectra of any interacting SN. 

In this paper, we did not base our model on any specific SN. When addressing the features with respect to specific objects, the parameters that we choose in our model needs to modified accordingly. In subsequent studies, we will be applying this approach to real SNe to account for their distinct observational manifestations.

 
\  \ \ 

We are very thankful to the referee for carefully going through the manuscript and providing very useful suggestions.  

We are extremely grateful to Dr. Isabelle Cherchneff for the chemistry of SNe, that Arka developed together with her for previous studies, which was also applied here.  


In addition, Arka is also very thankful for the useful discussions with Dr. Christa Gall and Professor Jens Hjorth at DARK, and Dr. Miguel Avillez  at University of Evora, Portugal.

This work was supported by a grant from VILLUM FONDEN, PI Jens Hjorth (project number 16599).

\ \ 

\software{CLOUDY \url{https://gitlab.nublado.org/cloudy/cloudy/-/wikis/home} \citep{ferland_2013}}


\bibliographystyle{aasjournal.bst}
\bibliography{Bibliography_sarangi}

\appendix 
\section{Appendix}
\label{appendix}
 \renewcommand{\theequation}{A\arabic{equation}}
Here all the relevant equations are derived for the evolution of the ejecta and the CSM. The density and velocity structure of the unshocked ejecta are taken from \cite{tru99} and the evolution of the shock in interacting SNe is derived with reference to the formalism by \cite{moriya_2013}. 

The ejecta is divided into an inner core and an outer envelope. The core is at constant density ($\rho_c$), while the density of the envelope ($\rho_e$) drops as a power-law of $n$. The density of a parcel of ejecta-gas, which is at velocity $v$~(= $r/t$) at a given time $t$, is assumed to be, 
\begin{equation}
\label{eqn_ejecta_rhodensity}
\rho_{c}(v,t) = \frac{M}{v_{M}^3} f_0 t^{-3}, \ \ \ \ \ \ \ \ \rho_{e}(v,t) = \frac{M}{v_{M}^3} f_n w^{-n} t^{-3},
\end{equation}
where $M$ is the mass of the ejecta, $v_M$ is the maximum ejecta velocity ($i.e.$, the velocity of the outermost layer of the ejecta), and $w = v/v_M$. From the continuity of the density at the core-envelope boundary (where core-velocity at the core-boundary is $v_c$ and $w_c = v_c/v_M$), we have, 
\begin{equation}
\label{eqn_f0fn}
f_0 = f_n w_c^{-n}, \ \ \ \ \ \ \ \ f_n = f_0 w_c^n, \ \ \ \ \ \ \  \rho_{e}(v,t) = \frac{M}{v_{M}^3} f_0 w_c^n w^{-n} t^{-3} = \frac{M}{v_{M}^3} f_0 \Big(\frac{v_c}{v}\Big)^n t^{-3} 
\end{equation}
The sum of the total ejecta therefore is, 
\begin{equation}
\label{eqn_totalmass}
\begin{gathered}
M = \int_0^{v_M t} 4 \pi r^2 \rho_{ej} d\mathrm{r} = \int_0^{v_c t} 4 \pi r^2 \rho_{c} d\mathrm{r} + \int_{v_c t}^{v_M t} 4 \pi r^2 \rho_{e} d\mathrm{r}  \\ 
= \frac{4 \pi}{3} \frac{M}{v_{M}^3} f_0 t^{-3} \big(v_c t \big)^{3} + \frac{4 \pi}{3-n} \frac{M}{v_{M}^3} v_c^n f_0 t^{n-3} \Big[\big(v_M t \big)^{3-n} - \big(v_c t \big)^{3-n}\Big]  \\
= \frac{4 \pi}{3} M f_0 w_c^{3} + \frac{4 \pi}{3-n} M f_0 w_c^{3} (w_c^{n-3}-1)
\end{gathered} 
\end{equation}
Here we have assumed the radius of the stellar remnant to be very small, thereby taking the lower limit of the integration as 0. Canceling $M$ from either side, $f_0$ can be expressed as, 
\begin{equation}
\label{eqn_f0_value}
\begin{gathered}
f_0 \frac{4 \pi}{3} w_c^{3} \Big[ 1 + \frac{3}{3-n} (w_c^{n-3}-1) \Big] = f_0 \frac{4 \pi}{3} w_c^{3} \frac{3w_c^{n-3} - n}{3-n} = 1 \\
f_0 = \frac{3}{4 \pi} w_c^{-3} \frac{n-3}{n - 3w_c^{n-3}} \ \ \ \ \ \rightarrow \ \ \ \ \ \ \ \ \ \ f_0 = \frac{3}{4 \pi} \frac{n-3}{n} w_c^{-3};  \ \ \ \ \ \ \ \ (w_c << 1) 
\end{gathered}
\end{equation}
Using the expression for $f_0$, we can rewrite the equation for density as, 
\begin{equation}
\label{eqn_f0_rho}
\rho_{c}(v,t) = \frac{3}{4 \pi} \frac{n-3}{n} \frac{M}{v_{c}^3} t^{-3}, \ \ \ \ \ \ \rho_{e}(v,t) = \frac{3}{4 \pi} \frac{n-3}{n} \frac{M}{v_{c}^3} \Big(\frac{v_c}{v}\Big)^n t^{-3}
\end{equation}
It is apparent that the density of the ejecta is therefore independent of $v_M$, in the limit that $v_M >> v_c$. 

The kinetic energy of the ejecta, which is equal to the explosion energy, $E$, can be expressed as the sum of the kinetic energy of the ejecta-core and the ejecta-envelope, $E = E_c + E_e$. 
\begin{equation}
\label{eqn_energy_core}
\begin{gathered}
E_c = \int_0^{v_c t} \frac{1}{2} v^2 d\mathrm{m} = \int_0^{v_c t} \frac{1}{2} 4 \pi r^2 \rho_{c} v^2 d\mathrm{r} = \frac{2 \pi}{5} \frac{M}{v_{M}^3} f_0 t^{-3} t^{-2} \Big| r^5 \Big|_0^{v_c t} = \frac{2 \pi}{5} M f_0 w_c^5 v_M^2 \\ 
E_e = \int_{v_c t}^{v_M t} \frac{1}{2} 4 \pi r^2 \rho_{e} v^2 d\mathrm{r} =  2 \pi \frac{M}{v_{M}^3} f_0 v_c^n t^{-3} t^{-2}  t^n \int_{v_c t}^{v_M t} r^2 v^{-n} v^2 d\mathrm{r}   \\ 
= \frac{2 \pi}{5-n}\frac{M}{v_{M}^3} f_0 t^{n-5} v_c^n \Big| r^{5-n} \Big|_{v_c t}^{v_M t} = \frac{2 \pi}{5-n} M f_0 w_c^5 v_M^2 (w_c^{n-5}-1) 
\end{gathered}
\end{equation}
The total energy therefore turns out to be, 
\begin{equation}
\label{eqn_corevelocity}
\begin{gathered}
E = E_c + E_e = \frac{2 \pi}{5} M f_0 w_c^5 v_M^2 \Big[1 + \frac{5}{5-n} (w_c^{n-5}-1)\Big] = \frac{2 \pi}{5} M f_0 w_c^5 v_M^2 \frac{5 w_c^{n-5} - n}{5-n}  \\
\frac{E}{M v_M^2} = \frac{2 \pi}{5} f_0 w_c^5 \frac{n}{n-5} = \frac{2 \pi}{5} w_c^5 \frac{n}{n-5} \frac{3}{4 \pi} \frac{n-3}{n} w_c^{-3} = \frac{3(n-3)}{10(n-5)}w_c^2;  \ \ \ \ \ \ \ \ (w_c << 1)  \\
v_c^2 = \frac{10(n-5)}{3(n-3)} \frac{E}{M} \ \ \ \ \ \rightarrow v_c = \Big[\frac{10(n-5)}{3(n-3)} \frac{E}{M}\Big]^{1/2} 
\end{gathered}
\end{equation}
To derive the density of CSM, $\rho_{csm}$, we assume a steady mass-loss rate, therefore the density declines with radius as a power-law with exponent $s = -2$. Taking the velocity of the pre-explosion wind to be $v_w$, we can write, 
\begin{equation}
\dot{m} = 4 \pi r^2 \rho_{csm} v_w \ \ \ \ \ \ \ \rightarrow \ \ \ \rho_{csm} = A r^{s} = A r^{-2} = \frac{\dot{m}}{4 \pi r^2 v_w} \ \ \ \ \ \ \ \rightarrow \ \ \ A = \frac{\dot{m}}{4 \pi v_w} 
\end{equation}

The FS propagates through the CSM and the RS through the ejecta. Let the mass of the shocked ejecta be $m'_{ej}$ and the shocked CSM be $m'_{csm}$. The equation for the conservation of momentum can be written as, 
\begin{equation}
\label{eqn_momentum_ejecta}
\begin{gathered}
\frac{d}{d\mathrm{t}}(mv) = \frac{d}{d\mathrm{t}}(m'_{ej} v_{RS} + m'_{csm} v_{FS}) \\
\frac{d}{d\mathrm{t}}(m'_{csm} v_{FS}) = m'_{csm}  \frac{d v_{FS}}{d\mathrm{t}} +  v_{FS} \frac{d m'_{csm}}{d\mathrm{t}} = m'_{csm} \frac{d v_{FS}}{d\mathrm{t}} +  v_{FS} \frac{d}{d\mathrm{t}}(4 \pi r_{FS}^2 \rho_{csm} d\mathrm{r_{FS}})  \\
m'_{csm} \frac{d v_{FS}}{d\mathrm{t}} +  v_{FS} 4 \pi r_{FS}^2 \rho_{csm} v_{FS} = m'_{csm} \frac{d v_{FS}}{d\mathrm{t}} +  4 \pi r_{FS}^2 \rho_{csm} v_{FS}^2 \\
\frac{d}{d\mathrm{t}}(m'_{ej} v_{RS}) =  m'_{ej}  \frac{d v_{RS}}{d\mathrm{t}} +  v_{RS} \frac{d m'_{ej}}{d\mathrm{t}} \\
= m'_{ej}  \Big[\frac{d v_{FS}}{d\mathrm{t}}-\frac{d v_{ej}}{d\mathrm{t}}\Big] +  (v_{FS} - v_{ej}) \frac{d}{d\mathrm{t}}(4 \pi r_{FS}^2 \rho_{ej} d\mathrm{r} - 4 \pi r_{FS}^2 \rho_{ej} d\mathrm{r_{FS}})  \\
= m'_{ej}  \frac{d v_{FS}}{d\mathrm{t}} + 4 \pi r_{FS}^2 \rho_{ej}(v_{FS} - v_{ej})(v_{ej} - v_{FS})
\end{gathered}
\end{equation}
Here we have used $dv_{ej}/dt = 0$, since a parcel of gas in the homologously expanding ejecta has a constant velocity ($r = vt$). The reverse shock velocity, $v_{RS} = v_{FS} - v_{ej}$. For the velocity of the wind, $v_w$, we have considered $v_{FS} >> v_w$, hence ($v_{FS}-v_{w}$) was taken as $v_{FS}$ when calculating the momentum of the CSM. Adding the two terms, we have, 
\begin{equation}
\label{eqn_momentum_csm}
\begin{gathered}
 m'_{csm} \frac{d v_{FS}}{d\mathrm{t}} +  4 \pi r_{FS}^2 \rho_{csm} v_{FS}^2 + m'_{ej}  \frac{d v_{FS}}{d\mathrm{t}} - 4 \pi r_{FS}^2 \rho_{ej}(v_{ej} - v_{FS})^2 = 0  \\
(m'_{csm} + m'_{ej})\frac{d v_{FS}}{d\mathrm{t}}  = M_{sh}\frac{d v_{FS}}{d\mathrm{t}} = 4 \pi r_{FS}^2 [\rho_{ej}(v - v_{FS})^2 - \rho_{csm} v_{FS}^2]  \ \ \ \ \ \ (v  \equiv v_{ej})
\end{gathered}
\end{equation}
Initially the RS travels through the envelope of the ejecta. Let us assume at $t = t_c$ the RS reaches the core of the ejecta. Therefore, for times $t <= t_c$ the mass of the shocked ejecta is, 
\begin{equation}
\label{eqn_momentum_env}
\begin{gathered}
 m'_{ej} = \int_{v t}^{v_M t} 4 \pi r^2 \rho_{e} d\mathrm{r} = 4 \pi \frac{3}{4 \pi} \frac{n-3}{n} \frac{M}{v_{c}^3} v_c^n t^{n-3} \int_{v t}^{v_M t} r^2 r^{-n} d\mathrm{r}  \\
= \frac{3 M (n-3)}{n (3-n)} v_c^{n-3} t^{n-3}[(v_M t)^{3-n} - (v t)^{3-n}] = \frac{3 M}{n} v_c^{n-3} [v^{3-n} - v_M^{3-n}]   \\
= \frac{3 M}{n} \Big[\Big(\frac{v_c}{v}\Big)^{n-3} -  w_c^{n-3}\Big] = \frac{3 M}{n}\Big(\frac{v_c}{v}\Big)^{n-3} =  \frac{3 M}{n} v_c^{n-3}\Big(\frac{t}{r_{FS}}\Big)^{n-3} \ \ \ \ \ \ \ \ \ (w_c << 1)  \\
m'_{CSM} = \int_{0}^{r_{FS}} 4 \pi r^2 \rho_{csm} d\mathrm{r} = \frac{\dot{m}}{v_w} r_{FS} \ \ \ \ \ \ \rightarrow \ \ \ M_{sh}(t <= t_c) = \frac{\dot{m}}{v_w} r_{FS} + \frac{3 M}{n} v_c^{n-3}\Big(\frac{t}{r_{FS}}\Big)^{n-3} 
\end{gathered}
\end{equation}
When the RS is interacting with the core of the ejecta, the mass of the shocked gas is, 
\begin{equation}
\label{eqn_momentum_core}
\begin{gathered}
m'_{ej} = M - \int_{0}^{v t} 4 \pi r^2 \rho_{c} d\mathrm{r} = M -  4 \pi  \frac{3}{4 \pi} \frac{n-3}{n} \frac{M}{v_{c}^3} t^{-3} \int_{0}^{v t} r^2 d\mathrm{r} = M - \frac{n-3}{n} \frac{M}{v_c^3} r_{FS}^3 t^{-3}  \\
M_{sh}(t > t_c) = \frac{\dot{m}}{v_w} r_{FS} + M - \frac{n-3}{n} \frac{M}{v_c^3} \Big(\frac{r_{FS}}{t}\Big)^{3}
\end{gathered}
\end{equation}
The equation for the conservation of momentum for $t <= t_c$ therefore can be written as, 
\begin{equation}
\label{eqn_diff_rFS}
\begin{gathered}
\Big[\frac{\dot{m}}{v_w} r_{FS} + \frac{3 M}{n} v_c^{n-3}\Big(\frac{t}{r_{FS}}\Big)^{n-3} \Big]  \dv[2]{r_{FS}}{t} = 4 \pi r_{FS}^2 \Big[\frac{3}{4 \pi} \frac{n-3}{n} \frac{M}{v_{c}^3} \Big(\frac{v_c}{v}\Big)^n t^{-3} (v - v_{FS})^2 - \frac{\dot{m}}{4 \pi r_{FS}^2 v_w}v_{FS}^2 \Big]  \\
= \frac{3(n-3)}{n} M v_c^{n-3} r_{FS}^{2-n} t^{n-3} \Big(\frac{r_{FS}}{t} -   \dv{r_{FS}}{t} \Big)^2 - \frac{\dot{m}}{v_w} \Big(\dv{r_{FS}}{t}\Big)^2   \\
\Omega r_{FS}\dv[2]{r_{FS}}{t}  + \beta r_{FS}^{3-n}t^{n-3} \dv[2]{r_{FS}}{t}  = (n-3) \beta r_{FS}^{2-n}t^{n-3} \Big(\frac{r_{FS}}{t} -   \dv{r_{FS}}{t} \Big)^2 - \Omega \Big(\dv{r_{FS}}{t}\Big)^2,  \\ 
 \Omega = \frac{\dot{m}}{v_w}, \beta = \frac{3 M}{n} v_c^{n-3}  \\ 
 \end{gathered}
 \end{equation}

 We can simplify the above differential equation using the following equalities, 
 \begin{equation}
\label{eqn_parts_of_differential}
 \begin{gathered}
 \dv{}{t} \Big(\Omega \ r_{FS} \dv{r_{FS}}{t}\Big) = \Omega r_{FS}  \dv[2]{r_{FS}}{t} + \Omega \Big(\dv{r_{FS}}{t}\Big)^2 \\ 
 \dv{}{t}  \Big[\beta \Big(\frac{r_{FS}}{t}\Big)^{3-n} \dv{r_{FS}}{t} \Big] = \beta \Big(\frac{r}{t}\Big)^{3-n} \dv[2]{r_{FS}}{t} + (3-n)\beta r_{FS}^{2-n} t^{n-3} \Big(\dv{r_{FS}}{t}\Big)^2 + (n-3)\beta r_{FS}^{3-n} t^{n-4} \dv{r_{FS}}{t}  \\ 
 = \beta \Big(\frac{r}{t}\Big)^{3-n} \dv[2]{r_{FS}}{t} - (n-3)\beta r_{FS}^{2-n} t^{n-3} \Big(\dv{r_{FS}}{t}\Big)^2 + (n-3)\beta r_{FS}^{2-n} t^{n-3} \frac{r_{FS}}{t} \dv{r_{FS}}{t} \\
  \dv{}{t}  \Big[\beta \frac{n-3}{n-4} \Big(\frac{r_{FS}}{t}\Big)^{4-n}\Big] = \beta \frac{n-3}{n-4} (4-n) r_{FS}^{3-n} t^{n-4} \dv{r_{FS}}{t} + \beta \frac{n-3}{n-4}(n-4) r_{FS}^{4-n} t^{n-5}  \\
  = - (n-3) \beta r_{FS}^{2-n} t^{n-3} \frac{r_{FS}}{t} \dv{r_{FS}}{t} + (n-3) \beta r_{FS}^{2-n} t^{n-3}\Big(\frac{r_{FS}}{t}\Big)^2 
 \end{gathered}
  \end{equation}

Therefore, Equation \ref{eqn_diff_rFS} can be written as, 
\begin{equation}
\label{eqn_omega_beta}
\dv{}{t} \Big[\Omega \ r_{FS} \dv{r_{FS}}{t} + \beta \Big(\frac{r_{FS}}{t}\Big)^{3-n} \dv{r_{FS}}{t} - \beta \frac{n-3}{n-4} \Big(\frac{r_{FS}}{t}\Big)^{4-n} \Big] = 0
\end{equation}
Assuming the solution to be of the form $r_{FS} = S t^{k}$, where $S$ and $k$ are constants, we can express $r_{FS} (t <= t_c)$, 
\begin{equation}
\label{eqn_sh_radius}
k = \frac{n-3}{n-2}, \ \ \ \ \ S = \Big[\frac{2 \beta}{\Omega (n-4)}\Big]^{\frac{1}{n-2}} \ \ \ \ \ \ \rightarrow  \ \ \ \ r_{FS} = \Big[\frac{6 M v_c^{n-3}}{n(n-4)}\frac{v_w}{\dot{m}}\Big]^{\frac{1}{n-2}} t^{\frac{n-3}{n-2}}
\end{equation}
The velocity in this regime ($t <= t_c$) is, 
\begin{equation}
v_{FS} = \dv{r_{FS}}{t} = \frac{n-3}{n-2} \Big[\frac{6 M v_c^{n-3}}{n(n-4)}\frac{v_w}{\dot{m}}\Big]^{\frac{1}{n-2}} t^{-\frac{1}{n-2}}
\end{equation}
The core-edge moves with the velocity $v_c$, so the shock will be at the core-edge at radius $r_{FS} = v_c t_c$. From Equation \ref{eqn_sh_radius}, we have, 
 \begin{equation}
v_c t_c = \Big[\frac{6 M }{n(n-4)}\frac{v_w}{\dot{m}}\Big]^{\frac{1}{n-2}} v_c^{\frac{n-3}{n-2}} t_c^{\frac{n-3}{n-2}} \ \ \ \ \ \ \rightarrow \ \ \ \ \ \ \  t_c = \frac{6 M}{n(n-4)v_c} \frac{v_w}{\dot{m}}
 \end{equation}
 When $t > t_c$, the equation for momentum conservation is expressed using the mass of the shocked ejecta as given by Equation \ref{eqn_momentum_core}. 
 \begin{equation}
\label{eqn_core_momentum}
\begin{gathered} 
\Big[\frac{\dot{m}}{v_w} r_{FS} + M - \frac{n-3}{n} \frac{M}{v_c^3} \Big(\frac{r_{FS}}{t}\Big)^{3}\Big]\dv[2]{r_{FS}}{t} = 4 \pi r_{FS}^2 \Big[\frac{3}{4 \pi} \frac{n-3}{n} \frac{M}{v_{c}^3} t^{-3} (v - v_{FS})^2 - \frac{\dot{m}}{4 \pi r_{FS}^2 v_w}v_{FS}^2 \Big]   \\
= \frac{3(n-3)}{n} \frac{M}{v_c^3} r_{FS}^{2} t^{-3} \Big(\frac{r_{FS}}{t} -   \dv{r_{FS}}{t} \Big)^2 - \frac{\dot{m}}{v_w} \Big(\dv{r_{FS}}{t}\Big)^2   \\
 \end{gathered}
  \end{equation}

There is not any easy analytical form for the solution of this equation, but we solve it numerically to find $r_{FS}$ and $v_{FS}$ when $t > t_c$.

\end{document}